%% file: main.tex
\documentclass[dutch,a4paper]{book}
\usepackage{babel,palatino,a4wide}
\usepackage{color,graphicx,hyperref}
\usepackage{style,links}

\advance\topmargin by -1cm

\begin{document}
\input{voorpagina}

\newpage
\tableofcontents

\input{introductie}
\input{waarom-vakgebied}
\input{waarbinnen-randvoorwaarden}
\input{wat-onderzoek}

\input{wat-opleiding}
\input{wat-onderwijs}

\input{waarmee-beoordeling}

\input{waarmee-kwaliteit}

\appendix
\input{brein}

\bibliography{refs}
\bibliographystyle{plain}
\end{document}

%% file: voorpagina.tex
\begin{titlepage}

\href{http://www.niii.kun.nl}{
{\ExtraHuge \NIIITekst} 
\begin{tabular}[b]{l}
 \Large \sf nijmeegs instituut\\[0.2mm]
 \Large \sf voor informatica en informatiekunde
\end{tabular}}

\begin{center}
  \vfill

  \vfill

  \textcolor{blue}{\ExtraHuge Informatiekunde}
 
  \vspace{0.5cm}

  \textcolor{red}{\Huge Visie 2003}

  \vfill

  \Image{width=7cm}{DaVinci}

  \vfill
\end{center}

\textcolor{blue}{\input{one-liner}}
\vfill

\begin{center}
  \textcolor{red}{\large Versie van: \today}
\end{center}

\end{titlepage}

%% file: one-liner.tex
\begin{tabular}{ll}
  {\large Vormgevers van de digitale samenleving:}\\[0.1cm]
  \mbox{--~}
  Informatiekundigen met een $\beta$-opleiding, een $\gamma$-feeling
  en een gezonde dosis creativiteit
\end{tabular}

%% file: introductie.tex
\chapter{Introductie}
\label{h:introductie}

\section{Doel van dit document}
Dit document heeft betrekking op de visies die ten grondslag liggen aan
het informatiekunde \emph{onderwijs} en \emph{onderzoek} binnen het Nijmeegs
Instituut voor Informatica en Informatiekunde (\NIII). Het uiteindelijke doel
van dit document is het bieden van een `repository' met betrekking tot deze
visies, en een basis voor de specifieke inrichting van het curriculum van
de opleiding en de onderzoeksplannen.

Daar informatiekunde voor het \NIII\ een relatief nieuwe opleiding en
onderzoeksgebied is, is er in de huidige (2003) versie van dit document primair
aandacht voor de informatiekunde opleiding. Het ligt in de lijn der
verwachtingen dat er in de komende jaren, in updates van dit document, ook meer
aandacht besteed zal worden aan het informatiekunde onderzoek.

Het feit dat dit document jaarlijks een update kan ondergaan betekent overigens
niet dat we verwachten dat er jaarlijks een koerswijziging zal plaatsvinden.
De ambitie met betrekking tot de stabiliteit van hetgeen in dit document word
is 5 \`a 6 jaar. In de huidige versie geldt dit specifiek voor de visie op de
informatiekunde opleiding. Het onderzoeksdeel van dit document zal in de
komende jaren nog specifieker ingevuld moeten worden.

\section{Doelgroep}
De doelgroep van dit document bestaat in eerste instantie uit de bestuurders
van de Faculteit NWI, de collega's binnen het \NIII\, en de collega's van
externe opleidingen, die bij informatiekunde opleidingen en/of onderzoek
betrokken zijn (de interne klankbordgroep, inclusief 4 studentleden).
In tweede instantie behoren daartoe ook vertegenwoordigers van
zusterfaculteiten en -opleidingen elders, en vertegenwoordigers uit het
werkveld (de externe klankbordgroep). Het stuk kan in een later stadium als
basis dienen voor het genereren van verschillende teksten voor andere
doelgroepen, zoals (potenti\"ele) studenten, de visitatiecommissie e.d. Op dit
moment wordt echter nog geen volledigheid/leesbaarheid voor die doelgroepen
nagestreefd.

\section{Ontstaan van dit document}
Het startpunt van dit document bestond uit:
\begin{itemize}
  \item Diverse presentaties en documenten met betrekking tot de
    informatiekunde opleiding en onderzoek, van de hand van diverse
    auteurs.
  \item Ervaringen met Curricula 2000, 2001 en 2002.
  \item Standaard curriculum Informatiemanagement van de ACM en de IEEE.
\end{itemize}
Het huidige document werd geproduceerd door een kernteam bestaande uit (in
alfabetische volgorde):
\begin{itemize}
  \item \ErikB\ (vertegenwoordiger van de afdeling \AfdGS\ en liaison informatica).
  \item \StijnH\ (editor).
  \item \VeraK\ (editor).
  \item \ErikP\ (editor en eindverantwoordelijke).
  \item \JanT\ (vertegenwoordiger van de afdeling \AfdST).
  \item \TvdW\ (vertegenwoordiger van de afdeling \AfdIRIS).
  \item \Hanno\ (vertegenwoordiger van de afdeling \AfdITT).
\end{itemize}
Het daadwerkelijke schrijfwerk is met name door \VeraK, \ErikP, en \StijnH\
gedaan. Tevens is gebruik gemaakt van twee klankbordgroepen.  Een (KUN) interne
klankbordgroep bestaande uit:
\begin{itemize}
  \item Informatiekundestudenten: Jeroen Groenen (1e jaars), Mark Jenniskens
  (2e jaars), Arnoud Vermeij (3e jaars), Wout Lemmens (HBO-instromer).
  \item \OCie\ (OpleidingsCommissie) liaison:
  Mark Jenniskens.
  \item \NSM: Bart Prakken.
  \item \MIK: Hans ten Hoopen.
  \item \IOWO: Bea Edlinger, G\'e Ophelders.
  \item \NIII: Jeroen Bruijning, Bas van Gils, Franc Grootjen, Stijn
  Hoppenbrouwers, Pieter Koopman, Martijn Oostdijk, Ger Paulussen, Rinus
  Plasmeijer, Eric Schabell, Frits Vaandrager, Gert Veldhuijzen van Zanten,
  Paul de Vrieze.
\end{itemize}
Tevens werd er een externe klankbordgroep ingeschakeld bestaande uit vertegenwoordigers
van zusterfaculteiten en vertegenwoordigers uit het werkveld:
\begin{itemize}
  \item Hans Bossenbroek, \Luminis.
  \item Jeroen Top, \Belastingdienst.
  \item Victor van Reijswoud, \UMU.
  \item Roel Wieringa, \UT.
\end{itemize}

In eerste instantie was dit visiedocument volledig ge\"{\i}ntegreerd met het document wat
het informatiekunde curriculum 2003 beschrijft. Wegens de uitgebreidheid van de visie op
het vakgebied, en de zelfstandige rol die deze visie kan vervullen richting opleiding \emph{en}
onderzoek, is uiteindelijk besloten deze documenten te splitsen. In~\cite{Curriculum2003}
is het curriculum van de informatiekunde opleiding (zowel de Bacherlorfase als de Masterfase)
te vinden.

\section{Gevolgde redeneerlijn}
Bij het structureren van dit document is er voor gekozen om een structuur
te gebruiken die het mogelijk maakt om:
\begin{itemize}
   \item in een aantal controleerbare stappen vanuit een visie op het vakgebied
         te komen tot de uiteindelijke inrichting van de opleiding,
   \item een duidelijke relatie te leggen tussen de structuur van de uiteindelijke opleiding
         en de rol (en bijbehorende vaardigheden) die studenten na afloop van
         hun studie kunnen vervullen.
\end{itemize}
Het laatste punt komt voort uit de hypothese dat de motivatie van een student
tijdens de studie hoger zal zijn als steeds duidelijk is hoe een vak bijdraagt
aan het doel waarmee de student gekozen heeft om de opleiding te volgen.
Zoals eerder aangegeven is het uiteindelijke curriculum te vinden in~\cite{Curriculum2003}.

De redeneerlijn die werd aangehouden bij het opstellen van dit visie document, en
het bijpassende curriculum, is daarom als volgt:
\begin{enumerate}
    \item Er werd een aantal visies geformuleerd: op het vakgebied van de
    informatiekunde, op de opleiding informatiekunde, het onderzoek, op
    onderwijs en leren, op beoordeling en op kwaliteitsbeheersing.
    \item Er werden randvoorwaarden beschreven: vaststaande factoren waar
    rekening gehouden zal moeten worden.
    \item Samen waren alle visies plus de randvoorwaarden input voor de
    inrichting van het curriculum.
    \item Vanuit de visies en de randvoorwaarden werden expliciet verbanden
    gelegd naar de inrichting. Dit gebeurde met behulp van:
    \begin{itemize}
       \item formulering van \emph{vereiste vaardigheden};
       \item formulering van \emph{inrichtingsprincipes}.
    \end{itemize}
\end{enumerate}
Dit document richt zich primair op stappen 1 en 2. De andere stappen zijn
terug te vinden in~\cite{Curriculum2003}.

We zullen nu eerst nader ingaan op wat we bedoelen met `vaardigheden' en
`inrichtingsprincipes'. In navolging van huidige didactische inzichten wordt
het wenselijk geacht het curriculum te richten op het aanleren van de
vaardigheden die een informatiekundige moet beheersen. Vaardigheden betreffen
`dingen die iemand moet \emph{kunnen}' (niet zo zeer \emph{kennen}). Het
formuleren van vaardigheden legt dus de nadruk op het actief toepassen van
kennis, en minder op het kunnen `beschrijven' van die kennis (dit wordt wel
eens omschreven als `knowledge how' versus `knowledge that'). De standaard
frase die bij het formuleren van vaardigheden steeds wordt gebruikt is `Een
informatiekundige moet in staat zijn om ...'. Zelfs vereisten op het gebied van
theoretische kennis worden daarbij in actieve zin verwoord, bijvoorbeeld `een
informatiekundige moet in staat zijn om theorie $X$ toe te passen op een casus,
en hierover op constructieve wijze te reflecteren'.

Er kunnen verschillen zijn in het \emph{ambitieniveau} waarop een bepaalde
vaardigheid moet worden aangeleerd. Met andere woorden, er moet worden
vastgesteld \emph{hoe goed} een informatiekundige bepaalde vaardigheden dient
te beheersen. Dat kan bijvoorbeeld uiteenlopen van oppervlakkige, passieve beheersing
tot aan diepgaande, actieve beheersing. Pas wanneer een vaardigheid gekoppeld wordt
aan een ambitieniveau kan bepaald worden wat de \emph{eindtermen} van het curriculum
zijn. Een eindterm kan dus beschreven worden als `vaardigheid + ambitieniveau'; het
is een duidelijk omschreven onderwijsdoel.

De vaardigheden mogen dan de basis zijn van de inrichting van het curriculum, ze zijn
niet het enige ingredient. Vanuit de verschillende visies en randvoorwaarden die aan
het curriculum ten grondslag liggen worden ook een aantal \emph{inrichtingsprincipes}
geformuleerd waaraan het curriculum moet voldoen. Die principes bepalen grotendeels
\emph{hoe} de onderwijsdoelen moeten worden bereikt. Inrichtingsprincipes dienen in het
algemeen te voldoen aan twee basisvoorwaarden:
\begin{itemize}
    \item Ze moeten zo SMART mogelijk zijn (Specific, Measurable, Attainable, Realisable,
    Timely).
    \item Ze mogen niet triviaal zijn: ze vertegenwoordigen een standpunt dat
    mogelijk discussie aantrekt.
\end{itemize}

Er is bijvoorbeeld een inrichtingsprincipe geformuleerd met
betrekking tot het doel van de propedeuse: `Het eerste jaar biedt,
naast een methodologische basis, een brede ori\"entatie op de
verschillende verbredingsgebieden, en illustreert de samenhang van
de verschillende kanten van de studie. Er is nog geen sprake van
specialisatie- of keuzevakken.'

Samengevat kunnen alle onderdelen van de visie en alle randvoorwaarden dus op
twee manieren input voor het curriculum vormen: in de vorm van vaardigheden van
de informatiekundige (\emph{wat} het curriculum moet bevatten) en in de vorm
van inrichtingsprincipes voor het curriculum (\emph{hoe} het ingericht moet
worden).  De vaardigheden en inrichtingsprincipes vormen een expliciete
verbinding tussen de visies en randvoorwaarden aan de ene kant, en de
eindtermen aan de andere. Zo kunnen de eindtermen van de inrichting altijd
herleid worden tot een bepaalde visie of een bepaalde randvoorwaarde.

\section{Structuur van dit document}
In dit document worden in een aantal stappen onze visie op de informatiekunde
opleiding, het informatiekunde onderzoek, en de achter\-lig\-gen\-de motivaties uiteengezet:
\begin{description}
  \item[Wat is nodig --] Eerst richten we ons op de vraag wat de behoeften zijn achter
        de opleiding informatiekunde, door een visie (en missie) te formuleren op
        de rol van een informatiekundige en de opleiding en het onderzoek wat hiervoor nodig
        is (hoofdstuk~\ref{h:waarom-vakgebied}).

  \item[Waarbinnen --] Alvorens de inrichting van de opleiding nader uit te werken,
        staan we in hoofdstuk~\ref{h:waarbinnen-randvoorwaarden} stil bij de
        kaderscheppende randvoorwaarden \emph{waarbinnen} we bij de inrichting van
        de opleiding dienen te opereren.
        Deze randvoorwaarden kunnen onder andere voortkomen uit wetgeving van locale
        of Europese overheden, universitair of facultair beleid, en beleid van het
        \NIII.

  \item[Wat doen --] Vervolgens buigen we ons over de vraag \emph{wat} er moet gebeuren
       in de informatiekunde opleiding en het onderzoek. Dit gedeelte bestaat uit drie onderdelen:
        \begin{itemize}
      \item Visie op onderzoek (hoofdstuk~\ref{h:wat-onderzoek}).
          \item Visie op de opleiding (hoofdstuk~\ref{h:wat-opleiding}).
          \item Visie op onderwijs en leren (hoofdstuk~\ref{h:wat-onderwijs}).
        \end{itemize}

  \item[Waarmee --] Dan komen we toe aan het beschrijven van enige middelen die nodig
       zijn om het curriculum operationeel uit te voeren.
       De volgende hoofdpunten komen aan de orde:
        \begin{itemize}
            \item Visie op beoordeling (hoofdstuk~\ref{h:waarmee-beoordeling}).
            \item Visie op kwaliteitsbeheersing (hoofdstuk~\ref{h:waarmee-kwaliteit}).
        \end{itemize}
\end{description}
Het \emph{hoe} van de opleiding informatiekunde is, tezamen met een samenvatting van het
bovenstaande, terug te vinden in~\cite{Curriculum2003}.

Merk op dat de visies en de inrichting met betrekking tot:
\begin{itemize}
   \item Onderwijs \& leren
   \item Beoordeling
   \item Kwaliteitsbeheersing
\end{itemize}
uiteindelijk vervat zullen moeten worden in een \NIII-breed `standaarddocument'
dat onverkort voor beide \NIII\ opleidingen zal gaan gelden.  Echter, zo lang
zo'n gemeenschappelijk document nog ontbreekt, is het noodzakelijk om deze
zaken expliciet in het huidige document op te nemen.

In een appendix wordt ook nog uitgebreidere informatie verschaft over enige punten
waar in de tekst naar verwezen wordt.

%% file: waarom-vakgebied.tex
\chapter{Visie op het vakgebied}
\label{h:waarom-vakgebied}
De introductie van de computer heeft in eerste instantie geleid tot het
wetenschapsgebied van de informatica. Naarmate dit wetenschapsgebied beter
begrepen werd, en computers meer en meer ge{\itr}ntegreerd werden in de
samenleving, ontstond de behoefte aan een dieper inzicht in de relatie tussen
computationele technologie en de context waarin deze wordt gebruikt. Zo
ontstond de informatiekunde: een wetenschapsgebied waarin, naast de
informatica, ook organisatorische en menselijke factoren centraal staan. Het
ontstaan van de informatiekunde als vakgebied heeft op zich ook weer invloed
op het vakgebied informatica.

In dit hoofdstuk worden de achtergronden van de informatiekunde en onze visie
hierop nader belicht. Achtereenvolgens staan we stil bij:
\begin{itemize}
  \item de maatschappelijke ontwikkelingen die het vakgebied van de informatica
  de informatiekunde be\"invloeden (paragraaf~\ref{p:InfMij});
  \item de basale verschillen tussen de vakgebieden informatica en
  informatiekunde en hun beider relatie tot informatiesystemen (paragraaf
  \ref{p:InfInf});
  \item het gemeenschappelijk speelveld van de informatica en informatiekunde
  (paragraaf \ref{p:speelveld})
  \item de symbiose van informatica en informatiekunde
  (paragraaf~\ref{s:symbiose});
  \item de marktpotentie voor afgestudeerde informatiekundigen
  (paragraaf~\ref{p:Markt});
  \item informatiekundige verbredingsgebieden (paragraaf~\ref{p:Toep});
  \item de missie en het profiel van informatiekunde binnen het \NIII\
  (paragraaf~\ref{p:Missie});
  \item de verankering van informatiekunde binnen de KUN
  (paragraaf~\ref{p:Veranker});
  \item de rol die informatiekundigen na hun studie gaan spelen en de
  vaardigheden die daarvoor nodig zijn (paragraaf \ref{p:InfVaard}).
\end{itemize}

\section{Maatschappelijke ontwikkelingen}
\label{p:InfMij}
We leven in een digitale samenleving!  Als vervolg op de algemene
`technologisering' en industrialisering van de samenleving die in de afgelopen
twee eeuwen plaatsvond kent onze huidige samenleving een sterk groeiende
afhankelijkheid van Informatie- en CommunicatieTechnologie (ICT). In veel
aspecten van ons leven neemt ICT (de digitale technologie) een voorname plaats
in. Hierbij gaat het niet alleen om voor de hand liggende voorbeelden van ICT,
zoals een tekstverwerker of een internetsite. Ook allerlei andere vormen van
technologie raken steeds meer `doordrenkt' met ICT. Van wasmachines tot
rolstoelen, van auto's tot vliegtuigen, van bibliotheek tot bushalte, van
elektronische agenda tot gemeentelijke bevolkingsadministratie, overal vinden
we technologische ondersteuning die niet meer zonder ICT kan.  Zelfs gebouwen
worden dankzij computers steeds slimmer (z.g.\ `smart buildings'), onder andere
door hoogwaardige beveiliging, voortdurende bewaking van het interne klimaat,
etc.  \emph{Daarom gebruiken we de term ICT in brede zin, waarbij nadrukkelijk
niet alleen `computertechnologie' bedoeld wordt, maar alle gevallen waarin
technologische oplossingen afhangen van van gecomputeriseerde onderdelen}. De
term dekt dus zowel informatievoorzienende en -verwerkende systemen die typisch
zijn voor administratieve organisaties (zoals banken, verzekeraars,
overheidsinstanties, etc.), als systemen die bedoeld zijn voor de besturing en
controle van andere vormen van technologie (zoals vliegtuigen, robots, etc.).
Kortom alle belangrijke aspecten van de digitale samenleving.

Traditioneel ligt de focus van de Nijmeegse informatica op het maken van
correcte softwaresystemen. Een softwaresysteem wordt als correct beschouwd
wanneer het voldoet aan alle vooraf gestelde kwaliteitseisen. Hierbij worden
expliciet niet alleen de zogenaamde \emph{functionele} kwaliteitseisen bedoeld,
maar juist ook de niet-functionele kwaliteitseisen zoals reactietijd,
onderhoudbaarheid, maakbaarheid, flexibiliteit, beheerbaarheid, bruikbaarheid,
robuustheid, etc. In praktische situaties blijkt het bepaald geen triviale zaak
om alle kwaliteitseisen waaraan een softwaresysteem moet voldoen, precies vast
te stellen. Een belangrijke reden hiervoor is dat softwaresystemen vaak dermate
sterk verweven zijn met het dagelijkse leven en werk, dat men bij het opstellen
van de kwaliteitseisen rekening dient te houden met een ruime schakering aan
aspecten, zoals bedrijfsvoering, mensen, organisatie en locatie.

Een extra probleem hierbij is dat veel van deze aspecten in onze huidige
maatschappij aan allerlei veranderingsprocessen onderhevig zijn. Dit stelt
extra eisen aan de softwarematige ondersteuning van zowel de processen als van
de veranderingen. Bovendien staan de verschillende klassen van kwaliteitseisen
vaak ook nog eens op gespannen voet met elkaar. Bij voorbeeld kan vanuit
bedrijfseconomische redenen de voorkeur worden gegeven aan een softwaresysteem
dat binnen \emph{\'e\'en maand} met 80\% betrouwbaarheid 90\% van de
functionaliteit kan bieden, boven een softwaresysteem dat over een \emph{half
jaar} met 99\% betrouwbaarheid 100\% van de functionaliteit kan bieden. Het op
een verantwoorde manier maken van dergelijke keuzes is doorgaans niet triviaal,
en de doorslaggevende overwegingen zullen vaak geen technologische zijn.
Allerlei vragen worden daarbij opgeroepen: \emph{Wat kosten die laatste procenten
functionaliteit en betrouwbaarheid in termen van gederfde omzet bij vijf
maanden vertraging? Hoe erg is het eigenlijk als het systeem fouten maakt?
Kunnen we ons daar niet tegen verzekeren? Is dat niet goedkoper?}. We zullen
dus verder moeten kijken dan de techniek alleen.

In onze westerse samenleving is ICT inmiddels alom aanwezig, en van vitaal
belang voor economie en maatschappij. Financi\"ele instellingen zouden binnen
enkele dagen, zo niet uren, failliet gaan als hun ICT systemen gedurende
langere tijd uit zouden vallen. Datzelfde geldt voor veel andere
bedrijfstakken. Het samenspel tussen mens, organisatie, en ICT wordt dan ook
steeds kritieker. Tot slot lijkt men ook in de zich nog ontwikkelende delen van
de wereld snel de sprong naar het ICT tijdperk te willen wagen.\footnote{Zie
bijvoorbeeld studies van het International Institute for Communication and
Development (IICD), \IICDURL.} ICT is dus geen luxe-artikel meer, het is een
noodzakelijk bestanddeel geworden van vele vormen van maatschappelijke
ontwikkeling.

In het licht van de genoemde ontwikkelingen wordt het belang van een goede
afstemming tussen mens, organisatie en technologie dan ook in toenemende mate
duidelijk.

\section{Informatica, informatiekunde en informatiesystemen}
\label{p:InfInf}
Hoewel in de bedrijfspraktijk de grens tussen het vakgebied van de
informatica en de informatiekunde vaak niet zo duidelijk getrokken wordt, is
het verstandig en noodzakelijk om een helder onderscheid te maken teneinde het
bestaansrecht van beide vakgebieden (en de Nijmeegse opleidingen die daarmee
verbonden zijn) te staven. In onze visie zijn de vakgebieden complementair. Ze
hebben een gemeenschappelijke inhoudelijke kern, maar verschillende
vertrekpunten. Hieronder gaan we nader in op de verschillen tussen de twee
vakgebieden.

\subsection{Verschillende vertrekpunten}
Zoals de hierboven gegeven schets van maatschappelijke ontwikkelingen duidelijk
maakt, heeft de toenemende inbedding van de ICT in onze maatschappij geleid tot
een situatie waarin expertise op het gebied van de ICT zelf niet langer het
optimaal functioneren ervan garandeert. Er is een groeiende behoefte ontstaan
aan een nieuw soort expertise, die \emph{aanvullend} is op het traditionele
wetenschapsgebied van de `computing science', en die deze in haar
organisatorische en menselijke context plaatst. Die nieuwe expertise behoort in
de visie van het \NIII\ in de eerste plaats toe aan het terrein van de
informatiekunde, terwijl computing science meer met informatica wordt
geassocieerd.

We moeten echter wel beseffen dat ook de informatica de laatste jaren meer is
geworden dan slechts \emph{computing} science. Ook binnen dit vakgebied is men
zich er steeds meer van bewust dat ICT in een menselijke en organisatorische
context ingebed dient te worden. Echter, waar de informatica de grondslagen van
de ICT als startpunt neemt om zich vervolgens af te vragen hoe deze optimaal is
in te passen in een organisatorische en/of menselijke context, beweegt de
informatiekunde zich juist in omgekeerde richting.

Dit verschil in focus komt ook tot uiting in de omschrijvingen die door de
verkenningscommissie informatica, resp.\ de VSNU worden gehanteerd. De
verkenningscommissie informatica~\cite{InformaticaVerkenning} geeft de volgende
omschrijving van het wetenschapsgebied der informatica:
\begin{quote} \it
  Informatica is de wetenschap die zich bezighoudt met de theorie\"en, methoden
  en technieken voor het voortbrengen en in stand houden van informatiesystemen
  met nadruk op de architectuur en de softwarecomponenten van zulke systemen.

  Informatiesystemen realiseren de informatievoorziening van organisaties,
  individuen en apparaten door middel van generatie, opslag, interpretatie,
  transformatie, transport en presentatie van gegevens, in de
  verschijningsvormen tekst, beeld of geluid.
\end{quote}

De definitie van het wetenschapsgebied der informatiekunde zoals deze door het
VSNU is opgesteld luidt:
\begin{quote} \it
  Informatiekunde richt zich op theorievorming en onderzoek naar het effectief
  structureren, verwerken en communiceren van informatie en de rol die de
  informatietechnologie daarbij speelt. Informatieprocessen bij individuen en
  organisaties worden niet alleen uit technisch, maar ook uit cognitief,
  sociaal en bedrijfskundig perspectief bezien.
\end{quote}

Kortweg zouden we kunnen stellen dat informatica als vertrekpunt heeft:
\begin{quote} \it
  Betrouwbare informatiesystemen, om mensen en organisaties te ondersteunen.
\end{quote}
terwijl de informatiekunde als vertrekpunt heeft:
\begin{quote} \it
  Mensen en organisaties ondersteunen met betrouwbare informatiesystemen.
\end{quote}

Een andere manier om dit verschil in insteek te verwoorden, is dat het in de
informatica allereerst gaat om

\begin{quote} \it
  Building the system right.
\end{quote}
terwijl de informatiekunde als insteek heeft:
\begin{quote} \it
  Building the right system.
\end{quote}

\subsection{Informatiesystemen}
Wellicht is het de lezer opgevallen dat we naast het technologie-gerelateerde
begrip `ICT' het meer algemene begrip `informatiesysteem' hanteren. Dit behoeft
enige verklaring. Informatiesystemen zijn het kernthema van zowel de
informatica als de informatiekunde. Het zijn systemen die als doel hebben het
structureren, verwerken, en communiceren van informatie. Deze
informatiesystemen kunnen op zich weer onderdeel zijn van andere systemen.
Bijvoorbeeld als uitvoerend systeem binnen administratieve organisaties, of als
een besturend systeem van een vliegtuig.

Merk op dat een informatiesysteem niet alleen uit ICT \emph{hoeft} te bestaan.
Hoewel technologie vaak een belangrijke rol speelt in informatiesystemen, hoeft
het hierbij niet eens perse om computertechnologie te gaan.  Zo is een
eenvoudige, klassieke kaartenbak een prima voorbeeld van een informatiesysteem,
en een zogenaamde `scheepstelegraaf' (vooruit - STOP - achteruit) van een eenvoudig
communicatiesysteem.  Belangrijker nog is het besef dat ook non-technologische
elementen deel uit kunnen maken van een informatiesysteem.  Onderzoek naar
informatiesystemen heeft ons geleerd dat technische systemen in hun
functioneren vaak zodanig zijn verweven met hun (menselijke) gebruikers dat een
informatiesysteem feitelijk de machinerie en menselijke actoren in zich
verenigt. Er is in die gevallen geen sprake van een louter technisch systeem,
maar van een \emph{socio-technisch systeem}: een combinatie van mensen en
machines die in onderlinge samenwerking een functioneel geheel vormen. Hierbij
spelen menselijke en organisatorische aspecten dus een even cruciale rol als de
ICT. Dat is op zich misschien geen verrassing, maar het heeft forse
consequenties voor de vakgebieden die zich met informatiesystemen bezighouden.
Wanneer we alle verschillende facetten willen combineren in een coherente en
gedetailleerde visie op informatiesystemen, ontstaat namelijk een zeer en
complex samenspel van diverse elementen en factoren --complexer en diverser
dan bij puur technische systemen.

Informatiesystemen kunnen vele verschillende doelen dienen, en vanuit
verschillende invalshoeken benaderd worden. Al eerder maakten we een
onderscheid maken tussen \emph{uitvoerende} systemen (bijv. een
loonadministratiesysteem of een voorraadbeheersysteem) en \emph{besturende}
systemen (bijv. een besturingssysteem voor een onbemenst vliegtuig of een
workflow management systeem). Verder kunnen we onderscheid maken tussen
\emph{productiesystemen} (bijv. een klantenregistratiesysteem of cockpit van
een vliegtuig) en \emph{productsystemen} (bijv. een DVD speler, DVD
bibliotheek, klantregistratiemodule of personal organiser). Het laatste
onderscheid is overigens afhankelijk van het perspectief van waaruit men kijkt.
Voor een \emph{producent} is een personal organiser bijvoorbeeld een
productsysteem, voor de \emph{gebruiker} is diezelfde organiser een
productiesysteem. Hoe een systeem bezien wordt kan grote invloed hebben op de
prioritering van systeemdoelen en op de manier waarop het systeem ontwikkeld
wordt.

\subsection{Kwaliteit van informatiesystemen}
Kwaliteit (met name de betrouwbaarheid) van software is al sinds jaren het
bindende thema binnen de Nijmeegse informatica. Voor het nieuwere vakgebied van
de informatiekunde kan dit thema worden verlegd naar kwaliteit van
informatiesystemen. Daarbij wordt de kwaliteit van ICT dus beschouwd in
samenspraak met haar menselijke en organisatorische context. Verschillende
kwaliteitseisen en verwachtingen `van buitenaf' spelen daarbij een grote rol --
groter nog dan bij de informatica. Daarnaast dienen de diverse menselijke,
organisatorische en technologische ingredi\"enten zorgvuldig op elkaar te
worden afgestemd zodat een soepel samenspel ontstaat. Hoe kritiek dit samenspel
soms kan zijn, en welke aspecten hierbij zoal een rol spelen, kunnen we
illustreren aan de hand van drie voorbeelden: de luchtvaart, het bedrijfsleven,
en de privacy van de burger.

\subsubsection{Met veilige technologie ben je er nog niet!}
In de luchtvaart is een juiste afstemming tussen mens, organisatie en ICT vaak
van levensbelang. De meeste nieuwe vliegtuigen maken gebruik van `fly-by-wire'
technologie, waardoor de `automatische piloot' steeds automatischer wordt. Zo
wordt bijvoorbeeld de traditionele hydraulische besturing vervangen door
elektrische systemen die worden aangestuurd door een boordcomputer, geholpen
door computers op de grond.  Het is niet voldoende de correctheid van
dergelijke software en de onderliggende elektrische systemen te bewijzen om te
kunnen concluderen dat de aansturing veilig is. Na vele vliegtuigongelukken
waaraan geen technisch falen ten grondslag lag is men er meer dan ooit van
overtuigd geraakt dat de rol van de `menselijke factor' zelfs niet ten dele
uitgevlakt mag worden.  Daarbij gaat het niet om klassieke `pilot error': een
onvergefelijke fout van de piloot; men gaat zich steeds meer afvragen wat er
redelijkerwijs van het menselijk functioneren verwacht mag worden.  Kortom,
zelfs als de software en hardware doen wat ze behoren te doen, betekent dat
niet dat het gehele systeem (het vliegtuig met piloten en passagiers)
betrouwbaar is.\footnote{Zie bijvoorbeeld: \BoeingURL\ of \AirbusURL.} Om
zekerder -- want absolute zekerheid bestaat niet -- te zijn van de veiligheid
van vliegtuigen, moet men niet slechts naar de technologie kijken, maar naar
het systeem \emph{als geheel}. Dat wil zeggen: inclusief piloten, overig
personeel, passagiers en grondpersoneel (zowel tijdens als v\'o\'or de vlucht).

\subsubsection{Afstemming tussen bedrijfsvoering en ICT; rempedaal of gaspedaal?}
Ook bij de bedrijfsvoering van bedrijven en organisaties zien we het belang van
een juiste afstemming tussen de ICT en haar menselijke en organisatorische
omgeving. Veel organisaties worstelen met de vraag hoe zij ICT optimaal
kunnen inzetten ten behoeve van hun bedrijfsvoering.

Moderne organisaties moeten opereren in een voortdurend veranderende
(evoluerende) omgeving. De liberalisering van markten, de vermindering van
protectionisme, de privatisering van staatsbedrijven, de toenemende wereldwijde
concurrentie, grensoverschrijdende bedrijfsfusies, het ontstaan van nieuwe
economische blokken, de invoering van gemeenschappelijke munteenheden; al deze
aspecten dragen bij aan de dynamiek van het huidige ondernemersklimaat.

De druk om fundamentele veranderingen aan te brengen in bestaande systemen
wordt steeds groter. Het `jaar 2000 probleem' en de invoering van de Euro waren
twee in het oog springende voorbeelden van ontwikkelingen die grootschalige
veranderingen van bestaande informatiesystemen vereisten om deze `in de running
te houden'. Schijnbaar minder dwingend maar ook cruciaal is de
technologie-gedreven vernieuwing. De invoering van nieuwe communicatiemiddelen
zoals call-centres, E-commerce en Mobile-commerce (Imode/Wap/WiFi) is een
typisch voorbeeld van een ontwikkeling die geen directe bedreiging vormt voor
organisaties, maar die deze in staat stelt (of zou moeten stellen) om nieuwe
vormen van commercie te exploreren -- hetgeen voor hen vaak van levensbelang
is, bijvoorbeeld in een concurrentiestrijd. Ook zijn vele organisaties
bezig te transformeren van een monolithische structuur naar een
netwerkstructuur; de zogenaamde netwerkorganisatie. Een dergelijke
organisatorische transformatie heeft doorgaans een enorme invloed op de in
gebruik zijnde informatiesystemen.

In het ideale geval biedt de ICT aan bedrijven de ondersteuning en de
stimulansen waarmee zij veranderingen teweeg kunnen brengen om nieuwe kansen en
uitdagingen aan te gaan. E\'en van de paradoxale problemen waar veel bedrijven
momenteel mee worstelen is echter het feit dat de in een bedrijf reeds
aanwezige ICT vaak eerder een remmende werking lijkt te hebben op het vermogen
van een organisatie om vernieuwingen door te voeren, dan dat deze de
veranderingsprocessen daadwerkelijk ondersteunt of stimuleert. Daarnaast blijkt
keer op keer dat de invoering van nieuwe technologie dermate veel onzekerheden
met zich meebrengt, dat de voortschrijdende technologische ontwikkelingen zelf
een constante bedreiging vormen voor de stabiliteit van een organisatie.

\subsubsection{Gegevensbescherming en privacy}
De ontwikkeling van een `digitale samenleving' stelt niet alleen grote eisen aan
de betrouwbaarheid en veiligheid van de gebruikte ICT als zodanig, maar ook
aan het praktische beheer van de ontstane gegevensbestanden. Hierbij speelt de
privacy-gevoeligheid van de informatie die wordt geregistreerd een belangrijke
rol.

Denk bijvoorbeeld aan de communicatie tussen centrale of gemeentelijke
overheden en burgers via het digitale loket. De belastingdienst, het
verzekeringswezen, grote advocatenkantoren, de rechterlijke macht, diverse
uitkerende instanties: voor al deze diensten is de correcte en veilige
verwerking van privacy-gevoelige gegevens essentieel. Basisfouten in de
informatisering leiden regelmatig tot grote kostenposten; daarnaast speelt ook
niet-materi\"ele schade een rol (het maatschappelijk leed en de hoeveelheid
frustratie en stress en die het gevolg kunnen zijn van de foute verwerking van
gegevens moeten niet worden onderschat). De juridische implicaties van deze
issues zijn groot.

\subsubsection{Kwaliteitseisen}
Voor de beoordeling van de kwaliteit van socio-technische systemen zijn
verschillende soorten kwaliteitseisen relevant. Allereerst zijn er de
\emph{globale kwaliteitseisen}, die (van buitenaf) kunnen worden gesteld aan
het systeem als geheel.  Bijvoorbeeld, in de geschetste `fly-by-wire' situatie
is de uiteindelijke wens een besturingssysteem te realiseren dat het vliegtuig
op betrouwbare wijze zal besturen. Typische globale klassen van kwaliteitseisen
zullen dan zijn:
\begin{itemize}
  \item Voldoen aan de door de International Air Transport Association
  (\href{http://www.iata.org}{IATA}) gestelde veiligheidseisen.
  \item De vliegervaring voor de passagiers zo prettig mogelijk maken
  (vermijden van onnodige turbulentie, vertragingen en gemiste aansluitingen).
  \item Het vliegen economisch zo effici\"ent mogelijk uitvoeren (besparen op
  brandstof, slijtage en vliegtijd).
\end{itemize}

Naast deze kwaliteitseisen die van buitenaf aan het systeem worden gesteld,
zijn er ook kwaliteitseisen die gelden voor de systeem-interne componenten
afzonderlijk. Zo zal de piloot capabel moeten zijn, en bijvoorbeeld niet
dronken terwijl hij het vliegtuig bestuurt; het vliegtuigmaterieel zal in goede
conditie moeten zijn, en de technologie als zodanig zal moeten werken. Tot slot
zullen de systeem- interne componenten ook eisen/verwachtingen aan elkaar
stellen. Voor de menselijke en ICT component kunnen we hierbij bijvoorbeeld
denken aan de volgende:
\begin{itemize}
  \item De piloot wordt geacht om te kunnen gaan met de interface.
  \item Het systeem dient een gebruikersvriendelijke interface te bieden.
  \item De piloot dient de computer te erkennen als co-piloot, en erop te durven
  vertrouwen.
  \item De computer dient de taken en de cognitieve processen van de piloot
  optimaal te ondersteunen.
\end{itemize}

Om de kwaliteit op deze verschillende vlakken te kunnen beoordelen, is het
derhalve niet alleen nodig dat we inzicht hebben in de wijze waarop de
afzonderlijke componenten functioneren, maar tevens in de wijze waarop zij --
inhoudelijk en temporeel -- optimaal op elkaar kunnen worden
afgestemd.\footnote{Het zal duidelijk zijn dat er aan een dergelijke
rolverdeling tussen mens en computer uitgebreide cognitieve en ethische
discussies ten grondslag liggen; discussies die maar beter tijdig gevoerd
kunnen worden.}

De drie genoemde voorbeelden illustreren twee belangrijke aspecten die van
belang zijn bij een juiste afstemming tussen mens, organisatie en ICT. De
\emph{taken} \emph{verantwoordelijkheden} binnen een systeem moeten optimaal zijn
verdeeld. Idealiter vullen mens, organisatie en technologie elkaar hierbij aan,
elkaars fouten corrigerend, waarbij elke component kan excelleren in haar eigen
kern-competentie. Bijvoorbeeld, waar ICT-ondersteuning de structuur kan
aanbrengen waardoor menselijke tekortkomingen of fouten bijtijds kunnen worden
gesignaleerd, moeten mogelijke onzekerheden in het functioneren van de
technologie worden opgevangen door `back-up' procedures in de menselijke
omgeving (het sociale systeem).

\subsection{Afstemming}
Het is met name op het gebied van de afstemming dat het verantwoord gebruiken
van ICT in veel opzichten nog in de kinderschoenen staat. Er valt hier nog veel
te leren, te onderzoeken en te ontwikkelen. Daarbij is overigens niet alleen
het criterium van betrouwbaarheid relevant. Ook andere aspecten zijn hierbij
van groot belang, zoals toegankelijkheid en gebruiksgemak.

De moderne westerse maatschappij begint bijvoorbeeld de consequenties te
ondervinden van het feit dat het het menselijk lichaam niet bedoeld is om 8 uur
per dag achter een beeldscherm te zitten en met een toetsenbord te werken. RSI
is een toenemend probleem. Taal- en spraaktechnologen hebben inmiddels een
alternatief voor het toetsenbord ontwikkeld in de vorm van spraakherkenners,
maar bij gebruikers van dergelijke programma's -- die hun stem nu gedurende
lange tijd op een enigszins onnatuurlijke wijze moeten gebruiken -- ontstaan
inmiddels klachten die wijzen op 'stem-RSI'. De menselijke natuur (fysiek,
emotioneel of mentaal) laat zich niet ongestraft negeren, en met de inzet van
ICT lopen wij tegen die grenzen aan.

Ook andere, met ICT samenhangende problemen illustreren dit punt. Computers
kunnen nog zulke zuivere informatie leveren met betrekking tot bijvoorbeeld de
schending van het luchtruim door een onbekend vliegtuig; wanneer een soldaat in
een oorlogssituatie deze informatie moet verwerken en bijvoorbeeld door de
emotionele spanning niet alle informatie goed registreert, of in een
`split-second' een verkeerde knop indrukt, kan dit fatale consequenties hebben.
De mate waarin de technologie ons van dienst kan zijn wordt dus niet alleen
bepaald door datgene wat die technologie voor ons kan doen, maar tevens door
verwerkingswijze en -capaciteit van de mens. Ook de zgn.\ `info-stress'
illustreert dit.

De oorzaak van de veelal gebrekkige afstemming tussen mens en technologie is
grotendeels gelegen in het feit dat tot nu toe nog vaak alleen of voornamelijk
wordt gedacht vanuit de mogelijkheden van de technologie, en niet vanuit de
manier waarop de mens hiermee zou kunnen of willen werken. Een technologisch
product dat voorbij gaat aan de menselijke gebruikscontext schiet echter zijn
doel voorbij. Evenzeer geldt dit voor een product dat voorbij gaat aan de
structuur en dynamiek van een organisatie waarin het wordt ingezet. Het is
daarom van belang dat inzichten in de werking van het menselijke organisme en
van de grotere 'organismen' die organisaties zijn, zo veel mogelijk kunnen
worden vertaald naar het ontwerp en de inzet van ICT.  Alleen z\'o wordt een
ICT-ondersteunde samenleving ook een \emph{gezonde} samenleving, waarin mens en
mensheid optimaal kunnen blijven functioneren. Hier ligt dan ook een niet
geringe taak en verantwoordelijkheid -- maar tegelijkertijd ook een prachtige
uitdaging -- voor de ontwikkelaars van technologische oplossingen.

Toch moeten we ook niet vergeten dat de toenemende rol van ICT intussen de
meest directe aanleiding is voor het optreden van de bovengenoemde problemen,
en dat ICT dus hoe dan ook een zeer voorname rol speelt in de informatiekunde --
die de problemen hoopt te helpen oplossen. Het is niet een kwestie van `know
your enemy', maar meer een complementaire hand-in-hand benadering waarmee we
onze samenleving op termijn hopen te verbeteren.

\subsection{Ontwikkeling van informatiesystemen}
\label{ss:ontwikkeling}
De omgevingen waarin ICT wordt gebruikt zijn vaak onderhevig aan veranderingen.
Als een gebruikscontext verandert, is het meestal wenselijk dat zowel het
sociale systeem als het ondersteunend technisch systeem mee verandert. Dit
veronderstelt co-evolutie van het sociale en technische systeem, zodanig dat
het socio-technische systeem als geheel blijft functioneren. Om dergelijke
evolutie mogelijk te maken is het noodzakelijk om verder te kijken dan
momentopnamen van systemen (verder dus dan een statisch ontwerp of blauwdruk):
het ontwikkel\emph{proces} dient ook te worden meegenomen in de overweging. Het
is daarom van groot belang om onze aandacht gaandeweg te verleggen van
\emph{ontwerpdenken} naar \emph{ontwikkeldenken}.

Ondanks de wederzijdse afhankelijkheden vormen vier aspecten van
informatiesystemen (mens, organisatie, informatie en technologie) qua aard en
eigenschappen elk een systeem op zich, met een \emph{eigen ontwikkelingstempo
en -wijze} (een eigen dynamiek). Bij de afstemming van deze componenten moet
met deze 'eigen-aardigheden' rekening gehouden worden. Zo roepen
veranderingsprocessen bij mensen bijvoorbeeld vaak weerstand op. Het overwinnen
van die weerstand heeft meestal alleen kans van slagen met een langzaam proces van
geleidelijke stappen. In organisaties werken veranderingsprocessen soms
druppelsgewijs door in allerlei afdelingen, en is een radicale omslag in alle
afdelingen tegelijk vaak niet aan de orde. De mate waarin en wijze waarop
technologie kan meeveranderen met de wensen van mens en organisatie is sterk
afhankelijk van het ontwerp.

\subsection{Vormgeven aan de 'Digitale Samenleving'}
In het licht van geschetste maatschappelijke ontwikkelingen komt vanuit de
praktijk een steeds luidere roep naar academisch geschoolde mensen die niet
alleen verstand van ICT hebben, maar deze juist ook goed weten in te bedden in
de menselijke, maatschappelijke en organisatorische context. Verwijzend naar
de architecten van de fysieke wereld, zou men deze mensen de `architecten van
de digitale samenleving' kunnen noemen.\footnote{Zie ook: \NAFURL,
\ArchForumURL, of \ItForHumansURL.}

Naast een behoefte aan specialisten op het gebied van afstemming is er ook
steeds meer behoefte aan adequate theorie\"en, aanpakken en methoden om de
afstemming tussen de ICT en haar menselijke en organisatorische context te
realiseren. Dit onderschrijft het belang van een degelijke, goed-doordachte
opleiding, waarin de vormgevers van de toekomstige digitale samenleving worden
opgeleid en in het kader waarvan de kennis over de realisatie van deze
afstemming wordt ontwikkeld.

Er zijn uiteraard niet alleen architecten nodig bij het vorm en richting geven
aan de digitale samenleving. Naast architecten moeten we ook denken
aan bouwmeesters (en daarbij denken we juist niet alleen aan software),
beleidsmakers, programmamanagers, analisten, adviseurs, etc. Tezamen kunnen
zij vanuit een sterk inhoudelijke achtergrond mede vormgeven aan de digitale
samenleving. We benadrukken hier opnieuw hoe belangrijk het is om te beseffen
dat deze vormgevers van de digitale samenleving niet alleen oog moeten hebben
voor het product, maar ook het proces waarin dat product tot stand komt. De
broodnodige afstemming vindt immers plaats in dat proces, en zal
\emph{uiteindelijk} zijn weerslag hebben in het product (doorgaans een
architectuur van een informatiesysteem).

\section{Gemeenschappelijk speelveld}
\label{p:speelveld}
Het zou intussen duidelijk moeten zijn dat in onze opvatting informatica en
informatiekunde zich bevinden in hetzelfde speelveld, maar daarbinnen een
andere nadruk leggen. In deze paragraaf brengen we dat speelveld beter in
kaart. Daarbij besteden we eerst enige aandacht aan de verschillende
deelgebieden binnen het speelveld, en vervolgens aan procesmatige aspecten die
de deelgebieden verbinden.

\subsection{Deelgebieden van het speelveld}
Het speelveld kent vier hoofdaspecten, oftewel deelgebieden die elk hun eigen
factoren, prioriteiten, en `requirements' voortbrengen. Twee van de vier
aspecten zijn we al enkele malen eerder tegengekomen: \emph{mens} en
\emph{organisatie}. Tot nog toe hebben we gesproken over ICT als derde aspect.
Het is, zeker vanwege onze brede kijk op informatiesystemen, belangrijk om dit
aspect op te delen in een informatie (\& communicatie) deel en een puur
technologisch deel.  In diagram~\ref{Aspecten} staan de deelgebieden afgebeeld
in relatie tot een aantal steekwoorden, die ruwweg aangeven welke onderwerpen
er binnen een deelgebied spelen.

De deelgebieden zijn nadrukkelijk gegroepeerd rond het centrale aandachtspunt:
informatiesystemen (dit zou men kunnen zien als een vijfde aspect in het
figuur). Het idee is dus dat de vier deelgebieden elk hun eigen soort van
invloed uitoefenen op de requirements voor en het functioneren van
informatiesystemen. Wat die invloed is hangt natuurlijk sterk af van de context
waarin een bepaald systeem wordt ingezet. Daarnaast gaat onze aandacht niet
zozeer uit naar de deelgebieden op zich, maar naar de impact die ze op elkaar
hebben, en in het bijzonder natuurlijk hun impact op informatiesystemen en de
ontwikkeling daarvan (zie figuur~\ref{Aspecten}).

\TexFigure{0.8}{Aspecten}{Verschillende aspecten van het speelveld}{Aspecten}

De deelgebieden zijn overigens niet altijd even scherp af te bakenen; ze zijn
op bepaalde punten sterk aan elkaar gerelateerd en vertonen soms zelfs enige
overlap. Ze zijn dan ook onderscheiden om het totale speelveld overzichtelijker
te maken, niet om het zeer strikt op te delen. Verder zijn er in potentie veel
meer aspecten denkbaar; de 4 + 1 deelgebieden die benoemd zijn beschouwen wij
als de `hoofdsmaken', die we dan ook terug zullen vinden in de inrichting van
het Informatiekunde programma.

Het is van groot belang in te zien dat het onderscheiden van aspecten zoals in
figuur~\ref{Aspecten} op zich al een bepaald gezichtspunt veronderstelt. We
moeten dus onderkennen dat er bij verschillende belanghebbenden in het
speelveld verschillende percepties zullen bestaan ten aanzien van de diverse
aspecten en deelgebieden. Een informatiekundige zal hiermee moeten kunnen
omgaan. Zij zal in de huid moeten kunnen kruipen van de verschillende
belanghebbenden teneinde bruggen te slaan tussen hun verschillende percepties.
Daarbij zijn informatiekundigen niet alleen neutrale toeschouwers: zij kunnen
ook de rol spelen van \emph{belangenbehartigers} of zelfs
\emph{onderhandelaars}.

\subsubsection{Mens-aspect}
Het mens-aspect richt zich op de individuele mens: haar capaciteiten, wensen,
en manier van denken, communiceren, werken, leren, etc. Traditioneel wordt
onderzoek naar dit aspect vooral vertegenwoordigd door velden als psychologie,
cognitiewetenschap, taalwetenschap, ge\-drags\-we\-ten\-schap, etc. De mens kan
echter vaak niet volledig los gezien worden van haar functioneren in een
sociaal verband; hier vertoont dit aspect enige overlap met zowel het
informatie-aspect als het organisatie aspect. Directe relaties met
informatiesystemen liggen bijvoorbeeld op het gebied van mens-machine
interactie, ergonomie, individuele customisation van systemen, individuele
drijfveren en motivatie.

\subsubsection{Organisatie-aspect}
Het organisatie-aspect betreft essentieel alles wat te maken heeft met
doelbewust vormgegeven menselijke samenwerkingsvormen. Onderliggende
mechanismen binnen het deelgebied betreffen algemene processen van interactie
tussen mensen (oftewel `sociale interactie'), maar de nadruk ligt sterk op die
aspecten van samenwerking die te `construeren', te `managen' of te
`organiseren' zijn. Concreet hebben we het daarbij meestal over bedrijven,
instellingen, regeringen etc.; kortom organisaties. Deze zijn vanouds al
onderwerp van onderzoek in velden als organisatiewetenschap, bedrijfswetenschap
en managementwetenschap. Wij zijn hier echter vooral ge\"interesseerd in hoe
dit aspect op directe wijze gerelateerd is aan informatiesystemen en de
ontwikkeling daarvan. Daarbij moet men voornamelijk denken aan de rol van ICT
in relatie tot organisatorische strategie, doelen, producten, diensten, en
procesondersteuning.

\subsubsection{Informatie-aspect}
Het informatie-aspect betreft een specifieke focus op de informatie die mensen
en groepen mensen uitwisselen, en de wijze waarop dit gebeurt. De vakgebieden
die zich hier van oudsher mee bezighouden zijn o.a. de communicatiewetenschap,
de taalwetenschap, en de tekstwetenschap, maar het aspect heeft tevens wortels
in de wiskunde en klassieke informatica. Merk op dat dit aspect niet alleen
informatieuitwisseling \emph{door middel van} informatiesystemen betreft, maar
ook informatieuitwisseling \emph{voor de ontwikkeling van} informatiesystemen.
Concrete termen gerelateerd aan informatiesystemen zijn o.a.\ informatie- en
datastructuur, documentstructuur, communicatieprotocollen, media,
gebruikerstalen, en programmeertalen.

\subsubsection{Technologie-aspect}
Tot slot onderscheiden we het technologie-aspect, dat geworteld is in
`enabling technologies' voor informatiesystemen. De wetenschapsgebieden die
hier traditioneel mee verbonden zijn zijn o.a.\ de computerwetenschap,
wiskunde, en elektrotechniek. Een steeds belangrijkere bijdrage wordt geleverd
door de combinatie van computertechnologie met telecommunicatietechnologie;
deze combinatie wordt ook wel aangeduid met de term `telematica'. Concreet
kan men bij enabling technology denken aan begrippen als software, hardware, en
netwerken. Binnen de technologie-geori\"enteerde vakgebieden is overigens al
jaren veel aandacht voor het slaan van bruggen naar de andere deelgebieden;
technologische ontwikkelingen hebben ontwikkelingen binnen die gebieden vaak
ook in een stroomversnelling gebracht (denk aan o.a. `engineering' van
datastructuur, bedrijfsprocessen en communicatieprotocollen, de ontwikkeling
van webgebaseerde werk- en communicatieprocessen, en mens-machine interface
ontwikkeling).

\subsection{Dynamiek tussen aspecten binnen het speelveld}
Zoals reeds geschetst in paragraaf \ref{ss:ontwikkeling} dienen de
verschillende aspecten niet alleen goed op elkaar zijn afgestemd, maar moet die
afstemming vaak \emph{dynamisch} zijn. De betrokken deelgebieden kennen elk hun
eigen aard en tempo van evolutie. Dit vergt \emph{continue} afstemming en
co-evolutie. In veel gevallen zullen hier expliciet en weloverwogen processen
voor moeten worden ingericht. Dergelijke ontwikkelingsprocessen spelen zich af
op twee vlakken. Allereerst is er de verandering op het gebied van de
verschillende aspecten: veranderingen m.b.t.\ mensen, organisaties, informatie
en communicatie, en technologie. Daarnaast dienen dergelijke veranderingen vaak
ook \emph{bestuurd} te worden, hetgeen weer aanleiding geeft voor het inrichten
van specifieke besturingsprocessen met wortels in de diverse aspecten (zie
figuur~\ref{Evolutie}).

\Figure{width=0.8\textwidth}{Evolutie}{Evolutie}{Co-evolutie van de
verschillende aspecten}

In de wereld van de ICT en de afstemming daarvan op de diverse andere aspecten
is een aanpak in opkomst die zich specifiek richt op de \emph{besturing van de
co-evolutie binnen de verschillende aspecten}. Deze aanpak wordt vaak de
`architectuuraanpak' genoemd. Architectuur, in de context van socio-technische
systemen, kan het best gedefinieerd worden
(conform~\cite{Report:99:IEEE:Architecture}) als:
\begin{quote} \it
  De fundamentele organisatie van een systeem zoals deze wordt vormgegeven
  door zijn componenten, hun onderlinge verbanden alsmede die met de
  omgeving, en de principes welke sturend zijn voor hun ontwerp en evolutie.
\end{quote}
Architectuur is een stuurmiddel voor het afstemmingsproces dat moet
plaatsvinden tussen mens, organisatie, informatie, en ICT. In dit
afstemmingsproces zal architectuur doorgaans als volgt worden ingezet (zie
wederom~\cite{Report:99:IEEE:Architecture}):
\begin{itemize}
   \item Als een communicatiemiddel tussen de verschillende belanghebbende
   partijen
   \item Voor het bieden van een kader waarbinnen het systeem in de toekomst
   kan evolueren
   \item Als basis voor het evalueren en vergelijken van alternatieve ontwerpen
   van een systeem
   \item Als plannings- en stuurinstrument voor de daadwerkelijke ontwikkeling
   en realisatie van het systeem
   \item Als ijkpunt om de daadwerkelijke realisatie van een systeem aan te
   verifi\"eren
\end{itemize}

\section{De symbiose tussen informatica en informatiekunde}
\label{s:symbiose}
Door de reeds eerder verschillende startpunten van de twee vakgebieden:
\begin{quote} \it \center
  $\Longrightarrow$ Betrouwbare ICT, om mensen en organisaties te
  ondersteunen.\\
  ~\\
  Mensen en organisaties ondersteunen met betrouwbare ICT. $\Longleftarrow$
\end{quote}
zullen informatici en informatiekundigen ook verschillende, aanvullende rollen
spelen bij het vormgeven van de digitale samenleving .\footnote{Hierbij is het
belangrijk te onderkennen dat in de \emph{huidige} praktijk veel mensen met een
informatica opleiding de rol vervullen die we hier aan de informatiekundige
toedichten. Daarbij is het tevens ontnuchterend om te zien dat in veel van de
gevallen dergelijke rollen worden vervuld door mensen die \emph{helemaal geen}
informatica of informatiekunde opleiding hebben.} We hebben al eerder een link
gelegd naar architecten van de fysieke wereld, en hierbij het begrip
`architectuur' ingevoerd. Informatici en informatiekundigen die de rol van
architect in de digitale samenleving gaan vervullen kunnen we, in lijn met deze
analogie, respectievelijk \emph{ICT-architect} (of software architect) en
\emph{informatie-architect} noemen. Dit zijn begrippen die momenteel in de
praktijk in toenemende mate gebruikt worden in een betekenis die redelijk
overeenkomt met de hierboven aangeduide betekenis.

De verschillende rollen die regelmatig zullen worden vervuld door informatici
en informatiekundigen kunnen helder worden gepositioneerd aan de hand van de
belangrijkste activiteiten binnen de levenscyclus van een informatiesysteem.
Deze zijn:
\begin{description}
    \item[Definitie --] dit betreft activiteiten met als doel het achterhalen
    van alle eisen (`requirements') waaraan het systeem en de
    systeembeschrijving zouden moeten voldoen.

    \item[Ontwerp --] hierbij gaar het om het proces dat als doel heeft het
    ontwerpen van een systeem conform de beschreven requirements. Het
    resulterende systeemontwerp kan vari\"eren van een ontwerp van de essentie
    op strategisch of achitectuur-niveau tot een detailontwerp dat raakt aan
    programmeer-statements of zeer specifieke handelingen die door een
    menselijke actor verricht moeten worden.

    \item[Constructie --] dergelijke activiteiten richten zich op het
    realiseren en testen van een systeem dat wordt beschouwd als een (mogelijk
    kunstmatig) samenhangend geheel van functionaliteiten dat \emph{nog niet
    operationeel is}.

    \item[Invoer --] hierbij gaat het om het operationeel maken van een
    systeem, m.a.w.\ het \emph{implementeren} van gebruik van het systeem door
    haar bedoelde gebruikers.

    \item[Bestendiging --] dit betreft activiteiten die bijdragen aan het
    ondersteunen, onderhouden of verder in de organisatie verankeren van een
    systeem. Men kan denken aan technisch onderhoud en het beter afstemmen van
    werkprocessen op het systeem, maar ook aan het geven van trainingen aan
    gebruikers of het schrijven of verbeteren van systeemdocumentatie.
\end{description}
In figuur~\ref{LifeCycle} staan deze activiteiten onderling gepositioneerd
binnen de levenscyclus van een informatiesysteem. Tezamen noemen we deze
activiteiten het \emph{veranderen \& bestendigen van informatiesystemen}.

\Figure{width=0.8\textwidth}{LifeCycle}{LifeCycle}{
  Hoofdactiviteiten in de levenscyclus van een informatiesysteem
}

De informatiekundige en de informaticus zullen vaak gezamenlijk verantwoording
dragen voor de diverse activiteiten in de cyclus, maar zullen daarbij
verschillende vertrek- en aandachtspunten hanteren. Deze verschillen zijn
weergegeven in figuur~\ref{Positionering}. Hieruit blijkt ook hoe de
informaticus en de informatiekundige ten opzichte van elkaar min of meer in een
symbiotische relatie staan.  Concreet, in termen van de benoemde
hoofdactiviteiten binnen de levenscyclus van een informatiesysteem, kunnen we
de rol van de informatiekundige hierin aanscherpen en als volgt formuleren:
\begin{description}
    \item[Definitie --] De belangrijkste zorg van een informatiekundige tijdens
    dit proces zal er op gericht zijn een evenwichtig pakket aan eisen op te
    stellen met betrekking tot de externe en interne relaties van het beoogde
    informatie systeem, waarbij `evenwichtig' inhoudt dat het resulterende
    systeem effectief is met betrekking tot de menselijke, organisatorische en
    technologische context. Informatiekundigen dienen het onderhandelingsproces
    dat zich hierbij doorgaans afspeelt tussen verschillende belanghebbende
    partijen te kunnen faciliteren en waar nodig bijsturen.

    \item[Ontwerp --] Het ontwerpen is een gezamenlijke verantwoordelijkheid
    van informatici en informatiekundigen. Daarnaast is de informatiekundige
    primair verantwoordelijk voor de bewaking van de belangen van de
    verschillende belanghebbenden.

    \item[Constructie --] In dit proces is de informatiekundige vooral
    verantwoordelijk voor het scheppen van de juiste voorwaarden in de
    menselijke en organisatorische context van het systeem als voorbereiding
    van de daadwerkelijke invoering. Denk hierbij aan de ontwerp \& invoering
    van nieuwe werkprocessen, ontwerpen \& verzorgen van de opleiding van
    toekomstige gebruikers, etc.

    \item[Invoer --] Gedurende het daadwerkelijke uitrollen van het nieuwe
    systeem blijft de informatie\-kun\-di\-ge primair verantwoordelijk voor een
    goede `landing' van het systeem in de menselijke en organisatorische
    context.

    \item[Bestendiging --] Informatiekundigen zijn bij een operationeel systeem
    specifiek verantwoordelijk voor de blijvende aansluiting tussen het
    informatiesysteem en haar menselijke en organisatorische context. Dit kan
    er toe leiden dat een discrepantie wordt ontdekt, waarna een nieuwe
    ontwikkelcyclus gestart moet worden. Maar denk in dit verband ook aan het
    up-to-date en beschikbaar houden van systeemdocumentatie (ontwerp \&
    gebruik), etc.
\end{description}

\Figure{width=0.8\textwidth}{Positionering}{Positionering}{
  De verschillende foci van informatiekundige en informaticus.
}

Voor het \NIII\ is het belangrijk de verschillende rollen van informaticus en
informatiekundige te onderkennen, en hieraan consequenties te verbinden in
relatie tot de inrichting van de respectievelijke opleidingen. De inrichting
van de opleiding informatiekunde biedt hierbij om meer dan \een\ reden een
insteek voor het aanboren van een groep potenti\"ele studenten (\een\ van de
drijfveren voor het \NIII\ om met deze opleiding te beginnen).

Enerzijds kan voor een informatiekunde opleiding de als zwaar ervaren wiskunde
eis verminderd worden. Daarnaast biedt het beeld van de behoefte-gedreven
versus de aanbod-gedreven insteek ook een bruikbaar perspectief op de werving
van niet-traditionele studenten. Traditioneel is informatica een studie die
studenten met een uitgesproken $\beta$-drijfveer zal aantrekken; dit vanwege de
ori\"entering op de ICT. Met informatiekunde krijgt het \NIII\
nu de mogelijkheid om ook studenten met een $\gamma$-drijfveer aan te trekken:
diegenen die meer in de vraagkant van ICT zijn
ge\"{\i}nteresseerd. Deze studenten dienen wel degelijk een exacte gerichtheid
te hebben; het is echter niet die $\beta$-gerichtheid die de initi\"ele
\emph{drijfveer} vormt (men is, met andere woorden, niet zozeer
ge{\itr}nteresseerd in de technologie om de technologie zelf).  Merk hierbij
overigens op dat het hebben van een $\beta$- of een $\gamma$- drijfveer op
zichzelf \emph{niets} zegt over het vermogen van een student zich te kunnen
verdiepen in typische $\beta$-kennis. Het zegt mogelijkerwijze wel iets over de
manier waarop het onderwijs ingericht kan/moet worden.

\section{Marktpotentie; De vraag naar Informatiekundigen}
\label{p:Markt}
Wij zijn buitengewoon optimistisch over de ontwikkeling van de ICT
arbeidsmarkt. Het is inderdaad zo dat de economische situatie momenteel niet
erg rooskleurig is, hetgeen uiteraard ook negatieve gevolgen heeft voor de ICT
sector. Er zijn diverse berichten in de nationale pers te vinden over ontslagen
in de ICT sector.

Wij denken dat de huidige dip in de landelijke vooraanmeldingcijfers voor de
Informatica en Informatiekunde studies slechts tijdelijk zal zijn, dat over 1 of hooguit 2
jaar de studentenaantallen voor Informatica zich weer zullen herstellen, en de
aantallen voor Informatiekunde potentieel zelfs flink kunnen groeien.  Ter
onderbouwing van ons optimisme citeren we een stukje uit een artikel dat recent
verschenen is in het tijdschrift FEM business (22 maart 2003):
\begin{quote} \it
  Het is nauwelijks voor te stellen dat automatiseerders vier jaar geleden
  leefden als muizen in een kaaswinkel. Het waren de tijden van de
  onwaarschijnlijke groeicijfers in de ict-branche. Het milleniumprobleem en de
  overschakeling naar de euro leverden een oneindige stroom aan opdrachten op.
  Bovendien ontdekten bedrijven de beloftes van internet en zochten massaal
  aansluiting bij deze hype. De automatiseerders waren niet aan te slepen. In
  1999 bleven er volgens branchevereniging Nederland-ICT ruim 16.000 vacatures
  onvervuld.

  In korte tijd is er veel veranderd. Na de milleniumwende, de invoering van de
  euro en de ineenstorting van de internethype, drogde de opdrachtenstroom op
  en gingen ict-bedrijven massaal saneren.

  De ontslaggolf heeft als een verfrissend regenbuitje gewerkt op de
  oververhitte arbeidsmarkt.  De personeelstekorten zijn verdwenen.
  Branchevereniging Nederland-ICT constateert dat er in 2002 ruim 2800
  automatiseerders teveel waren. Tot opluchting van de ondernemingen. Die
  hoeven niet meer tegen elkaar op te bieden om de schaarse professionals te
  lokken.

  Die opluchting zal van korte duur zijn. Nederland-ICT voorspelt dat de markt
  volgend jaar met zo'n anderhalf tot drie procent groeit. Dat is niet meer dan
  een licht herstel, maar toch voldoende om terug te keren naar een tekort van
  zo'n 2900 automatiseerders.  Henk Broeders, voorzitter van Nederland-ICT en
  directievoorzitter van Cap Gemini Ernst \& Young, constateert dat het huidige
  personeelsoverschot conjunctureel is en niet structureel. ``Ik denk dat er
  zeker weer een tekort zal ontstaan.''

  De ineenstorting van de economie heeft de spanning op de arbeidsmarkt even
  gemaskeerd.  En klein beetje groei en het gaat alweer mis. Nederland kampt
  met een paar fundamentele problemen. Zo melden te weinig jongeren zich aan
  voor de studie informatica.  Daardoor is de uitstroom uit het onderwijs
  continu veel te laag om de banengroei op te vullen. ``Het aantal aanmeldingen
  is altijd laag geweest'', zegt Miranda Valkenburg van HBO-I, een
  samenwerkingsverband van informaticaopleidingen bij de hogescholen. ``Maar de
  laatste twee jaar is het verder afgenomen. In 2010 heeft de branche
  tienduizend nieuwe mensen per jaar nodig. De hogescholen en universiteiten
  leveren er maar vierduizend.''
\end{quote}
In aanvulling op het artikel uit FEM Business is het relevant om te melden dat
alhoewel er op dit moment sprake is van een overschot aan automatiseerders,
heel veel organisaties nog steeds naarstig op zoek zijn naar goede ICT-ers.
Door de krappe ICT arbeidsmarkt in het recente verleden, zijn er vogels van
divers pluimage in de ICT sector terechtgekomen. Momenteel is het bepaald niet
zo dat ``de gemiddelde ICT-er'' iemand met een formele Informatica of
Informatiekunde opleiding is.  Tot voor kort leek het er in ICT sector op alsof
iedereen die ``een leuk hondehok'' kon ontwerpen en/of maken, per direct
architect of hoofdaannemer kon worden van ``het nieuwe hoofdkantoor''. De
huidige situatie op de ICT arbeidsmarkt lijkt gelukkig een einde te maken aan
deze onwenselijke situatie.

Ondanks de huidige economische situatie kunnen we naar onze mening dus veilig
stellen dat de vraag naar goede opgeleide ICT-ers zal stijgen. Zeker wanneer
we, los van de diverse voorspellingen, met grote zekerheid kunnen vaststellen
dat de meeste bedrijven in Nederland sterk afhankelijk zijn van ICT. Men zou
kunnen stellen dat vele bedrijven in Nederland inmiddels sterker afhankelijk
van een goede inrichting van de ICT dan van het beschikken over goede
transportmogelijkheden.  Als bij de KLM het centrale boekingssysteem uitvalt,
is de KLM binnen een dag failliet. ICT dringt ook duidelijk door tot in de
directiekamers. Bedrijven als Ahold en Schiphol trekken speciaal een
directielid aan voor het ICT-beleid.  Rapporten van gerenommeerde
managementadviesbureaus zoals Gartner, Forrester en de MetaGroup onderstrepen
dit belang nog eens.

\section{Verbredingsgebieden}
\label{p:Toep} Nu we een beeld geschetst hebben van het speelveld
van de informatiekunde en informatica, en tevens van de manier
waarop deze vakgebieden zich tot elkaar verhouden, kunnen we ons
blikveld uitbreiden naar de diverse \emph{verbredingsgebieden} van
de informatiekunde, en naar het \emph{onderzoek} wat binnen het
\NIII\ op informatiekundig gebied gedaan zal worden.

Wanneer we praten over verbredingsgebieden, dan kunnen we een
onderscheid maken tussen gebieden in de zin van:
\begin{description}
   \item[Informatiekunde toegepast op ...] waarbij de informatiekunde
   wordt toegepast op een een bepaald \emph{toepassingsgebied}. Bijvoorbeeld:
   \begin{description}
      \item[Medische informatiekunde] richt zich op het toepassing van ICT in
      een medische setting.
      \item[Beslissingsondersteuning] betreft de inzet van ICT ter ondersteuning van
      besluitvormingsprocessen; meestal in de context van bedrijfsprocessen.
      \item[Rechtsinformatica] betreft het gebruik van ICT methoden, technieken en
      hulpmiddelen, zoals ``Kunstmatige Intelligentie'', ten behoeve van de juridische
      praktijk en theorie.
   \end{description}
   \item[Informatiekunde toepassen onder ....] waarbij de informatiekunde
   bekeken wordt \emph{vanuit} een bepaald ander domein. Bijvoorbeeld:
   \begin{description}
      \item[Informaticarecht] betreft de juridische aspecten van ICT.
      Informaticarecht betreft de juridische aspecten van informatietechnologie.
      \item[Kennis- \& informatiemanagement] richt zich op de vraag hoe organisaties
      het beste kunnen omgaan met het besturen van hun kennis- en informatiestromen.
      Een belangrijk aspect hierbij is uiteraard de besturing (Governance) van de inrichting
      van de ICT middelen.
   \end{description}
\end{description}

In het kader van het toenemende belang van een optimale afstemming
van ICT op de specifieke context van mens en organisatie wordt het
voor informatiekundigen steeds belangrijker ook iets af te weten
van de verschillende verbredingsgebieden en de specifieke
afstemmingsproblemen die daaromtrent spelen. De verschillende
contexten waarin de ICT kan worden ingezet, zoals
verzekeringsmaatschappijen, banken, de advocatuur, de medische
wereld, zijn vaak zodanig gespecialiseerd dat hieruit vaak
specifieke wensen ten aanzien van de ICT-ondersteuning ontstaan.
Bijvoorbeeld, om een medisch specialist bij de diag\-nostisering
optimaal te ondersteunen met ICT is het nuttig als het systeem zo
goed mogelijk aansluit bij de denk- en redeneerwijze en de manier
van informatie-verwerken die in een dergelijke context
gebruikelijk is. Specialisten uit de diverse verbredingsgebieden
hebben meestal niet genoeg kennis van ICT om zelf een optimaal
systeem te ontwikkelen. De experts op het gebied van de
technologie hebben weer niet de expertise om te kunnen denken
vanuit het specifieke vakgebied. Er ontstaat daarom meer en meer
behoefte aan mensen die enerzijds voldoende kennis hebben van
diverse verbredingsgebieden om mee kunnen denken met de
specialist, diens informatiseringswensen te kunnen begrijpen en
hierin te adviseren, maar die anderzijds ook voldoende expertise
hebben op het gebied van de ICT om deze wensen te kunnen vertalen
naar het juiste ontwerp en de juiste ontwikkelaanpak.

Er zijn binnen het wetenschapsgebied van de informatiekunde (en
het sterk gerelateerde gebied van de informatica) gaandeweg al
allerlei verbredende deelgebieden ontstaan. Voorbeelden hiervan
zijn o.a.\ juridisch kennisbeheer, medische informatiekunde,
bio-informatica, en taal- en spraaktechnologie. Met de toenemende
specialisatie binnen verschillende disciplines en de toenemende
rol van ICT binnen deze specialisaties, zal ook het aantal
verbredingsgebieden van de informatica en de informatiekunde in de
komende jaren alleen maar groeien. Aangezien hierbij steeds sprake
zal zijn van het slaan van een brug tussen twee in eerste
instantie verschillende disciplines, zal hierbij vooral de
informatiekunde een belangrijke rol spelen.

\section{Missie en profiel}
\label{p:Missie}
Kort samengevat zien wij informatiekunde als het wetenschapsgebied wat zich
bezighoud (op macro en micro niveau) met de afstemming tussen de vier
essentiele elementen van de digitale samenleving: \emph{mens},
\emph{organisatie}, \emph{informatie} en \emph{technologie} waarbij de
technologie bedoelt is als ICT intensieve technologie (digitale technologie).
Hier liggen voor het wetenschappelijk onderzoek grote uitdagingen.

De ambitie van het \NIII\ met betrekking tot de verdere ontwikkeling van
het onderzoek, en de opleiding, op informatiekundig gebied kan samengevat worden
in de volgende twee-ledige missie:
\begin{quote} \it
  \begin{itemize}
     \item Uitgroeien tot een nationaal en internationaal vooraanstaand
     instituut op het gebied van afstemmingsproblemen (tussen mens, organisatie,
     informatie en technologie) in de digitale samenleving, waarbij dergelijke
     afstemmingsproblemen primair onderzocht worden met een exact-wetenschappelijke
     bril maar met oog voor de empirische en de zachte aspecten.
     \item Studenten opleiden die gewaardeerd worden door zowel het
     bedrijfsleven als de wetenschappelijke wereld:
     \begin{itemize}
        \item Vanuit het bedrijfsleven wegens hun theoretisch onderbouwbare, doch
        praktisch relevante, kennis \& vaardigheden;
        \item Vanuit de wetenschap wegens de academische kwaliteit van hun
        kennis \& vaardigheden.
     \end{itemize}
  \end{itemize}
\end{quote}

Om deze missie te realiseren, heeft het \NIII\ inmiddels de beschikking over
de volgende belangrijke ingredi\"enten:
\begin{itemize}
  \item Inbedding in een sterke beta-traditie.
  \item Gezonde relaties met de menswetenschappelijke en
        organisatiewetenschappelijke faculteiten van de KUN.
  \item Gezonde relatie, in termen van inhoudelijke samenwerking, met
        enkele van de belangrijke spelers uit de praktijk van afstemming in de
        digitale samenleving. Denk hierbij onder andere aan bijzondere leerstoelen
        en het opzetten van een avondopleiding met een partner uit de praktijk.
  \item Een actieve, en \emph{sturende}, deelname in nationale projecten/activiteiten
        op relevante gebieden. Denk hierbij ondermeer aan: het ArchiMate
        onderzoeksproject, mede oprichting van het Nederlands Architectuur
        Forum, deelname aan de organisatie van het Landelijke (ICT) Architectuur Congres,
        en het initi\"eren van architectuuravonden op de KUN.
\end{itemize}

\section{Verankering van Informatiekunde binnen de KUN}
\label{p:Veranker}
Informatiekunde draait om het beter afstemmen van mens, organisatie,
informatievoorziening en technologie. Vanuit die inhoud biedt de
Informatiekunde een goede springplan om samen te werken met andere faculteiten
binnen de KUN. De informationele en technologische aspecten van de
Informatiekunde zijn uiteraard sterk vertegenwoordigd binnen het \NIII. Echter,
de menselijke en organisatorische aspecten vinden we vooral terug bij andere
faculteiten.

Daarnaast is het voor de Informatiekunde, meer nog dan bij de
Informatica, belangrijk om contact te onderhouden met de
verbredingsgebieden. Deels vinden we dit terug bij de samenwerking
met het werkveld, maar daarnaast wordt ook expliciet gekeken naar
samenwerking met andere faculteiten binnen de KUN die specifieke
verbredingsgebieden van de Informatiekunde vertegenwoordigen.
Momenteel richt het opbouwen van samenwerking zich met name op het
onderwijs. Het doel voor de komende jaren zal zijn dit te
verbreden tot onderzoek.

Concreet zijn er vanuit Informatiekunde inmiddels de volgende activiteiten
opgestart om tot een nauwere samenwerking te komen met andere faculteiten
binnen de KUN:
\begin{itemize}
  \item De Informatiekunde opleiding maakt gebruik van diverse vakken van
        andere faculteiten. Dit biedt aanleiding tot het voeren van een
        dialoog om nauwer samen te werken, zowel qua onderwijs als qua onderzoek.
  \item Diverse specialisatierichtingen van de opleiding Informatiekunde:
        \begin{itemize}
          \item Medische Informatiekunde
      \item Juridisch kennisbeheer
      \item Recht \& informatica
          \item Informatie- \& kennismanagement
          \item Taal- \& spraaktechnologie
          \item Cognitiewetenschap
        \end{itemize}
  \item Voorstudie naar een avondopleiding `Architectuur van de Digitale
        Wereld', in samenwerking met het ABK (Avondopleiding Bedrijfskunde
        KUN), de NSM (Nijmegen School of Management) en CGEY (Cap-Gemini Ernst \&
        Young).
  \item Bij de ontwikkeling van het nieuwe Informatiekunde curriculum is
        gewerkt met een (KUN-)interne klanbordgroep. Hierin zaten ook
        vertegenwoordigers van de Nijmegen School of Management en Medische
        Informatiekunde.
\end{itemize}

\section{De vaardigheden van de informatiekundige}
\label{p:InfVaard}
Om te kunnen bepalen wat de vaardigheden zijn die een informatiekundige moet
beheersen, redeneren we vanuit de rol die zij naar verwachting gaan spelen met
betrekking tot de ontwikkeling van informatiesystemen. Daarom gaan we eerst
verder in op deze rol, om pas daarna de expliciete vaardigheden onder de loupe
te nemen.

Echter, om het beeld van de werkzaamheden van informatiekundigen in de praktijk
wat nader te illustreren volgen belichten we de werkzaamheden van twee
voormalige studenten.  Omdat informatiekunde aan de KU Nijmegen pas sinds het
jaar 2000 bestaat, zijn er nog geen afgestudeerden om te vertellen van hun
werkervaringen. Er zijn echter wel afgestudeerde informatici die
in de praktijk werkzaam op een positie waar informatiekundigen op kunnen
terechtkomen. We gebruiken hiervoor twee interviews die zijn afgenomen in
het kader van voorlichtingsmateriaal voor de opleiding Informatiekunde.

{\bf Neem nu ... Araminte Bleeker (afgestudeerd in 1995)}
\begin{quote} \it
  ``Helder krijgen wat mensen nu echt aan ICT ondersteuning nodig hebben voor
  hun dagelijkse werk, en dan zorgen dat ze die ondersteuning ook echt krijgen.
  Dat is voor mij de grootste uitdaging.''

  Araminte voltooide haar studie Informatica in 1995, haar afstudeeronderzoek
  verrichtte ze in Itali\"e. Hierna begon ze aan een onderzoek naar de rol van
  mensen bij en hun relatie met informatiesystemen.  Al snel werd haar echter
  duidelijk dat ze veel liever met dit probleem in de praktijk bezig was dan er
  vanuit een theoretisch standpunt naar te kijken.

  Ze werkte achtereenvolgens bij de ICT bedrijven Origin, Panfox, en Escador.
  In die periode werkte ze zich op tot informatiearchitect. ``Een goede
  informatiearchitect'', legt Araminte uit, ``moet niet alleen in staat zijn de
  taal en wereld van haar klanten te begrijpen en deze in kaart te brengen,
  maar moet ook de vertaling kunnen maken van hun wensen naar de mogelijkheden
  en onmogelijkheden van ICT. Bovendien moet de requirements engineer gedurende
  het gehele ontwikkelingstraject zelf de kwaliteit en voortgang van haar werk
  bewaken.'' Araminte werkt daarom voortdurend nauw samen met software
  architecten, software engineers en de projectleiding. ``Om elkaar aan te
  vullen en scherp te houden.''

  Waarna ze besluit met: ``En als de ICT ondersteuning dan uiteindelijk
  gerealiseerd is en ook nog eens van hoge kwaliteit is, dan pas ben ik echt
  tevreden!''
\end{quote}

{\bf Neem nu ... Jeroen Top (afgestudeerd in 1991)}
\begin{quote} \it
  ``Voordat je goed kunt managen is het essentieel te weten waaraan je leiding geeft.''

  In 1991 rondde Jeroen zijn studie informatica af. Na zijn afstuderen heeft
  hij ``om het vak echt te leren'' eerst gewerkt als programmeur en database
  Beheerder bij Info Support. Na enkele jaren trad hij in dienst bij zijn
  huidige werkgever, het Belastingdienst/Centrum voor ICT in Apeldoorn.

  Na verloop van tijd klom hij van teamleider, via projectmanager op tot
  afdelingshoofd. In zijn projectmanagement tijd heeft Jeroen onder andere
  gewerkt aan het inningssysteem. ``Het uitdagende aan dat project'', legt Jeroen
  uit, ``was dat er nogal veel raakvlakken waren met andere systemen en diverse
  organisatie-onderdelen. Als projectmanager moet je dan samen met de
  informatiearchitect behoorlijk uiteenlopende belangen van de diverse
  belanghebbenden met elkaar in overeenstemming brengen.''

  Sinds kort is Jeroen manager startarchitectuur. Hierover zegt Jeroen:
  ``Voordat we een nieuw systeem ontwikkelen, of een bestaand systeem aanpassen,
  doen we eerst een voorstudie naar de haalbaarheid en de bedrijfsmatige kosten
  en baten hiervan. Deze voorstudie noemen we de startarchitectuur. Het
  afstemmen van de belangen van de betrokken partijen speelt hierbij ook weer
  een belangrijke rol.''

  Als we Jeroen vragen wat hij zou willen meegeven aan studenten die zich
  uiteindelijk willen ontwikkelen tot manager in de ICT wereld, zegt hij: ``Een
  goede basiskennis van Mens, Organisatie, Informatie en IT vormen de hoekstenen waarop je
  zo'n carri\"ere kunt bouwen. Voordat je goed kunt managen is het onontbeerlijk
  om te weten waaraan je leiding geeft.''
\end{quote}

\subsection{De rol van de informatiekundige}
We baseren ons bij het beschouwen van de voorziene rol van de
informatiekundige op de vijf kernactiviteiten die zijn genoemd in
paragraaf \ref{s:symbiose}, meer bepaald de rolverdeling tussen
informaticus en informatiekundige zoals aangegeven in
figuur~\ref{Positionering}. Daarnaast kijken we ook naar de rol
van de informatiekundige zoals bezien vanuit procesperspectief en
productperspectief, vanuit de verbredingsgebieden, en vanuit het
wetenschappelijk onderzoek.

\subsubsection{Ontwikkeling en bestendiging}
Zoals we hebben gezien zijn de basisactiviteiten waarbij de informatiekundige
is betrokken. Merk op dat de vijf basisactiviteiten verder ingedikt kunnen
worden tot twee soorten: verandering (met als sub-activiteiten
\emph{definitie}, \emph{ontwerp}, \emph{constructie} en \emph{invoer}) en
bestendiging (hiervan worden geen subtypes onderscheiden in ons model). Zoals
te zien was in figuur~\ref{Positionering} neemt de informatiekundige ten
opzichte van de informaticus duidelijk het voortouw bij definitie, en treedt
zij wat meer in de achtergrond bij constructie (waarbij de informaticus meer
het heft in handen neemt). Ontwerp, invoering en bestendiging zijn typisch
co-producties tussen informaticus en informatiekundige, waarbij beiden een
ongeveer even grote rol spelen.

Het moet worden benadrukt dat in alle geval sprake moet zijn van op zijn minst
enige betrokkenheid van zowel informaticus als informatiekundige. Alle
ervaringen in het vakgebied wijzen erop dat het volledig onderbrengen van een
bepaalde activiteit bij een enkele partij problemen oproept (met als klassiek
voorbeeld het in handen leggen van de constructie bij een systeembouwer die
verder niet betrokken is bij de andere activiteiten). Bij alle genoemde
activiteiten kan overigens zowel sprake zijn van betrokkenheid bij feitelijke
\emph{uitvoering} daarvan, als van \emph{uitbesteding}. In beide gevallen kan
door de informatiekundige een rol worden gespeeld bij \emph{begeleiding,
bewaking, of aansturing} van de systeemontwikkeling.

Belangrijk is ook te beseffen dat informatiekundigen in staat moeten zijn
oplossingen te ontwikkelen die verdergaan dan bestaande oplossingen. Moderne ICT
staat bijna garant voor nieuwe toepassingen en nieuwe toepassingsgebieden. Dit
vereist van informatiekundigen dat ze in staat zijn om bestaande oplossingen en
paradigmas los te laten en met werkelijk nieuwe (innovatieve) oplossingen te
komen. Dit is dan ook meteen een belangrijke reden waarom informatiekundigen een
academische opleiding behoeven.

\subsubsection{Proces en subject}
Een informatiekundige moet dus kaas hebben gegeten van het \emph{veranderen} en
\emph{bestendigen} van informatiesystemen, maar dan wel \emph{in context}. Deze
context wordt, zoals aangegeven in figuur~\ref{Aspecten}, afgedekt door de
aspecten mens, organisatie, informatie en technologie. Dit vereist vaardigheden
op het gebied van het uitvoeren van het \emph{proces} van verandering en
bestendiging, maar ook inzicht in het \emph{subject} van dat proces (dus het
socio-technische system dat er het resultaat of product van is), waarbij
aandacht moet zijn voor alle aspecten die in een bepaalde context relevant
zijn.

\subsubsection{Verbredingsgebieden}
Het is verder ook wenselijk dat informatiekundigen in staat zijn
zich in te werken in verbredingsgebieden. Binnen de Nijmeegse
universiteit bestaan mogelijkheden voor het uitdiepen van kennis
op het gebied van diverse verbredingsgebieden. Het is echter ook
heel belangrijk dat de informatiekundige zich snel redelijk goed
kan inwerken in willekeurig welk nieuw verbredingsgebied. Het is
daarom noodzakelijk dat zij kan reflecteren op de verschillen en
overeenkomsten tussen verbredingsgebieden, en dat men snel inzicht
kan verwerven in de specifieke aspecten die bij een bepaald
verbredingsgebied komen kijken.

\subsubsection{Onderzoek en reflectie}
Het vakgebied van de informatiekunde is sterk in beweging. Er is sprake van een
voortdurende stroom van nieuwe technologie\"en, methoden, aanpakken, contexten,
inzichten, en gezichtspunten. Een informatiekundige reflecteert daarom
structureel over de status van het vakgebied en verbeterpunten daarin, en neemt
deel aan onderzoek naar mogelijke verbeteringen. Laten we hierbij ook niet vergeten
dat een academisch opgeleide informatiekundige in staat moet zijn om met echt
vernieuwende oplossingen te komen.

\subsection{Vaardigheden}
\label{ss:vaardigheden}
Op basis van de rollen en hoofdactiviteiten van de informatiekundige zoals
geschetst in de vorige paragraaf, kunnen we nu in meer detail ingaan op de
vaardigheden die de informatiekundige dient te bezitten. Vanzelfsprekend
weerspiegelen de vaardigheden de visie op het vakgebied zoals deze is neergezet
in de voorgaande paragrafen.

\subsubsection{Proces van verandering en bestendiging}
Een informatiekundige moet in staat zijn om:
\begin{description}
   \item[Probleemoplossend vermogen --] ... op verantwoorde wijze
   informatiekundige problemen op te lossen, in het bijzonder:
   \begin{itemize}
      \item ... een accurate diagnose te stellen, die te vertalen naar
      probleemstellingen, en de maatschappelijke relevantie van de onderkende
      problemen vast te stellen;

      \item ... problemen te analyseren, een synthese van oplossingsrichtingen te
      maken, en een solide oplossing te construeren;

      \item ... zich bewust te zijn van en kunnen reflecteren over het
      \emph{proces} van formuleren van probleemstellingen en het ontwikkelen van
      oplossingen, en over de rol die verschillende belanghebbenden hierin spelen;

      \item ... keuzes te maken voor geschikte onderzoeksmethoden, en op basis
      hiervan een onderzoeksplanning kunnen maken en uitvoeren;

      \item ... resultaten te verantwoorden en te presenteren.
   \end{itemize}
\end{description}

Met betrekking tot de sub-activiteiten van veranderings- en bestendigingsprocessen
moet een informatiekundige in staat zijn om:
\begin{description}
   \item[Defini\"eren --] ... een evenwichtig pakket van eisen op te kunnen stellen
   met betrekking tot de relaties van een informatiesysteem met haar omgeving en
   met betrekking tot de relaties tussen de systeemcomponenten onderling;

   \item[Ontwerpen --] ... een ontwerp van de essentie van een informatiesysteem te
   maken dat voldoet aan de gestelde eisen;

   \item[Constru\"eren --] ... de daadwerkelijke constructie van een
   informatiesysteem te begeleiden en te bewaken;

   \item[Invoeren --] ... te kunnen meewerken aan de invoering van een
   informatiesysteem in een gegeven context, en deze te begeleiden en bewaken;

   \item[Bestendigen --] ... mee te kunnen werken aan de bestendiging van een
   bestaand informatiesysteem, en deze te begeleiden.
\end{description}

Met betrekking tot verschillende wijzen van uitvoering van processen ter
verandering of besten\-di\-ging van een informatiesysteem moet de informatiekundige
in staat zijn om:
\begin{description}
   \item[Aanbesteding van verandering en bestendiging --] ... te kunnen meewerken
   aan de uitvoering, begeleiding of bewaking van de \emph{aanbesteding} van
   delen van het proces van het defini\"eren, ontwerpen, constru\"eren,
   invoeren of bestendigen van informatiesystemen;

   \item[Besturing van verandering en bestendiging --] ... voor een gegeven
   situatie een adequaat \emph{projectplan} op te stellen voor een project
   waarbinnen een proces van verandering of bestendiging van (dan wel
   aanbesteding hiervan) zal plaatsvinden, en de daadwerkelijke uitvoering van
   een dergelijk project te kunnen \emph{begeleiden}.
\end{description}

Binnen processen voor verandering en bestendiging van informatiesystemen moet
de informatiekunige verder in staat zijn om:
\begin{description}
   \item[Analyseren en modelleren --] ... in een gegeven probleemsituatie een voor
   de informatiekunde relevant domein te:
   \begin{itemize}
       \item ... \emph{analyseren};
       \item ... en de belangrijkste kenmerken van het domein met betrekking tot
       die probleemsituatie in kaart te brengen in termen van een geschikt \emph{model},
       \item ... door te \emph{abstraheren} van irrelevante details/aspecten;
       \item ... tevens dient men het resulterende model te kunnen \emph{valideren}.
   \end{itemize}
\end{description}

De informatiekundige moet in staat zijn om:
\begin{description}
   \item[Belangen behartigen --] ... de belangen van de verschillende
   belanghebbenden te behartigen;
   \item[Onderhandelen --] ... de voor het defini\"eren noodzakelijke
   onderhandelingen met de verschillende belanghebbende partijen te voeren, te
   faciliteren en waar nodig bij te sturen;
   \item[Leven met vaagheden --] ... om te gaan met `vaagheden' en al dan niet
   schijnbare tegenstrijdigheden, en hier toch (op het juiste moment) en
   compleet en precies (formeel) pakket van eisen uit af te leiden;
   \item [Communiceren --] ... effectief en op gepaste wijze te communiceren, meer
   concreet:
   \begin{itemize}
      \item ... verschillende communicatie-rollen aan te nemen, zoals leiding
      geven aan een discussie, actief luisteren, open luisteren, van gedachten
      wisselen;
      \item ... vakinhoudelijke informatie op een heldere manier mondeling en
      schriftelijk te presenteren.
   \end{itemize}
\end{description}

Tot slot moet een informatiekundige binnen het proces van verandering en
bestendiging van informatiesystemen in staat zijn om:
\begin{description}
   \item[Balans tussen product en proces --] ... een gemotiveerde afweging te
   maken tussen kwaliteit en compleetheid van de bij systeemontwikkeling op te
   leveren producten en van de voortgang en haalbaarheid van het
   daadwerkelijke ontwikkelings- en invoeringsproces.
\end{description}

\subsubsection{Subject van verandering en bestendiging}
Met betrekking tot de subjecten van processen van verandering en bestendiging
moet de informatiekundige in staat zijn om:
\begin{description}
   \item[Gezichtspunten --] ... op basis van een gedegen kennis van de
   \emph{organisatorische, menswetenschappelijke, informationele,
   technologische en systemische} gezichtspunten op informatiesystemen, de
   bijbehorende \emph{theorie\"en, methoden, technieken en hulpmiddelen}:
   \begin{itemize}
      \item ... te beoordelen op hun mogelijkheden en gedrag in een concrete
      toepassingssituatie;
      \item ... deze op een adequate wijze in te zetten.
   \end{itemize}
\end{description}

Een informatiekundige moet dus in staat zijn om:
\begin{description}
   \item[Integrale visie --] ... vanuit de verschillende gezichtspunten een
   integrale visie op informatiesystemen te hebben, en te redeneren over de
   onderlinge impact tussen en samenhang van de verschillende gezichtspunten.
\end{description}

\subsubsection{Verbredingsgebieden}
Met betrekking tot het omgaan met verschillende
verbredingsgebieden van de informatiekunde moet de
informatiekundige in staat zijn om:
\begin{description}
   \item[Inwerken in verbredingsgebieden --] ... zich in een verbredingsgebied in
   te werken teneinde in ieder geval in staat te zijn om:
   \begin{itemize}
      \item ... het gedachtegoed van dat verbredingsgebied te kunnen waarderen, en
      het betreffende domein object van analyse en modellering te maken;
      \item ... met domeinexperts te communiceren over, en zich in te leven in, de
      voor de informatiekundige essenti\"ele eigenschappen van het
      verbredingsgebied;
   \end{itemize}
   \item[Reflecteren over verbredingsgebieden --] ... te reflecteren over de
   \emph{verschillen} en \emph{overeenkomsten} tus\-sen diverse
   verbredingsgebieden.
\end{description}

\subsubsection{Onderzoek en ontwikkeling}
Met betrekking tot onderzoek en ontwikkeling moet de informatiekundige in staat
zijn om:
\begin{description}
   \item[Onderzoeksvragen --] ... voor de maatschappelijke omgeving relevante
   onderzoeksvragen te kunnen formuleren met betrekking tot het
   informatiekundige vakgebied;
   \item[Besturen van onderzoek --] ... een voor een gegeven onderzoeksvraag
   passende onderzoeksaanpak te formuleren in termen van een projectplan, en de
   uitvoering van dit onderzoek te begeleiden;
   \item[Uitvoeren van onderzoek --] ... conform een opgestelde onderzoeksaanvraag
   onderzoek uit te voeren naar een voor de informatiekunde relevante
   onderzoeksvraag.
\end{description}

%% file: waarbinnen-randvoorwaarden.tex
\chapter{Randvoorwaarden}
\label{h:waarbinnen-randvoorwaarden}

Het doel van dit hoofdstuk is het geven van een inventarisatie van
de randvoorwaarden waarbinnen de informatiekunde opleiding en/of het onderzoek
gestalte dienen te krijgen. Deze randvoorwaarden kunnen voortkomen uit:
\begin{itemize}
   \item Wetgeving
   \item Beleidskeuzes op hoger niveau
   \item Praktische omstandigheden \& ervaringen
\end{itemize}
Elke randvoorwaarde kan implicaties hebben ten aanzien van de vaardigheden
die studenten informatiekunde aan het einde van hun studie dienen
te hebben en/of de inrichtingsprincipes voor de opleiding en/of het onderzoek.

In de huidige versie van dit visiedocument is er vooral aandacht voor de
randvoorwaarden met betrekking tot de opleiding informatiekunde.  De eigenlijke
randvoorwaarden kunnen in drie klassen worden verdeeld, op basis van hun
`herkomst':
\begin{itemize}
   \item Europees \& Nationaal
   \item Universitair \& Facultair
   \item Nijmeegs Instituut voor Informatica \& Informatiekunde
\end{itemize}

De onderstaande tekst bevat per herkomstklasse een `droge' opsomming van
de randvoorwaarden zoals die ons tot op heden bekend zijn, gecombineerd
met de implicaties hiervan in termen van vaardigheden en/of 
inrichtingsprincipes.

\section{Europese \& Nationale randvoorwaarden}

\subsection{Internationalisering}
Als gevolg van de internationalisering van de opleiding,
zowel op Europees als mondiaal niveau, is gekozen voor
een overgang naar het Bachelor-Master stelsel zoals dit
reeds in gebruik was in de Angel-Saksische wereld.

Deze keuze heeft er concreet toe geleid dat er een
ontkoppeling tot stand is gekomen tussen de bachelorfase (3 jaar)
en de masterfase (1 \'a 2 jaar).
Deze ontkoppeling heeft tot gevolg dat er een
extra punt in de opleiding is ge\"introduceerd waarop instroom
en/of uitstroom van studenten kan plaatsvinden.
Daarnaast kan deze in- en uitstroom zich ook over
de landsgrenzen, of zelfs de EU-grenzen, heen begeven.

In termen van inrichtingsprincipes hebben deze ontwikkelingen
de volgende gevolgen op de inrichting van de informatiekunde
opleiding:
\begin{description}
   \item[Ba-Ma structuur --] De opleiding dient conform het bachelor \& master
   stelsel te worden opgezet
   \item[Instroom in Master --] Er dienen in het masterfase van de opleiding
   instroom mogelijkheden te zijn voor bachelors van andere universitaire
   studies. De masterfase van de opleiding dient tevens toegankelijk te zijn
   voor studenten uit het buitenland.
\end{description}

\subsection{Gebruik van de architectentitel in de ICT wereld}
Er wordt momenteel reeds informeel gebruikgemaakt van het label
`informatie-architect', `ICT-architect', etc., in de ICT wereld.  Wettelijk
gezien is dat \emph{nog} niet toegestaan.  Er zijn momenteel gesprekken gaande
tussen het Nederlands Architectuur Forum (NAF), de Stichting Certificatie
Informatie-Architecten (SCIA), het Genootschap voor Informatie-Architecten
(GIA) en de wereld van bouwkundig-architecten om tot over\'e\'enstemming te
komen met betrekking tot het gebruik van de titel architect in de ICT wereld.
Deze ontwikkelingen zullen, naar verwachting, ook met zich meebrengen dat er in
de loop van 2003 en 2004 offici\"ele eindtermen benoemd kunnen worden waar een
informatie-architect aan zal moeten voldoen.  Het uitstroomprofiel van de
Master Informatiekunde sluit naar verwachting heel nauw aan bij de eisen die
aan de theoretische bagage van dergelijke architecten gesteld zal worden.  Voor
de inrichting van de informatiekunde opleiding geldt
daarom het volgende principe:
\begin{description}
   \item[In lijn met certificering --] Waar mogelijk en relevant zal de
   informatiekunde opleiding zoveel mogelijk worden afgestemd op de eisen die
   voortvloeien uit de certificering van architecten in de ICT wereld
\end{description}

\section{Universitair \& Facultair}

\subsection{Katholieke grondslag}
Als direct uitvloeisel van de Katholieke grondslag van de KUN, is het
verplicht dat studenten een minimaal aantal ECTS aan vakken volgen
die gericht zijn op filosofische en ethische verdieping, middels
reflectie op het eigen vakgebied.
Dit gaat concreet om 3 ECTS in de Bachelorfase en 3 ECTS in de 
masterfase.
Met andere woorden:
\begin{description}
   \item[Filosofie \& ethiek --] Er dienen zowel in de masterfase als in de bachelorfase 3
   ECTS besteed te worden aan vakken die zijn gericht op filosofische en
   ethische verdieping, middels reflectie op het eigen vakgebied.
\end{description}

\subsection{Algemene facultaire eisen}
De faculteit heeft zelf een aantal algemene inrichtingsprincipes
geformuleerd waar de opleidingen aan dienen te voldoen
Deze vallen uiteen in vier klassen: algemeen, propedeuse, bachelor en master

Algemene inrichtingsprincipes:
\begin{description}
   \item[Koppeling onderwijs \& onderzoek --] Er dient een consequente koppeling
   tussen onderwijs en onderzoek te zijn.
   \item[Aansluiting bij zwaartepunten --] De opleiding moet aansluiten aan bij c.q.\
   integreert zwaartepunten van het facultair onderzoek, i.c.\ het onderzoek van het 
   \NIII.
   \item[Afgestemd op beroepsperspectief --] De opleiding moet inhoudelijk zijn
   afgestemd op de beroepsperspectieven en -profielen van de afgestudeerde.
   \item[Transparante opbouw --] De opleiding dient transparant te zijn qua
   opbouw. Met andere woorden, er dient een samenhangende, cumulatieve,
   inhoudelijke opbouw door de opleiding heen te zijn, die de ontwikkeling van
   de student weerspiegelt.
   \item[Wenselijke inhoud --] De inhoud van de opleiding moet actueel, aantrekkelijk
   en uitdagend zijn.
\end{description}

Inrichtingsprincipes voor de propedeuse:
\begin{description}
   \item[Rol propedeuse --] De propedeuse dient een \emph{selecterende},
   \emph{ori\"enterende} en \emph{verwijzende} functie te hebben. Studenten
   dienen daarom in het eerste jaar ook een tijdige terugkoppeling te krijgen
   t.a.v. hun prestaties, bijvoorbeeld middels een studieadvies.
\end{description}

Inrichtingsprincipes voor de bachelor:
\begin{description}
   \item[Breed; doch diepe focus --] De bachelorfase van de opleiding dient
   breed ge\"orienteerd te zijn, maar met een sterke vakdisciplinaire
   component.
   \item[Academisch vormend --] De bachelorfase dient tevens academisch
   vormend te zijn, hetgeen de academische bacheloropleiding moet onderscheiden
   van een HBO bachelor.  
   \item[Aansluiten op voorkennis --] Het opleidingsniveau sluit inhoudelijk
   aan op het voorkennis- en abstractieniveau van de vwo-eindprofielen.
   \item[Smaakt naar meer --] Het derde jaar van de bachelor dient te worden
   opgezet als aanzet tot het doen van een master, en niet als uitstroommoment.
\end{description}

Inrichtingsprincipes voor de master:
\begin{description}
   \item[Afstudeervarianten --] Er zijn drie afstudeervarianten
      \begin{itemize}
         \item Onderzoek
         \item Communicatie \& educatie
         \item Management \& toepassing
      \end{itemize}
\end{description}

\subsection{Algemene vaardigheden}
De faculteit heeft tevens een aantal algemene vaardigheden verwoord
waar de opleiding aan dient tegemoet te komen. 
Een student is na afloop van de studie bij de FNWI in staat om:
\begin{description}
   \item[Academisch  --] ... op een academisch niveau te werken en te denken;
   \item[Zelf leren  --] ... zelfstandig en onder eigen verantwoordelijkheid te
   leren.
\end{description}

\subsection{Onderwijsinnovatie}
De faculteit innoveert haar onderwijs. Dit heeft uiteraard consequenties voor
de informatica en informatiekunde opleidingen.  Echter, informatiekunde is een
nieuwe opleiding. Hierbij is het belangrijk om er voor de zorgen dat eerst de
opleiding inhoudelijk goed in elkaar zit. Als het er dus om gaat waar de
informatiekunde de laatste spreekwoordelijke cent aan uit zou moeten geven,
dient de prioriteit vooralsnog bij de inhoud van de opleiding te liggen, en
niet de onderwijsvorm.  Dat impliceert niet dat de informatiekunde opleiding
tegen onderwijsinnovatie is. Het impliceert alleen dat informatiekunde
vooralsnog alleen \emph{re-actief} met informatica zal mee-innoveren, en totdat
de opleiding inhoudelijk goed in elkaar steekt niet \emph{pro-actief}
investeren in onderwijsinnovatie.

Dit vertaalt zich naar het volgende principe:
\begin{description}
   \item[Volgen in innovatie --] Het informatiekunde curriculum dient zich te
   conformeren aan de uitkomsten van de facultaire onderwijsinnovatie.
   Echter, het zorgen dat de (inhoudelijk) juiste vakken worden aangeboden
   binnen het informatiekunde curriculum heeft een \emph{hogere} prioriteit dan
   het inzetten van de juiste onderwijsvorm.
\end{description}

\section{Instituut}
Kwaliteit staat hoog in het vaandel bij het \NIII.
Denk hierbij aan: studeerbaarheid, roosterbaarheid, robuustheid en migreerbaarheid.
Met andere woorden:
\begin{description}
   \item[Essenti\"ele kwaliteiten --] 
   De inrichting van de opleiding dient zodanig te zijn dat deze
   (in volgorde van prioriteit) een verantwoord kwaliteitsniveau heeft met betrekking
   tot:
   \begin{enumerate}
      \item Studeerbaarheid
      \item Roosterbaarheid
      \item Robuustheid
      \item Migreerbaariheid
   \end{enumerate}
\end{description}

Samenwerking tussen het \NIII\ en de andere KUN Instituten \& Faculteiten is
noodzakelijk.  Dit geldt in het bijzonder voor informatiekunde, daar dit
vakgebied, meer nog dan informatica, een interdisciplinair vakgebied is met
diverse inhoudelijke specialisaties.  Dit laatste leidt in potentie al snel tot
complexe situatie met betrekking tot de roosterbaarheid van onderwijs en brengt
tevens het risico met zich mee dat de invulling van de resulterende opleiding
als los zand aan elkaar zit, omdat er weinig tot geen integratie tussen de
vakken is (o.a.\ over de (sub-)facultaire muren heen).
Om hier bewust mee om te springen zijn de volgende principes verwoord:
\begin{description}
   \item[Regie in eigen handen --] Informatiekunde onderwijs zoveel mogelijk onder
   directe (roostering \& inhoudelijke) controle van het \NIII\ georganiseerd
   te worden. Dit geldt met name voor onderwijs dat tot de kern van het 
   vakgebied behoord.

   \item[Gedeelde verantwoordelijkheid --] Als onderwijs niet onder directe
   \NIII\ controle georganiseerd kan worden, dan dienen de studenten hiervan op
   de hoogte te zijn. Ook van studenten wordt een actieve houding gevraagd om in de
   loop van het studiejaar de roosterbaarheid en studeerbaarheid optimaal te houden.
   Denk hierbij heel concreet aan het vroegtijdig signaleren van clashes in de 
   roostering.
\end{description}

De vakgebieden der informatiekunde en informatica vullen elkaar aan.
Dit dient dus ook in het onderwijs naar voren te komen. Na afloop van hun
studie zullen informatici en informatiekundigen nauw met elkaar samenwerken.
Het is verstandig de studenten deze samenwerking vanaf dag \'e\'en van de studie,
en bij voorkeur zelfs al tijdens de voorlichting, te laten ervaren.

Diverse vakken zullen qua inhoud overlap vertonen. Er zullen accentverschillen
zijn tussen beide opleidingen, maar diverse vakken zullen zich goed lenen voor
gemeenschappelijk onderwijs. Het gezamenlijk aanbieden van vakken verhoogt niet
alleen de efficiency waarmee het \NIII\ kan opereren, maar verhoogt ook het
wederzijds begrip van informatici en informatiekundigen. Dit vergt wel
expliciet een roostertechnische afstemming tussen beide curricula. Deze
intenties zijn vastgelegd in het volgende principe:
\begin{description}
   \item[Samen; met wederzijds respect --] Waar mogelijk moet het informatica en
   informatiekunde onderwijs gezamenlijk aangeboden worden.
    Dit moet juist niet alleen uit efficiency overwegingen gedaan worden.
    De beide vakgebieden moeten wel in hun waarde gelaten worden.
\end{description}

Bij de overgang van het studiepunten meetsysteem naar het ECTS meetsysteem is
enige onduidelijkheid ontstaan hoe de 60 ECTS die per jaar beschikbaar zijn
worden opgedeeld in tijdseenheden (semesters, trimesters of kwartalen?) en
vakken (3, 4, 5 of 6 ECTS?).  Het \NIII\ heeft er voor gekozen om in principe
een semestersysteem te gebruiken, waarbij een semester wel expliciet wordt
opgedeeld in twee kwartalen, zodat er ruimte is voor mid-semester tests en
kwartaalvakken.  Deze flexibiliteit is met name voor de informatiekunde
opleiding essenti\"eel daar er redelijkerwijze altijd sprake zal zijn van
vakken die door andere faculteiten gegeven worden. Qua omvang van vakken heeft
het \NIII\ besloten dat er gewerkt zal worden met vakken ter grootte van 3
of 6 ECTS. Er is momenteel geen Universitaire standaard met betrekking tot
de omvang van vakken. Dit betekent dat de informatiekunde opleiding zich
soms geconfronteerd zal zien met een afwijkende omvang van vakken. 
Samenvattend:
\begin{description}
   \item[Semester als standaard --] Het \NIII\ gebruikt semesters, onderbroken
   met een `rustperiode' ten behoeve van mid-semester tests, als standaard.
   Kwartaalvakken zijn toegestaan.
   \item[Standaardomvang --] Het \NIII\ gebruikt 3 en 6 ECTS als standaardomvang 
   voor vakken.
\end{description}

Voor beide \NIII\ opleidingen geldt dat er in de bachelorfase ruimte dient te
zijn voor studenten om op basis van hun persoonlijke interesses een eigen kleur aan
hun vakkenpakket te geven. Binnen het \NIII\ is
er voor gekozen om dit te realiseren door het benoemen van 30 ECTS aan wisselvakken
verspreid over de laatste anderhalf jaar van de opleiding. Studenten mogen deze vakken 
inwisselen tegen andere vakken binnen het \NIII\ cluster van opleidingen (Informatica
en Informatiekunde). 
\begin{description}
  \item[Wisselvakken --] Er wordt naar gestreefd om 30 ECTS aan wisselvakken te hebben
  in de bachelorfase. Dit geldt voor beide \NIII\ opleidingen.
\end{description}

%% file: wat-onderzoek.tex
\chapter{Visie op onderzoek}
\label{h:wat-onderzoek}

Het doel van dit hoofdstuk is om ons, gegeven de bovenstaande beschrijving van
het vakgebied, nader te bezinnen op het onderzoek wat in dit kader uitgevoerd
zou moeten worden.  Zoals eerder aangegeven richt de huidige versie van dit
document zich nog primair op de opleiding informatiekunde.  Door de beperkte
omvang van Informatiekunde staf, en als gevolg van het feit dat men zich
allereerst moest richten op het opzetten van de opleiding, is het
informatiekunde onderzoek nog niet echt van de grond gekomen. Daar staat
tegenover dat inmiddels al wel een en ander in de stijgers is gezet.

Binnen het informatiekunde-onderzoek zal onderzoek naar theorie\"en, methoden,
technieken en hulpmiddelen voor een betere afstemming tussen mens, organisatie,
informatievoorziening en technologie een centrale plek krijgen.  Adequate
theorie\"en, aanpakken en methoden met betrekking tot deze afstemming ontbreken
tot op heden. De tot dusverre ontwikkelde benaderingen komen doorgaans
rechtstreeks voort uit bestaand onderzoek op het gebied van menswetenschappen,
organisatiekunde, bedrijfskunde en informatica. De expliciete `brugfunctie'
ontbreekt daarbij echter nog. In de praktijk wordt de brug meestal nog op een
ad-hoc manier geslagen, op basis van persoonsgebonden visies en ervaringen.
Kort samengevat wordt informatiekunde gezien als het wetenschapsgebied wat zich
bezighoud (op macro en micro niveau) met de afstemming tussen de vier
essentiele elementen van de digitale samenleving: \emph{mens},
\emph{organisatie}, \emph{informatie} en \emph{technologie} waarbij de
technologie bedoelt is als ICT intensieve technologie (digitale technologie).
Hier liggen voor het wetenschappelijk onderzoek grote uitdagingen.  Als
belangrijke aandachtsgebieden voor onderzoek op dit terrein zien we:
\begin{itemize}
  \item Informatiearchitectuur als stuurmiddel
  \item Validatie \& Verificatie technieken/strategie\"en van systeemeisen
  \item Systeemtheoretische grondslagen
  \item Communicatietheoretische en informatietheoretische grondslagen
\end{itemize}
Tenslotte, citeren we, ter illustratie van het belang wat het \NIII\ hecht aan
het Informatiekunde onderzoek, de volgende passage uit het \NIII\
onderzoeksvisitatie rapport:
\begin{quote}\it
   In our view, Information Science is primarily an exact science.
   Even though a lot of the issues involved in Information Science are
   indeed issues that are traditionally attributed to business or human
   sciences, it is still necessary to approach the task of clearly
   identifying what the right software system for a given human and
   organisational context \emph{is} in a precise and unambiguous manner.
   
   The languages and models needed to analyse the human and organisational
   context, and to express \emph{all} requirements of the software system,
   require the rigour of an exact science.
   By the same token, however, the \emph{process} of gathering, eliciting
   and negotiating the requirements from/with the different stakeholders
   of a software system, do require skills and knowledge that
   can traditionally be found in human and/or business sciences.
   In other words, in our opinion, Information Science has a strong
   rooting in exact sciences, but should be enriched with aspects from
   human and business sciences.
   This also implies that Information Science research at the NIII should
   be conducted in close cooperation with other disciplines within the
   University of Nijmegen.
   Information Science can therefore also provide a bridge between 
   research at the NIII and other disciplines at the University; 
   a bridge that should have a very strong footing in the exact sciences.
   
   Active research in Information Science has only started in late 2002, 
   with the appointment of Prof.dr.\ H.A.\ Proper in IRIS and the appointment 
   of the UHD (associate professor) dr.ir.\ J.\ Tretmans in ST. 
   We expect that Information Science research will grow
   considerably over the next years. This belief
   is strengthened by some recent successes in this field:
   \begin{itemize}
     \item The start, in early 2002, of the NWO supported project: PRONIR
     (Profile based Retrieval Of Networked Information Resources). This project
     aims to define the need for information, as well as the supply of
     information, from a user centered point of view.

     \item The start up of a collaborative research project (ArchiMate) on
     animation and validation of software/information-system architectures.
     This project has a duration of 2 years, and involves a consortium of
     industrial and academic partners: ABN-Amro, ABP, Belastingdienst, Ordina,
     Telin, CWI, LIACS, and the NIII.
   
     \item The start up of the `Netherlands Architecture Forum', a foundation
     involving three classes of organisations involved in architectures of
     information systems: 
     \begin{itemize}
        \item Users of ICT, such as ABN-Amro, ABP, Belastingdienst,
        Essent, Shell, ING, Wehkamp, etc.
        \item ICT services providers, such as CGEY, Ordina, Atos,  
        Sogeti, HP, IBM, etc.
        \item Academic institutions, such as UT, Telin, NIII, UvT, etc.
     \end{itemize}
     The University of Nijmegen was one of the co-founders, and
     Prof.dr.\ H.A.\ Proper serves a member on the forum's programme
     committee.
   
     \item The SIKS research school has requested Prof.dr.\ H.A.\ Proper to 
     become the theme leader of the theme: Architecture-driven
     system development.
   \end{itemize}
\end{quote}

%% file: wat-opleiding.tex
\chapter{Visie op de opleiding}
\label{h:wat-opleiding}
Het doel van dit hoofdstuk is om ons, gegeven het vakgebied en de rol
van de informatiekundige daarbinnen, nader te bezinnen op de opleiding
die hiervoor nodig is.
Als zodanig richt dit hoofdstuk zich qua visie op het \emph{wat}
van de opleiding informatiekunde.
Achtereenvolgens zullen we stilstaan bij:
\begin{itemize}
  \item de uitdagingen met betrekking tot de opleiding die samenhangen
        met de inhoud van het vakgebied van de informatiekunde
        (paragraaf~\ref{p:uitdagingen});
  \item de voorbereiding van informatiekundigen op hun toekomstige rol in de
        maatschappij
        (paragraaf~\ref{p:beroep});
  \item het Bachelor/Master stelsel, met aspecten zoals internationalisering,
        instroom \& uitstroom, relatie tussen de twee \NIII\ opleidingen, etc.\
        (paragraaf~\ref{p:stelsel});
  \item en tenslotte een samenvatting van de inrichtingsprincipes voor
        de opleiding zoals deze volgen uit de visie op de opleiding
        (paragraaf~\ref{p:principes}).
\end{itemize}
In elke van de eerste drie paragrafen zullen we explicit aangeven wat
de inrichtingsprincipes zijn die volgen uit de specifieke deel-visie.

\section{Vakinhoudelijke uitdagingen}
\label{p:uitdagingen}
Kunnen de inhoudelijke kanten van een wetenschapsgebied om specifieke
ingredi\"enten in een opleiding vragen? Het lijkt misschien vreemd om een
dergelijke vraag te stellen, maar wanneer we het profiel van de
informatiekundige zoals beschreven in het vorige hoofdstuk nader beschouwen,
valt een aantal zaken op dat wel degelijk in deze richting zou kunnen gaan.
Laten we dit profiel nogmaals bekijken.

Zoals in hoofdstuk~\ref{h:waarom-vakgebied} is besproken dient, in termen van
de levenscyclus van een informatiesysteem, een informatiekundige een
substanti\"ele bijdrage te leveren aan de volgende soort activiteiten:
\begin{description}
    \item[Definitie --] dit betreft activiteiten met als doel het achterhalen van
    alle eisen (`requirements') waaraan het systeem en de systeembeschrijving
    zouden moeten voldoen.

    \item[Ontwerp --] hierbij gaar het om het proces dat als doel heeft het
    ontwerpen van een systeem conform de beschreven requirements. Het
    resulterende systeemontwerp kan vari\"eren van een ontwerp van de essentie op
    strategisch of achitectuur-niveau tot een detailontwerp welke raakt aan
    programmeer-statements of zeer specifieke handelingen die door een
    menselijke actor verricht moeten worden.

    \item[Constructie --] dergelijke activiteiten richten zich op het realiseren
    en testen van een systeem dat wordt beschouwd als een (mogelijk kunstmatig)
    samenhangend geheel van functionaliteiten dat \emph{nog niet operationeel
    is}.

    \item[Invoer --] hierbij gaat het om het operationeel maken van een systeem,
    m.a.w.\ het \emph{implementeren} van gebruik van het systeem door haar
    bedoelde gebruikers.

    \item[Bestendiging --] dit betreft activiteiten die bijdragen aan het
    ondersteunen, onderhouden of verder in de organisatie verankeren van een
    systeem. Men kan denken aan technisch onderhoud en het beter afstemmen van
    werkprocessen op het systeem, maar ook aan het geven van trainingen aan
    gebruikers of het schrijven of verbeteren van systeemdocumentatie.
\end{description}
Hierbij zijn informatiekundigen specifiek verantwoordelijk voor:
\begin{description}
    \item[Definitie --] De belangrijkste zorg van een informatiekundige tijdens
    dit proces zal er op gericht zijn een evenwichtig pakket aan eisen op te
    stellen met betrekking tot de externe en interne relaties van het beoogde
    informatie systeem, waarbij `evenwichtig' inhoudt, dat het resulterende systeem
    effectief is met betrekking tot de menselijke, organisatorische en
    technologische context.

    Informatiekundigen dienen het onderhandelingsproces dat zich hierbij
    doorgaans afspeelt tussen verschillende belanghebbende partijen, dienen te
    kunnen faciliteren en waar nodig bijsturen.

    \item[Ontwerp --] Het ontwerpen is een gezamenlijke verantwoordelijkheid van
    informatici en informatiekundigen. Daarnaast is de informatiekundige
    primair verantwoordelijk voor de bewaking van de belangen van de
    verschillende belanghebbenden.

    \item[Constructie --] In dit proces is de informatiekundige vooral
    verantwoordelijk voor het scheppen van de juiste voorwaarden in de
    menselijke en organisatorische context van het systeem als voorbereiding
    van de daadwerkelijke invoering.  Denk hierbij aan de ontwerp \& invoering
    van nieuwe werkprocessen, ontwerpen \& verzorgen van de opleiding van
    toekomstige gebruikers, etc.

    \item[Invoer --] Gedurende het daadwerkelijke uitrollen van het nieuwe systeem
    blijft de informatie\-kun\-di\-ge primair verantwoordelijk voor een goede `landing'
    van het systeem in de menselijke en organisatorische context.

    \item[Bestendiging --] Informatiekundigen zijn bij een operationeel systeem
    specifiek verantwoordelijk voor de blijvende aansluiting tussen het
    informatiesysteem en haar menselijke en organisatorische context. Dit kan er toe
    leiden dat een discrepantie wordt ontdekt, waarna een nieuwe ontwikkelcyclus
    gestart moet worden. Maar, denk in dit verband ook aan het up-to-date en beschikbaar
    houden van systeemdocumentatie (ontwerp \& gebruik), etc.
\end{description}

Als we dit globale profiel van informatiekundigen vertalen naar concretere
eigenschappen van een informatiekundige zelf, dan moet deze een academisch
gevormd persoon zijn die, meer specifiek, in staat is om:
\begin{itemize}
   \item een concrete vraag of situatie met betrekking tot een
   informatiesysteem systeem te \emph{analyseren}, de verschillende aspecten
   ervan kan zien evenals hun samenhang, daarover kan \emph{abstraheren} en
   deze kan \emph{modelleren}. Dat wil zeggen dat een informatiekundige niet
   alleen de vertaling moet kunnen maken van concreet naar abstract, maar
   tevens in staat moet zijn om voldoende afstand te nemen tot de concrete
   situatie om deze vanuit verschillend perspectief te kunnen zien en
   beoordelen. Men moet als het ware in staat zijn om een begrip ontwikkelen op
   meta-niveau.  Een en ander vraagt vaardigheden die over het algemeen meer
   liggen op het vlak van het geheel, de samenhang, dan op het vlak van de
   details.

   \item niet alleen het uiteindelijke \emph{product} voor ogen te hebben, maar
   die tevens oog heeft voor de essenti\"ele afstemmings- en
   onderhandelings\emph{processen} die hiervoor nodig zijn.

   \item met \emph{creatieve oplossingen} te komen voor situaties waarin veel
   verschillende perspectieven een rol spelen. Dit vraagt dat men \emph{los kan
   komen} van vertrouwde kaders, en nieuwe, onverwachte combinaties van
   mogelijkheden kan en durft te poneren. Dit punt is minder triviaal dat het
   in eerste instantie misschien lijkt; verderop komen we hierop nog
   uitgebreider terug.
\end{itemize}

Uit deze beschrijving komt de informatiekundige naar voren als iemand die een
$\beta$-opleiding met een goed ontwikkelde $\gamma$-feeling nodig heeft, met een
gezonde dosis creativiteit.
De $\beta$-opleiding is nodig om een exacte, analytische en modelmatige kijk en
attitude aan te kweken, terwijl het benodigde afstemmings\emph{proces} een
duidelijk $\gamma$-feeling vereist. Het behoeft geen discussie dat het vormgeven
van de digitale samenleving en het oplossen van de problemen die daarbij
onderweg gegarandeerd boven zullen komen drijven, een gezonde dosis aan
creativiteit vereist.

Bovenstaande profielschets van de informatiekundige heeft op verschillende
manieren consequenties voor de opleiding. Hieronder staan we stil bij de,
naar onze mening, belangrijkste consequenties.

\subsection{Focus op verbanden}
Er zal veel meer aandacht besteed moeten worden aan de \emph{verbanden} tussen
verschillende vakgebieden en perspectieven die bij de problematiek van het
ontwerpen van een adequaat informatiesysteem een rol spelen. De
interdisciplinariteit van de opleiding geeft al aan dat deze verbanden voor het
vakgebied een belangrijke rol spelen; echter, ook binnen de opleiding tussen de
vakken onderling zullen ze duidelijk naar voren moeten worden gebracht. Vakken
mogen niet in isolement \emph{naast} elkaar (kunnen) blijven staan; het
onderwijs zal een manier moeten vinden om de samenhang tussen verschillende
vakken duidelijk naar voren te brengen.
Met andere woorden:
\begin{description}
   \item[Focus op verbanden --] Er dient in de opleiding veel aandacht te zijn voor
   verbanden tussen gezichtspunten, vakgebieden \& verbredingsgebieden.

   \item[Geen verzuiling van de gezichtspunten --] De 4+1 gezichtspunten dienen
   eveneens expliciet in de opleiding naar voren te komen. Echter, hierbij
   dient verzuiling te worden voorkomen!  De nadruk moet liggen op een
   integrale visie op informatiesystemen vanuit de gezichtspunten, en de
   onderlinge impact van die gezichtspunten.
\end{description}
Bij de zoektocht naar werkvormen die zich hiervoor lenen is onder andere het
probleem-gestuurd onderwijs relevant (zie paragraaf \ref{s:Onderwijs}).

\subsection{Product- \& procesgericht}
Naast productgerichte onderdelen zullen
ook processen inhoudelijk onderwerp van studie moeten zijn. Studenten zullen
hierin meer inzicht moeten krijgen, en bijvoorbeeld de dynamiek en de
verschillende fases van processen moeten leren herkennen. Het gaat
hier om processen die met de inhoudelijke kant van het vakgebied te maken
hebben; op de aandacht voor het eigen \emph{leerproces} komen we in
paragraaf \ref{s:Onderwijs} terug. Dit leidt tot de volgende principes:
\begin{description}
   \item[Processen ook inhoud van studie --] De ontwikkelings-, bestendigings-,
   en aanbestedingspro\-ces\-sen dienen zelf ook inhoudelijk ontwerp van studie te
   zijn. Zowel in het onderwijs als in het onderzoek.

   \item[Taakgericht --] Er moeten in de opleiding vakken zijn die zowel vanuit
   theoretisch als praktisch perspectief het toekomstige takenpakket benaderen,
   waarbij informatica en informatiekunde studenten geacht worden nauw samen te
   werken.
\end{description}

\subsection{Linker- \`en rechterhersenhelft aanspreken}
Een derde punt is van een geheel andere orde, maar niet minder belangrijk.
Wanneer we de hierboven beschreven eigenschappen nader beschouwen, valt op dat
we hier over het algemeen te maken hebben met vaardigheden die worden
geassocieerd met het functioneren van de \emph{rechterhersenhelft} van de
neocortex (zie appendix \ref{h:brein}). Over de verschillende werking van
linker- en rechterhersenhelft is veel bekend
\cite{Book:98:Carter:HetBreininKaart,
Book:89:Edwards:DrawingontheRightSideoftheBrain,
Book:98:Brandhof:GebruikjeHersens}; gedurende de laatste jaren wordt echter ook
steeds meer gezien en onderkend dat deze verschillen belangrijke implicaties
hebben ten aanzien van de manier waarop wij informatie verwerken en leren, en
worden hieraan consequenties verbonden, ook richting onderwijs
\cite{Blase}.\footnote{Zie ook: \BrainStoreURL, \BrainStudioURL\ of
\LearningRevolURL\ voor een aantal recente publicaties over `breinvriendelijk
leren'.}

In onze Westerse maatschappij, en zeker in ons onderwijssysteem en in de
wetenschap, heeft echter-jarenlang de op analyse (segmentatie) en details
gerichte verwerkingswijze van de linkerhersenhelft voorop gestaan. De meer op
analogie\"en gerichte, holistische verwerkingswijze van de rechterhersenhelft
is hierbij op de achtergrond gebleven.\footnote{Vgl.\ de volgende uitspraak van
Roger Sperry \cite{Sperry}, wiens baanbrekend onderzoek eind zestiger en begin
zeventiger jaren de basis heeft gelegd voor het in kaart brengen van de
verschillen tussen linker- en rechterhersenhelft: `The main theme to emerge
{\ldots} is that there appear to be two modes of thinking, verbal and
nonverbal, represented rather separately in left and right hemispheres,
respectively, and that our educational system, as well as science in general,
tends to neglect the nonverbal form of intellect. What it comes down to is that
modern society discriminates against the right hemisphere.'}

In relatie tot bovengenoemde profielschets heeft dit belangrijke gevolgen.
Het betekent ten eerste, dat wij er niet automatisch van kunnen uitgaan dat
studenten de genoemde vaardigheden hebben, of dat zij weten hoe zij deze kunnen
ontwikkelen en toepassen. Een nog belangrijker probleem is, dat met name het
loslaten van de vertrouwde kaders (noodzakelijk voor het vinden creatieve
oplossingen) iets is dat door de verwerkingswijze van de linkerhersenhelft,
waarop in ons onderwijs juist zo'n sterk beroep wordt gedaan, juist wordt
\emph{belemmerd}~\cite{Book:98:Carter:HetBreininKaart,
Book:97:deBono:LeerUzelfDenken,
Book:89:Edwards:DrawingontheRightSideoftheBrain}.

Voor een opleiding die mensen wil voorbereiden op een rol als architect van de
digitale samenleving hebben deze observaties belangrijke consequenties.  Zij
impliceren dat aan deze `andere' (rechterhersenhelft) vaardigheden en denkwijze
expliciet aandacht zal moeten worden besteed, en dat wij deze moeten stimuleren
en mede helpen ontwikkelen waar nodig.  De uitdaging voor onze opleiding is dan
ook de mogelijkheden te cre\"eren waarop deze `andere'
informatieverwerkingswijze wordt gestimuleerd en benut, zonder dat de
wetenschappelijke werkwijze die iedere (toekomstige) academicus zich eigen moet
maken, verloren gaat. Zo wordt de informatiekundige iemand die het beeldende
vermogen van de architect verenigt met de analytische en exacte geest van de
exacte wetenschapper.

Kort samengevat leidt dit tot het volgende principe:
\begin{description}
   \item[Brede denkers --] In de opleiding dient er ook aandacht besteed te
   worden aan holistische en niet-verbale denkprocessen.
\end{description}

\subsection{Aandacht voor veranderingen}
\label{ss:Veranderingen}
Als er \een\ gebied is waarin veranderingen zich in een razendsnel tempo
voltrekken, dan is het wel dat van de informatisering en digitalisering van de
samenleving. Deze snelle ontwikkelingen en veranderingen in wetenschap en
maatschappij maken een nieuwe focus voor ons onderwijs noodzakelijk. In een
passage onder de titel `Een nieuw doel voor het onderwijs: een klimaat voor
verandering.'\ zegt Carl Rogers \cite{rogers} hierover het volgende:
\begin{quote}\small
   De wereld is in een veelvoudig versneld tempo aan het veranderen. Wil onze
   samenleving het hoofd bieden aan de uitdaging van de duizelingwekkende
   veranderingen in wetenschap, technologie, communicatie en sociale verhoudingen,
   dan kunnen wij niet steunen op de ons in het verleden verschafte
   \emph{antwoorden}, maar moeten wij ons vertrouwen stellen in de
   \emph{processen} waarmee men het hoofd biedt aan nieuwe problemen. Want de
   veranderingen halen ons z\'o snel in dat antwoorden, `kennis', methoden en
   vaardigheden bijna op hetzelfde moment dat ze worden gevonden alweer verouderd
   zijn.

   Dit impliceert niet alleen nieuwe onderwijstechnieken maar ook (\ldots)
   een nieuwe doelstelling. In de wereld op wier drempel wij ons reeds
   bevinden moet het doel van het onderwijs zijn de ontwikkeling van mensen
   die openstaan voor verandering. Alleen zulke mensen kunnen op
   constructieve wijze de verwarringen tegemoet treden van een wereld waarin
   de problemen veel sneller uit de grond schieten dan de antwoorden daarop.
   Het doel van het onderwijs moet zijn een samenleving tot ontwikkeling te
   brengen waarin de mensen gemakkelijker met \emph{verandering} dan met
   \emph{starheid} kunnen leven. In de komende wereld is het vermogen om het
   nieuwe op gepaste wijze onder ogen te zien van meer belang dan de
   vaardigheid om het oude te kennen en te herhalen. (\ldots)

   Er moet een weg gevonden om, binnen de onderwijsorganisatie als geheel en
   in elke component ervan, een klimaat tot ontwikkeling te brengen dat
   bevorderlijk is voor persoonlijke groei, een klimaat waarin vernieuwingen
   niet afschrikken, waarin de creatieve capaciteiten van bestuurders,
   leraren en leerlingen niet verstikt, maar veeleer gevoed en tot uiting
   gebracht worden.  Er moet een weg gevonden worden om in het
   \emph{systeem} een klimaat te ontwikkelen waarin de aandacht zich niet
   concentreert op het \emph{onderwijzen}, maar op de bevordering van het
   zelf geleide \emph{leren}. Alleen z\'o kunnen wij de creatieve mens tot
   ontwikkeling brengen die openstaat voor en zich bewust is van al zijn
   ervaringen, ze aanvaardt in het voortdurende proces van verandering. En
   alleen op deze manier kunnen wij geloof ik de creatieve
   onderwijsorganisatie tot stand brengen, die ook voortdurend in
   verandering zal zijn.
\end{quote}
Rogers' analyse (die men eveneens kan terugzien in de idee\"en van de
Zwitserse leerpsycholoog Jean Piaget \cite{Elkind}) maakt de veranderde
context van het onderwijs duidelijk, en laat zien welke belangrijke taak
haar wacht. Deze nieuwe taak werkt niet alleen door op de \emph{uitvoering}
van ons onderwijs (deze consequenties op algemeen onderwijskundig vlak
worden meer uitgebreid besproken in hoofdstuk~\ref{h:wat-onderwijs}
(zie paragraaf~\ref{s:veranderingen-onderwijs}). Studenten dienen
gestimuleerd te worden om zich te ontwikkelen tot creatieve mensen die
openstaan voor, en kunnen omgaan met veranderingen. Ook in de \emph{inhoud
en doelstellingen} van ons onderwijs zullen veranderingsprocessen daarom
expliciet punt van aandacht moeten zijn.

Bovenstaande discussie resulteert in het volgende principe:
\begin{description}
   \item[Ingebouwde dynamiek --] De inrichting van de opleiding dient zo
   gekozen te zijn dat vakken up-to-date (moeten en) kunnen blijven zonder dat
   dit gezien moet worden als een curriculum wijziging.

   \item[Aandacht voor trends --] Hoewel het voor een academische opleiding
   essenti\"eel is zich vooral te focussen op de onderliggende theorie\"en,
   dient er in de opleiding toch aandacht te zijn voor hedendaagse
   trends in het vakgebied en hun relatie naar de dieperliggende theorie\"en.
   Op technologisch vlak zal zich dit onder andere uiten in aandacht voor
   actuele ontwikkelingen, zoals web-services, XML, middleware, etc.
\end{description}

\subsection{Aandacht voor verbredingsgebieden}
Voor informatiekundigen is het, meer nog dan bij informatica het
geval is, belangrijk om enige kennis te hebben van de vele
verbredingsgebieden.  Omdat de Katholieke Universiteit Nijmegen
een vrij brede universiteit is, kunnen studenten de mogelijkheid
bieden zich op diverse verbredingsgebieden te \emph{ori\"enteren}.
Concreet kunnen we denken aan:
\begin{itemize}
   \item Medische informatiekunde i.s.m.\ medische wetenschappen.
   \item Beleidsondersteunende informatiesystemen i.s.m.\
   managementwetenschappen.
   \item Duurzame ontwikkeling van ICT in ontwikkelingslanden i.s.m.\
   ontwikkelingsstudies.
   \item Juridisch kennisbeheer i.s.m.\ rechten.
   \item Bio-informatica i.s.m.\ natuurwetenschappen.
   \item Kunstmatige intelligentie i.s.m.\ sociale wetenschappen.
   \item Taal- en spraaktechnologie i.s.m.\ taalwetenschappen.
\end{itemize}
Hierbij is het zeker niet de bedoeling dat studenten zich
exclusief op \'e\'en verbredingsgebied toeleggen. De reden
hiervoor is, dat (met name) de informatiekundige ook in staat zal
moeten zijn op een abstracter niveau te kijken naar de afstemming
van een ontwerp op de menselijke en organisatorische context. Door
studenten met meerdere verbredingsgebieden te confronteren stellen
we hen in staat ook te leren reflecteren op de \emph{verschillen}
en de \emph{overeenkomsten} tussen deze gebieden, en de
consequenties hiervan voor de optimale afstemming van een
socio-technisch systeem. Het verwerven van deze meer abstracte
inzichten en het kunnen vertalen van deze inzichten naar de
praktijk zal de student in staat stellen zich relatief makkelijk
in te werken in nieuwe verbredingsgebieden, waarmee hij optimaal
wordt voorbereid op een flexibele inzetbaar\-heid in de latere
beroepspraktijk.

Samenvattend volgt hieruit het volgende inrichtingsprincipe:
\begin{description}
   \item[Verbredingsgebieden --] De opleiding moet stil staan bij de
   diversiteit aan verbredingsgebieden. De focus moet hierbij liggen op het
   kunnen reflecteren over de \emph{verschillen} en de \emph{overeenkomsten}
   tussen de diverse verbredingsgebieden, evenals het kunnen inwerken in nieuwe
   verbredingsgebieden.

   De specifieke verbredingsgebieden, zoals die op de KUN reeds worden
   onderzocht en gedoceerd (medische informatiekunde, taal \&
   spraaktechnologie), dienen hierbij een illustrerende rol te hebben.
\end{description}

\section{Beroepsvorming}
\label{p:beroep}
In de opleiding informatiekunde zullen studenten de kennis van relevante
theorie\"en, methoden, technieken en hulpmiddelen tot zich moeten nemen
en hiermee moeten leren werken.
Met betrekking tot het beroep dat een informatiekundige kan uitvoeren na de
opleiding voorzien we in essentie drie primaire richtingen:
\begin{description}
  \item[Onderzoeker --] Informatiekundigen die onderzoek gaan verrichten naar de
        grondslagen van het informatiekunde vakgebied.
  \item[Opleider --] Informatiekundigen die op een universiteit, hogeschool,
        of in het bedrijfsleven anderen het vak leren.
  \item[Vakman --] Informatiekundigen die in het bedrijfsleven gaan werken en
    het vakgebied in de praktijk uitvoeren.
\end{description}

De vaardigheden die nodig zijn voor deze verschillende beroepsvormen zijn niet
exclusief voor deze vormen. Iemand die als vakman in het bedrijfsleven werkt
heeft wel degelijk onderzoeks\-vaar\-dig\-heden nodig, en moet ook een cursus kunnen
geven.  Een onderzoeker zal presentaties/cursussen moeten kunnen geven over het
onderzoek, en zal ook in staat moeten zijn om een gezonde inhoudelijke dialoog
aan te gaan met vakmensen om wederzijds theoretische/empirische kennis uit te
wisselen.  In de praktijk moeten informatiekundigen daarom een aantal
vaardigheden beheersen die voor alle beroepsrichtingen relevant zijn.

Om die reden kunnen we stellen dat \emph{alle} afgestudeerden in principe
gedurende hun opleiding een zekere basis moeten hebben gekregen op het gebied
van
\begin{itemize}
   \item onderzoeksvaardigheden,
   \item doceer- \& presentatievaardigheden en
   \item praktijkgerichte vaardigheden.
\end{itemize}
In relatie tot de Bachelor-Master structuur (zie paragraaf~\ref{p:stelsel})
kunnen we hieraan een concretere duiding geven. Meer specifiek betekent dit:
\begin{description}
   \item[Brede beroepsorientering --] In de bachelorfase zal er geen
   expliciete voorsortering zijn op een specifieke beroepsrichting. In de
   bachelorfase zal er voor alle studenten aandacht zijn voor:
   \begin{itemize}
      \item Onderzoeksvaardigheden
      \item Doceer- \& presentatievaardigheden
      \item Praktijkgerichte vaardigheden
   \end{itemize}
\end{description}

Orthogonaal op de beroepsvormen kunnen we ook een onderscheid maken in het
`bestu\-rings\-ni\-veau' waarmee men deze beroepen uitvoert, met andere woorden, of
men op een uitvoerend-, besturend- (manager) of beleidsmatig niveau werkzaam
is. In de opleiding zal daarom naast uitvoerende vaardigheden ook aandacht besteed
worden aan besturende en beleidsmatige vaardigheden:
\begin{description}
   \item[Uitvoeren \& besturen --] In de opleiding zal er zowel aandacht zijn
   voor uitvoerende, besturende, als beleidsmatige aspecten van het vakgebied.
\end{description}
In de masterfase zal er wel een vorm van specialisatie naar beroepsvorm kunnen
plaatsvinden. Echter, doordat bij informatiekunde de masterfase beperkt is tot
\'e\'en jaar, zal ook die specialisatie noodgedwongen een beperkt karakter
hebben.

\section{Bachelor-Master stelsel}
\label{p:stelsel}
De invoering van het Bachelor-Master stelsel brengt diverse nieuwe kansen en
bedreigingen met zich mee.
In deze paragraaf gaan we hier nader op in.

\subsection{Instroom}
De instroom van studenten is te splitsen in een bachelor instroom en een Master
instroom. Bij de bachelor instroom gaat het vooral om VWO studenten die een
full-time studie gaan beginnen. Voor deze groep blijft in eerste instantie
alles (min of meer) bij het oude, met dien verstande dat de nieuwe
vooropleiding in de vorm van de tweede fase in het middelbaar onderwijs een
zelfstandiger en gemotiveerder student lijkt op te leveren.\footnote{Uit
onderzoek van de KUN bleek recentelijk dat de tweede fase student in het eerste
jaar beter presteert en een groter aantal studiepunten behaalt als zijn collega
oude stijl.} Het onderwijs zal op deze veranderde studiehouding en -achtergrond
moeten worden afgestemd.

Voor de instroom van studenten richting de masterfase zien we twee
hoofdgroepen:
\begin{description}
   \item[Doorstromers --] Studenten die een informatiekunde bachelor hebben
   gedaan bij het \NIII;
   \item[Zij-instromers --] Studenten die geen informatiekunde bachelor hebben
   gedaan bij het \NIII.
\end{description}
Studenten die reeds een bachelor studie informatiekunde volgen zullen zoveel
mogelijk gestimuleerd worden om die te bekronen met een Master-jaar.  Conform
facultair beleid, dient de inrichting van het onderwijs hier ook op in te
spelen. Het is hierbij natuurlijk ook onze taak te zorgen dat de bachelor
opleiding dermate goed is dat de studenten het vertrouwen hebben dat ook het
Master-jaar van voldoende kwalitatief niveau is.  We lopen echter het risico
dat zodra de arbeidsmarkt weer aantrekt, er een zuiging vanuit het
bedrijfsleven zal ontstaan waardoor studenten al na hun bachelor studie
uitstromen. Het is belangrijk om van de huidige luwte op de arbeidsmarkt
gebruik te maken op bedrijven te overtuigen dit niet te doen, en hun
verantwoordelijkheid te nemen en studenten stimuleren eerst hun opleiding
\emph{echt} af te ronden in de vorm van een Master-jaar. Daarnaast moet de
mogelijkheid van een part-time masterfase, naast de reguliere full-time
opleiding, op voorshands zeker niet uitgesloten worden.\footnote{Met enige
regelmaat bereiken ons ook verzoeken van belangstellenden voor een dergelijke
part-time variant.}

Voor de zij-instromers geldt dat er afhankelijk van de vooropleiding sprake zal
moeten zijn van een schakelprogramma. Hierbij moet opgemerkt worden dat de
informatiekunde opleiding in eerste instantie een \'e\'enjarige masterfase zal
kennen. Aangezien ook voor deze zij-instromers een half jaar is gereserveerd
voor afstudeerwerkzaamheden, betekent dit dat er effectief slechts een half
jaar tijd is voor het volgen van vakken. Dat betekent dat het vrijwel
onmogelijk is om grote delen van een schakelprogramma `op te vangen' in dat
halve jaar (met name gezien het feit dat vakken vaak op elkaar volgen qua
voorkennis). In de praktijk kan hierin een iets grotere flexibiliteit worden
verkregen door het schakelpakket uit te smeren over de totale periode van 1,5
jaar die de zij-instromer tot zijn beschikking heeft. Niettemin zal het
voornaamste deel van het schakelpakket om inhoudelijke gronden vooraf moeten
gaan aan de onderdelen uit het masterprogramma.  Een realistisch schakelpakket
van een half jaar kan daarom alleen voor zij-instromers van relevante
vooropleidingen worden gedefinieerd. Hieronder verstaan we:
\begin{quote}
   Een afgeronde HBO Bachelor (4 jaar) of Academische Bachelor (3 jaar) in:
   informatica, informatiekunde, bedrijfs-informatietechnologie of bedrijfs-
informatiekunde.
\end{quote}
Het is de bedoeling ervoor te zorgen dat studenten die uit een dergelijke
studie instromen binnen 1.5 jaar na aanvang hun Masters kunnen afronden.
Met andere woorden:
\begin{description}
   \item[Doorstroomprogramma --] Voor zij-instromers dienen er
   doorstroomprogramma beschikbaar te zijn. Uitgaande van een relevante
   vooropleiding mag dit programma niet meer dan een ${1}\over{2}$ jaar aan
   studietijd kosten.
\end{description}
Hierbij zal er zoveel mogelijk naar gestreefd worden afspraken te maken met
`toeleverende' opleidingen met betrekking tot de afstemming tussen de
programma's. Denk hierbij met name aan HBO instellingen in de omgeving van
Nijmegen. In het kader van dergelijke afspraken is het wellicht mogelijk het
schakelprogramma in omvang af te laten nemen.

Studenten die geen relevante vooropleiding hebben stromen in principe in in de
bachelorfase van de opleiding informatiekunde, en dienen deze af te ronden
Hierbij zal per individueel geval gekeken worden naar mogelijke vrijstellingen
op basis van vakken uit de gevolgde vooropleidingen.

\subsection{Schakelen tussen informatica \& informatiekunde}
Momenteel zal een overgang van informatica naar informatiekunde maximaal een
half jaar vertraging opleveren na een afgeronde bachelor informatica. Het ligt
in de bedoeling om het schakelen tussen de \NIII\ opleidingen met zo min
mogelijk studievertraging te laten verlopen, en ook tussentijds schakelen
mogelijk te maken.
\begin{description}
   \item[Makkelijk schakelen --] De \NIII\ informatica en informatiekunde
   opleidingen dienen zodanig opgezet te worden dat het schakelen tussen de
   twee studies zo min mogelijk impact heeft op het studieverloop van de
   studenten.
\end{description}
Dit streven wordt natuurlijk vergemakkelijkt door het feit dat
informatiekunde vakken deels samen gegeven worden met informatica.

\subsection{Uitstroom}
Het beleid van de faculteit is er duidelijk op gericht studenten na afloop van
een bachelorfase zoveel mogelijk te laten instromen in de bijbehorende
masterfase.  Voor de informatiekunde opleiding wordt dit beleid overkort
overgenomen. Hierbij is een duidelijke Nijmeegse focus essenti\"eel, waarbij
het voorstel is om in eerste instantie te kiezen voor een profilering op
Informatiearchitectuur.
\begin{description}
   \item[Profilering van de Master --] Voor de \NIII\ Master Informatiekunde
   zal in eerste de specialisatie op Informatiearchitectuur expliciet
   geprofileerd worden.  Wellicht dat er op termijn nog specialisaties bijkomen,
   maar het is belangrijk ons eerst goed te specialiseren in \'e\'en
   specialisatierichting.
\end{description}
Merk op dat het verstandig is om in het komende academische jaar goed af te
stemmen wat betreft doorstroommogelijkheden naar verwante Master's binnen de
KUN. Denk aan:
\begin{itemize}
  \item Informatica
  \item Medische Informatiekunde
  \item Informatiemanagement
  \item Cognitiewetenschappen
\end{itemize}
Zeker als de \NIII\ Master informatiekunde zich specialiseerd op
Informatiearchitectuur, is het goed om heldere afspraken te maken
voor alternatieve paden. Hiermee kunnen we ons vervolgens als KUN
beter profileren in `de markt'.

Daarnaast zal het bachelorfase zodanig vormgegeven worden dat er een duidelijke
inhoudelijke verwachting naar het laatste jaar ontstaat:
\begin{description}
   \item[Cliff-hanger --] De bachelorfase dient zodanig ingericht te zijn dat
   er als vanzelf een `honger naar meer' ontstaat bij de studenten. De
   masterfase dient te voldoen aan die honger.
\end{description}

\subsection{Internationalisering van Bachelor-Master studies}
De internationalisering is van toenemende betekenis in het wetenschappelijk
onderwijs in Nederland~\cite{BeleidsbriefInternationaal}. In de een
wetenschappelijke opleiding is een krachtige internationale ori\"entatie
onontbeerlijk om studenten voor te bereiden op Europees en mondiaal
burgerschap. Voor zowel informatiekundigen als informatici geldt dit nog eens
extra omdat veel van de toekomstige werkgevers van een trans-nationale aard
zullen zijn.

Daarnaast biedt de Bachelor-Master structuur de gelegenheid om, met name op
Master's niveau, studenten van buiten Nederland te werven. Traditioneel kijkt
de de Katholieke Universiteit Nijmegen, gelegen in het midden tussen
`Holland' en `Ruhr', hierbij naar Duitsland.  Er is echter geen enkele
reden om niet verder te kijken dan Duitsland en Nederland als het gaat om het
werven van studenten.

Het is hierbij onvermijdelijk dat Engels als voertaal in een wetenschappelijke
opleiding een steeds belangrijker plaats inneemt. In de meeste Europese
samenwerkingsverbanden is Engels de voertaal. Dit geldt eveneens voor veel
trans-nationale bedrijven; zelfs wanneer deze bedrijven hun hoofdzetel in
Nederland hebben.

Met het oog op deze ontwikkelingen zal binnen de informatiekunde opleiding het
Engels dan ook een belangrijke rol gaan vervullen. Als principe willen we bij
de invoering van het Curriculum 2003, er daarom naar streven dat:
\begin{description}
   \item[Engels --] De engelse en nederlandse taal als volgt gebruiken:
   \begin{itemize}
      \item Leerstof -- Bachelor: optie, Master: Engels
      \item Tentamens -- Bachelor: Nederlands, Master: Engels
      \item Voorlichtingsmateriaal -- Bachelor: Nederlands, Master: beide
      \item Offici\"ele reglementen -- Bachelor: Nederlands, Master: beide
      \item Colleges -- Bachelor: optie, Master: Engels
   \end{itemize}
\end{description}

\section{Samenvatting inrichtingsprincipes}
\label{p:principes}
Tot slot van dit hoofdstuk noemen we ook hier weer even de kort de
inrichtingsprincipes voor de inrichting van de opleiding informatiekunde zoals
die volgen uit de bovenstaande visie op de opleiding.
\begin{description}
   \item[Focus op verbanden --] Er dient in de opleiding veel aandacht te zijn voor
   verbanden tussen gezichtspunten, vakgebieden \& verbredingsgebieden.

   \item[Geen verzuiling van de gezichtspunten --] De 4+1 gezichtspunten dienen
   eveneens expliciet in de opleiding naar voren te komen. Echter, hierbij
   dient verzuiling te worden voorkomen!  De nadruk moet liggen op een
   integrale visie op informatiesystemen vanuit de gezichtspunten, en de
   onderlinge impact van die gezichtspunten.

   \item[Processen ook inhoud van studie --] De ontwikkelings-, bestendigings-,
   en aanbestedingspro\-ces\-sen dienen zelf ook inhoudelijk ontwerp van studie te
   zijn. Zowel in het onderwijs als in het onderzoek.

   \item[Taakgericht --] Er moeten in de opleiding vakken zijn die zowel vanuit
   theoretisch als praktisch perspectief het toekomstige takenpakket benaderen,
   waarbij informatica en informatiekunde studenten geacht worden nauw samen te
   werken.

   \item[Brede denkers --] In de opleiding dient er ook aandacht besteed te
   worden aan holistische en niet-verbale denkprocessen.

   \item[Ingebouwde dynamiek --] De inrichting van de opleiding dient zo
   gekozen te zijn dat vakken up-to-date (moeten en) kunnen blijven zonder dat
   dit gezien moet worden als een curriculum wijziging.

   \item[Aandacht voor trends --] Hoewel het voor een academische opleiding
   essenti\"eel is zich vooral te focussen op de onderliggende theorie\"en,
   dient er in de opleiding toch aandacht te zijn voor hedendaagse
   trends in het vakgebied en hun relatie naar de dieperliggende theorie\"en.
   Op technologisch vlak zal zich dit onder andere uiten in aandacht voor
   actuele ontwikkelingen, zoals web-services, XML, middleware, etc.

   \item[Verbredingsgebieden --] De opleiding moet stil staan bij de
   diversiteit aan verbredingsgebieden. De focus moet hierbij liggen op het
   kunnen reflecteren over de \emph{verschillen} en de \emph{overeenkomsten}
   tussen de diverse verbredingsgebieden, evenals het kunnen inwerken in nieuwe
   verbredingsgebieden.

   De specifieke verbredingsgebieden, zoals die op de KUN reeds worden
   onderzocht en gedoceerd (medische informatiekunde, taal \&
   spraaktechnologie), dienen hierbij een illustrerende rol te hebben.

   \item[Brede beroepsorientering --] In de bachelorfase zal er geen
   expliciete voorsortering zijn op een specifieke beroepsrichting. In de
   bachelorfase zal er voor alle studenten aandacht zijn voor:
   \begin{itemize}
      \item Onderzoeksvaardigheden
      \item Doceer- \& presentatievaardigheden
      \item Praktijkgerichte vaardigheden
   \end{itemize}

   \item[Uitvoeren \& besturen --] In de opleiding zal er zowel aandacht zijn
   voor uitvoerende, besturende, als beleidsmatige aspecten van het vakgebied.

   \item[Doorstroomprogramma --] Voor zij-instromers dienen er
   doorstroomprogramma beschikbaar te zijn. Uitgaande van een relevante
   vooropleiding mag dit programma niet meer dan een ${1}\over{2}$ jaar aan
   studietijd kosten.

   \item[Makkelijk schakelen --] De \NIII\ informatica en informatiekunde
   opleidingen dienen zodanig opgezet te worden dat het schakelen tussen de
   twee studies zo min mogelijk impact heeft op het studieverloop van de
   studenten.

   \item[Profilering van de Master --] Voor de \NIII\ Master Informatiekunde
   zal in eerste de specialisatie op Informatiearchitectuur expliciet
   geprofileerd worden.  Wellicht dat er op termijn nog specialisaties bijkomen,
   maar het is belangrijk ons eerst goed te specialiseren in \'e\'en
   specialisatierichting.

   \item[Cliff-hanger --] De bachelorfase dient zodanig ingericht te zijn dat
   er als vanzelf een `honger naar meer' ontstaat bij de studenten. De
   masterfase dient te voldoen aan die honger.

   \item[Engels --] De engelse en nederlandse taal als volgt gebruiken:
   \begin{itemize}
      \item Leerstof -- Bachelor: optie, Master: Engels
      \item Tentamens -- Bachelor: Nederlands, Master: Engels
      \item Voorlichtingsmateriaal -- Bachelor: Nederlands, Master: beide
      \item Offici\"ele reglementen -- Bachelor: Nederlands, Master: beide
      \item Colleges -- Bachelor: optie, Master: Engels
   \end{itemize}
\end{description}

%% file: wat-onderwijs.tex
\chapter{Visie op onderwijs en leren}
\label{h:wat-onderwijs}

Bij het concretiseren van de visie op het vakgebied en de visie op de opleiding
in termen van een concreet opleidingsprogramma, krijgen we ook te maken met
keuzes ten aanzien van de specifieke inrichting van het onderwijs en het leren.
In dit hoofdstuk worden deze 
achtergronden en onze visie vanuit de informatiekunde opleiding hierop nader 
belicht. 
Achtereenvolgens staan we stil bij:
\begin{itemize}
  \item algemene maatschappelijke ontwikkelingen die van invloed zijn op 
        onderwijs en leren (paragraaf~\ref{s:Maatschappij});
 
  \item algemene onderwijskundige overwegingen en achtergronden 
        (paragraaf~\ref{s:Onderwijs});

  \item onderwijsstrategie\"en die relevant zijn voor de informatiekunde 
opleiding
        (paragraaf~\ref{s:Strategie});

  \item en tenslotte een samenvatting van de inrichtingsprincipes 
        voor de opleiding, en de gewenste vaardigheden van informatiekundigen,
        zoals deze volgen uit bovenstaande achtergronden 
        (paragraaf~\ref{s:Principes}).
\end{itemize}

\section{Maatschappelijke ontwikkelingen}
\label{s:Maatschappij}
In hoofdstuk~\ref{h:waarom-vakgebied} is duidelijk gemaakt dat de behoefte aan
informatiekundigen, in de rol van architect van de digitale samenleving, en
daarmee de noodzaak voor een opleiding daartoe, in sterke mate is gebaseerd op
maatschappelijke en technologische ontwikkelingen van de laatste decennia. Ook
met betrekking tot de feitelijke inrichting van het onderwijs zijn deze
ontwikkelingen relevant. 

\subsection{ICT en leren}
De ontwikkeling van de ICT werkt niet alleen inhoudelijk door
in het onderwijs.  Ook voor het geven van onderwijs hebben de ICT
ontwikkelingen grote consequenties.  Het inbedden van ICT in
het onderwijs kan bijvoorbeeld leiden tot een heel andere opzet van een les of
college. Maar het gebruik van ICT technologie als onderwijshulpmiddel
vraagt ook andere vaardigheden van de leerder. Dit is in praktisch opzicht het
geval (`Hoe vind ik de juiste informatie in de enorme hoeveelheid beschikbare
gegevens op het Internet? Welke zoekmachines kan ik gebruiken?  Hoe werken
ze?'), maar is ook van toepassing op het cognitieve vlak. De beschikbaarheid
van ICT, internet en de enorme hoeveelheid informatie die daarmee samenhangt,
leidt tot andere prioriteiten. Het is niet meer doenlijk en ook niet meer nodig
om al deze informatie in het hoofd op te slaan; veel belangrijker wordt het om
snel te kunnen \emph{selecteren} en de informatie te \emph{integreren}. Het
onderwijs kan hierop inspelen door andere prioriteiten te stellen, en
bijvoorbeeld minder nadruk te leggen op het hebben van parate kennis, maar
juist meer op het kunnen vinden van de juiste kennis voor specifieke situaties,
en deze vervolgens ook goed te kunnen toepassen in de gegeven situatie. Ook in
de tentaminering zou deze verandering van focus door moeten werken.  Laten we
hierbij vooral niet vergeten dat door de dynamische aard van de ICT,
informatiekundigen en informatici gedoemd zijn tot een loopbaan van blijvend
leren. Dit leidt tot de wens dat een informatiekundige iemand is die in staat 
is om:
\begin{description}
   \item[Kennis ontsluiten --] ... kennis- en ervaringsbronnen te ontsluiten, 
         voorzover deze aansluiten bij hun reeds bestaande kennis;
   \item[Kennis inzetten --] ... voor informatiekundige problemen, relevante 
         kennisgebieden aan te geven en hun mogelijke bijdrage aan de 
         oplossing van het probleem te identificeren, en waar relevant 
         deze kennis inzetten bij het oplossen van het probleem.
\end{description}

Het gebruik van moderne ICT technologie in het onderwijs biedt tevens goede
mogelijkheden om de toekomstige architecten van de digitale samenleving zelf
`bloot te stellen' aan diezelfde ICT.  Het experimenteren met het gebruik en de
inzet van de modernste ICT ontwikkelingen, zoals een `wireless classroom' via
een campus-wide wireless netwerk\footnote{Zie bijvoorbeeld: \WirelessClassURL},
groupware tools, digitale werkplaatsen\footnote{Zie bijvoorbeeld: \DigiWerkURL\
of \DigiUniURL}, etc, zullen daarom, mits er een duidelijke bijdrage aan het
onderwijsproces is aan te wijzen, worden gestimuleerd.  Voor de inrichting van
de opleiding leidt dit tot de volgende principes:
\begin{description}
   \item[Ondervinden van ICT --] De snelle opmars van ICT moet
         ook doorklinken in de opleiding. Studenten dienen daarom
         dan ook ICT aan den lijve te ondervinden, bijvoorbeeld door
         het inzetten van moderne ICT in het onderwijs zelf.
\end{description}

\subsection{Het snelle tempo van veranderingen}
\label{s:veranderingen-onderwijs}
Het is niet alleen de beschikbaarheid van ICT die aanleiding 
geeft tot deze verandering in focus in het onderwijs. Ook de snelle ontwikkelingen 
en veranderingen in wetenschap en maatschappij maken een dergelijke verandering
noodzakelijk. In paragraaf~\ref{ss:Veranderingen} hebben we hier al bij
stilgestaan, door middel van het citaat van Rogers~\cite{rogers}.  Rogers'
analyse (die men eveneens kan terugzien in de idee\"en van de Zwitserse
leerpsycholoog Jean Piaget \cite{Elkind}) maakt de veranderde context van het
onderwijs duidelijk, en laat zien welke belangrijke taak haar wacht. Het geeft
aanleiding tot een taakverdeling waarin ICT een ondersteunende
rol speelt voor de creatieve mens die oplossingen zoekt. 
Dit kan samengevat worden tot het volgende principe:
\begin{description}
   \item[Aandacht voor creativiteit en verandering --] Studenten dienen gestimuleerd te 
         worden om zich te ontwikkelen tot creatieve mensen die openstaan voor, en 
         kunnen omgaan met veranderingen.
\end{description}

Het ontwikkelen van de creatieve mens die openstaat voor veranderingen heeft
niet alleen consequenties voor de inhoud van ons onderwijs. Ook op meer
algemeen onderwijskundig vlak werkt dit door.  Wanneer wij mensen willen
opleiden die inderdaad kunnen omgaan met de vele veranderingen in maatschappij,
en met de snelle groei van wetenschappelijke inzichten, dan is het allereerst
van belang dat wij hen in staat stellen te leren hoe zij zichzelf kunnen
ontplooien; te leren hoe zij hun eigen (en evt.\ andermans) capaciteiten
optimaal kunnen benutten en onder welke omstandigheden zij optimaal
functioneren. Hiervoor is een \emph{bewustwordingsproces} nodig dat zij kunnen
doormaken wanneer zij zelf actief betrokken zijn bij hun eigen leerproces, door
daarin eigen verantwoordelijkheid te dragen en eigen keuzes te maken. Hiervoor
moet ons onderwijssysteem de ruimte bieden. Dit betekent qua vaardigheden dat
een informatiekundige ook in staat moet zijn om:
\begin{description}
   \item[Reflectie op leren --] ... te reflecteren op het eigen leerproces (of
   dat van een college) en de daarin gebruikte leerstrategie\"en/-stijlen, en
   indien nodig, in staat op deze leerprocessen bij te sturen;
   \item[Reflectie op handelen --] ... te reflecteren op hun (of die van een
   collega) potenti\"ele rol en kunnen participeren in een maatschappelijk
   debat over kwesties die samenhangen met het eigen vakgebied.
\end{description}
Dit wordt ondersteund door de observatie dat de behoefte van de mens aan
structuur `aan de buitenkant' (buiten zichzelf) omgekeerd evenredig lijkt te
zijn met de mate van controle en flexibiliteit die iemand `aan de binnenkant'
(in zichzelf) ervaart. Naarmate iemand meer controle ervaart en daardoor meer
flexibel in zichzelf kan zijn, is er minder behoefte aan vaste (want
veiligheid-biedende) patronen om in te passen. In deze context leidt dit tot de
volgende constatering.  Wanneer de maatschappelijke veranderingsprocessen
doorgaan in het tempo waarin zij nu plaatsvinden, is er steeds minder
mogelijkheid te rekenen op het bestaan van vaste patronen waar men deel van
uitmaakt. Dit brengt dus de \emph{noodzaak} met zich mee meer aandacht te
besteden aan de `binnenkant'. Met andere woorden, het wordt in deze context
noodzakelijk dat ons onderwijs meer ruimte en aandacht biedt aan innerlijke
processen bij de mens, \`en aan \emph{bewustwording} van deze processen.  Dit
is de enige manier waarop wij mensen tot ontwikkeling kunnen brengen die in
staat zijn om te gaan met de veranderingen in onze maatschappij. De veranderde
maatschappelijke situatie betekent voor ons onderwijs derhalve dat kennis en
vaardigheden alleen niet meer voldoen, en dat ruimte geschapen moet worden voor
een (meta-)niveau van verwerking waarin de student (en overigens ook op lagere
onderwijsniveaus, de leerling) het eigen leren en functioneren leert begrijpen.
Aandacht voor en evaluatie van het eigen leerproces speelt hierbij een
belangrijke rol.

\section{Algemeen onderwijskundige achtergronden}
\label{s:Onderwijs}
In deze paragraaf worden een aantal algemene onderwijskundige overwegingen en 
achtergronden besproken die relevant zijn voor de informatiekunde opleiding.

\subsection{Stimuleren van de rechterhersenhelft}
\label{s:Hersenhelften}
Zoals eerder besproken dient een informatiekundige te beschikken over
vaardigheden die over het algemeen te maken hebben met eigenschappen die worden
geassocieerd met het functioneren van de \emph{rechterhersenhelft} van de
neocortex (zie appendix~\ref{h:brein}).\footnote{Merk op dat de phrase
`\emph{architecten} van de digitale samenleving' dit ook goed weergeeft: een
architect is immers typisch iemand voor wie het beeldende en ruimtelijke
vermogen van het brein -- een rechterhersenhelft-functie -- van groot belang
is. Voor de architect van de digitale samenleving zouden we kunnen stellen dat
dit ruimtelijke vermogen betrekking heeft op de abstracte (virtuele) objecten
en operaties die deel uitmaken van het ontwerp van een computerprogramma.} Over
de verschillende werking van linker- en rechterhersenhelft is veel bekend
\cite{Book:98:Carter:HetBreininKaart,
Book:89:Edwards:DrawingontheRightSideoftheBrain,
Book:98:Brandhof:GebruikjeHersens}; gedurende de laatste jaren wordt echter ook
steeds meer gezien en onderkend dat deze verschillen belangrijke implicaties
hebben ten aanzien van de manier waarop wij informatie verwerken en leren, en
worden hieraan consequenties verbonden, ook richting onderwijs.\footnote{Zie
bijvoorbeeld: \BrainStoreURL, \BrainStudioURL\ of \LearningRevolURL\ voor een
aantal recente publicaties over `breinvriendelijk leren'.} Ook de variatie aan
leerstijlen moet in dit verband worden genoemd
\cite{Book:93:Gardner:MultipleIntelligences}. 

In onze Westerse maatschappij, en zeker in ons onderwijssysteem en in de
wetenschap, heeft echter jarenlang de op analyse (segmentatie) en details
gerichte verwerkingswijze van de linkerhersenhelft voorop gestaan. De meer op
analogie\"en gerichte, holistische verwerkingswijze van de rechterhersenhelft
is hierbij op de achtergrond gebleven.\footnote{Vgl.\ de volgende uitspraak van
Roger Sperry \cite{Sperry}, wiens baanbrekend onderzoek eind zestiger en begin
zeventiger jaren de basis heeft gelegd voor het in kaart brengen van de
verschillen tussen linker- en rechterhersenhelft: `The main theme to emerge
{\ldots} is that there appear to be two modes of thinking, verbal and
nonverbal, represented rather separately in left and right hemispheres,
respectively, and that our educational system, as well as science in general,
tends to neglect the nonverbal form of intellect. What it comes down to is that
modern society discriminates against the right hemisphere.'}

In relatie tot bovengenoemde profielschets heeft dit belangrijke gevolgen. Het
betekent ten eerste, dat wij er niet automatisch van kunnen uitgaan dat
studenten de genoemde vaardigheden hebben, of dat zij weten hoe zij deze kunnen
ontwikkelen en toepassen. Een nog belangrijker probleem is, dat met name het
loslaten van de vertrouwde kaders (noodzakelijk voor het vinden creatieve
oplossingen) iets is dat door de verwerkingswijze van de linkerhersenhelft,
waarop in ons onderwijs juist zo'n sterk beroep wordt gedaan, juist wordt
\emph{belemmerd}~\cite{Book:98:Carter:HetBreininKaart,
Book:97:deBono:LeerUzelfDenken,
Book:89:Edwards:DrawingontheRightSideoftheBrain}. Daarnaast zullen studenten die
juist w\`el aanleg hebben voor deze andere verwerkingswijze, in ons huidige
onderwijssysteem sneller het risico lopen uit de boot te vallen: een dergelijke
leerstijl past immers veel minder goed bij de
lingui{\itr}stisch/logisch-mathematische benadering die in het onderwijs (zeker
in de $\beta$-discplines) gebruikelijk is. 

Voor een opleiding die mensen wil voorbereiden op een rol als architect van de
digitale samenleving hebben deze observaties belangrijke consequenties. Zij
impliceren dat aan deze `andere' (rechterhersenhelft) vaardigheden en denkwijze
in ons onderwijs expliciet aandacht zal moeten worden besteed, en dat wij deze
moeten stimuleren en mede helpen ontwikkelen waar nodig.  (Technieken als
\emph{mind-mapping} \cite{Book:96:Buzan:MindMapBook} kunnen hierbij
bijvoorbeeld worden ingezet.\footnote{In het bedrijfsleven en de wereld van
management winnen dergelijke technieken in snel tempo veld; zij zullen dan ook
zeker een bijdrage leveren aan de aansluiting van de architect-in-sp\'e op zijn
toekomstige werkgebied.}) Om ervoor te zorgen dat deze vaardigheden niet op
zichzelf blijven staan maar zo goed mogelijk worden ge{\itr}ntegreerd door, en
beklijven bij de student,\footnote{Vgl.\ het zgn.  \emph{transfer-probleem},
dat zich vaak voordoet wanneer vaardigheidstraining in isolement plaatsvindt.}
zal het gebruik ervan tevens zo veel mogelijk moeten worden ge{\itr}llustreerd
aan, en ge\"{\i}ntegreerd in vakken die zich daarvoor lenen.  Een
wetenschappelijke ontwikkeling die in dit verband van belang is, is
bijvoorbeeld het \emph{systeemdenken}~\cite{Book:56:Ashby:Cybernetics}; een
kader dat bijzonder goed aansluit bij de thematiek van de
architectuur~\cite{Book:02:Maier:ArtOfSystemArchitecture,
Book:91:Rechtin:SystemArchitecture}. Tot slot zullen wij manieren moeten vinden
om de hindernissen en blinde vlekken die door de gebruikelijke
(linkerhersenhelft) benaderingswijze worden opgeworpen, zo veel mogelijk te
voorkomen of overwinnen.  Naast het werk van bijvoorbeeld Edward de Bono (zie
o.a.\ \cite{Book:90:deBono:LateralThinking, Book:97:deBono:LeerUzelfDenken}) en
Betty Edwards \cite{Book:89:Edwards:DrawingontheRightSideoftheBrain,
Book:99:Edwards:TheNewDrawingontheRightSideoftheBrain} zijn er in de literatuur
over brein-trainingsmethodes verschillende andere aanknopingspunten te vinden.

De uitdaging voor onze opleiding is om de mogelijkheden te cre\"eren waarop de
`andere' informatieverwerkingswijze wordt gestimuleerd en benut, zonder dat de
wetenschappelijke werkwijze die iedere (toekomstige) academicus zich eigen moet
maken, verloren gaat. Zo wordt de informatiekundige iemand die het beeldende
vermogen van de architect verenigt met de analytische en exacte geest van de
exacte wetenschapper. Samengevat leidt het bovenstaande tot de volgende
inrichtingsprincipe:
\begin{description}
   \item[Rechter hersenhelft --] Naast de typische linkerhersenhelft
   benaderingswijze dient het onderwijs ook de rechter\-hersenhelft manier van
   denken, informatie verwerken en leren te stimuleren, echter zonder dat de
   wetenschappelijke werkwijze die iedere (toekomstige) academicus zich eigen
   moet maken, verloren gaat. 
\end{description}

\subsection{Student-activerend onderwijs}
In de afgelopen jaren is aan onze universiteit gezocht naar onderwijsmethodes
die `student-activerend' werken. Dat het vinden van dergelijke methodes geen
eenvoudige zaak is, bleek o.a.\ uit een artikel in KU-nieuws nr. 34 van 16 juni
2000. Het student-activerend onderwijs bij psychologie zou, volgens een
enqu\^ete onder eerstejaars studenten aldaar, vooral de passiviteit stimuleren.
Studenten merkten op: \emph{`Ik had gedacht dat je op de universiteit een
kritische houding aangeleerd zou krijgen. Nu gaat het vaak om cijfers in plaats
van dat je leert denken.' `Bij methoden krijg je een kruisje als je je huiswerk
gemaakt hebt, dat vind ik toch wel behoorlijk zielig. Ik verwacht van
wetenschappelijk onderwijs dat je toch wel enigszins zelf verantwoordelijkheid
hebt.' `Je wordt aan het handje genomen.' `Als iets l\'euk is, dan ben ik ook
bereid om daarover na te denken en actief mee te doen, dat motiveert me, zo'n
beloningssysteem niet.'}

Onderwijs student-activerend maken blijkt dus zo eenvoudig niet. Nu zijn er
sinds genoemd artikel ruim twee jaar verstreken, en er zijn belangrijke
leerervaringen opgedaan met hoe het \emph{niet} werkt. Een benadering waarin de
docent de momenten van activatie van de student bepaalt en vooral ook de aard
van de activatie, werkt passiviteit bij de studenten in de hand. Immers, de
student kan hierbij in essentie nog steeds een afwachtende houding aannemen
(een houding die door deze benadering zelfs mede kan worden veroorzaakt, en
versterkt). De zoektocht naar goede onderwijsmethodes die de student motiveren
en activeren blijft daarom van belang, ook in relatie tot de instroom van de
tweede-fase-student (die in de laatste jaren van zijn opleiding gewend is
geraakt aan een veel zelfstandiger rol). 

Ideaalgesproken zou het in student-activerend onderwijs moeten gaan om meer dan
alleen `actief bezig zijn' met de stof. De passieve uitgangshouding van
studenten, zo deze aanwezig is, zou moeten worden doorbroken. Goed onderwijs
activeert zichzelf.  Maar hoe doet het dat?

\subsubsection{Leren van binnenuit}
Meer duidelijkheid over de vraag wat onderwijs activerend maakt, krijgen we
misschien wanneer we als uitgangspunt aannemen dat de impuls tot
\emph{werkelijk} leren altijd van binnenuit komt. `Leren leren' is een frase
die tegenwoordig nogal vaak wordt gebruikt in onderwijsland -- maar stelt dit
niet de zaken op zijn kop? Heeft immers niet iedere mens deze impuls \`en
mogelijkheid tot leren, als innerlijke drijfveer voor zijn ontwikkeling? Elk
kind (mits niet gehandicapt) leert zitten, kruipen, lopen, praten en wat al
niet meer, en dat doet het -- zij het in interactie met zijn omgeving --
helemaal zelf. Je hoeft een kind niet te `leren leren'; leren is ons natuurlijk
erfgoed. Wel kunnen wij in ons onderwijs zoeken naar manieren om die impuls tot
leren die in iedere mens aanwezig is, aan te spreken. Een onderwerp, een
situatie, een docent of een onderwijsinstantie kan deze impuls wakker maken, en
zo als katalysator voor het leerproces fungeren.\footnote{Ter overweging de
volgende uitspraak van John Holt (\cite{kindzijn}[p.293]): `Ik kan in vijf tot
zeven woorden samenvatten wat ik als leraar uiteindelijk leerde. De versie in
zeven woorden is: Leren is niet het product van onderwijs. De versie in vijf
woorden is: Onderwijs leidt niet tot leren.'} Dit is, idealiter, wat goed
onderwijs kan zijn. (Als ons onderwijssysteem deze impuls soms eerder lijkt te
onderdrukken en mensen passief maakt, moet er iets heel ernstigs aan de hand
zijn!)

E\'en manier om deze impuls tot leren aan te wakkeren is door de student meer
verantwoordelijkheid te geven in het leerproces zelf. Dit kan bijvoorbeeld door
de student aan het begin van een leertraject aan te sporen om na te denken over
de eigen achtergrondkennis die hij/zij op dit gebied heeft, en vragen te
formuleren die aan de hand van deze beginsituatie opkomen. Nog maar al te vaak
zijn studenten verbaasd om te horen dat het leren stellen van de juiste vragen
veel belangrijker is dan het kennen van alle antwoorden!\footnote{Helaas moet
worden gezegd dat de gangbare evaluatiesystemen dit doorgaans juist in de hand
werken.} Door de student aan te moedigen zelf vragen te formuleren wordt
hij/zij in staat gesteld met meer initiatief de stof tegemoet te treden. Van
belang is hierbij overigens wel dat met het resultaat van deze reflectie in het
kader van het leerproces ook iets wordt gedaan. Wanneer de student van meet af
aan meedraagt in de verantwoordelijkheid voor zijn eigen leerproces, en
betrokken wordt bij de keuzes die in dat proces een rol kunnen spelen, kan
echte betrokkenheid ontstaan. 

Voor de opleiding informatiekunde wordt daarom het volgende inrichtingsprincipe
gehanteerd:
\begin{description}
   \item[Leren onder eigen verantwoordelijkheid --] Studenten dienen zelf
   verantwoordelijk gesteld te worden voor het leerproces. Een zelfstandigheid
   welke in de loop van de opleiding geleidelijk zal moeten groeien (groeiende
   zelfsturing).
\end{description}

\subsubsection{Cognitief leren} 
Alhoewel leren tot het natuurlijk erfgoed van de 
mens behoort en ieder mens `vanzelf' kan leren, wordt binnen het kader van een 
universitaire opleiding onder `leren' een specifiekere, meer op de cognitie 
gerichte activiteit verstaan, die niet voor iedereen even natuurlijk is, en 
evenmin voor iedereen gelijk. Leren in deze engere vorm speelt in een 
universitaire studie een centrale rol. Het aanleren van leer- en cognitieve 
strategie\"en, en het bewust maken van de manieren van d\`{\i}t leren
en van de manier waarop het eigen leerproces verloopt, kan daarom een 
belangrijke bijdrage leveren aan de succesvolle afronding van een studie. 

In het leerproces zijn verschillende leer- en denkactiviteiten te
onderscheiden.  Er kan verschil gemaakt worden tussen verwervingsactiviteiten
(die te maken hebben met het zich eigen maken van de leerinhouden) en
regulatieactiviteiten.  Voorbeelden van deze activiteiten zijn \cite{Boer93,
BS93, Rijswijk92, Vermunt92}: 
\begin{itemize}
   \item \emph{Verwervingsactiviteiten}
     \begin{itemize}
	\item het opdoen van kennis en vaardigheden (bijv. luisteren,
	selecteren, herhalen);
	\item het integreren van kennis en vaardigheden (bijv. relaties leggen
	met eerder verworven kennis/vaardigheden, voorbeelden zoeken);
	\item het toepassen van kennis en vaardigheden (bijv. oefenen,
	problemen oplossen).
    \end{itemize}

  \item \emph{Regulatie-activiteiten}
    \begin{itemize} 
	\item afstemmingsactiviteiten (gaan vooraf aan de
	verwervingsactiviteiten: de bereidheid ontwikkelen om te leren, bekend
	zijn met de doelen, op de hoogte zijn van de criteria, beschikken over
	een plan van aanpak);
	\item bewakingsactiviteiten (gaan gepaard met de
	verwervingsactiviteiten: bewaken van het plan, toetsen, diagnostiseren,
	bijsturen);
	\item evaluatieactiviteiten (na de verwervingsactiviteiten: beoordelen
	van de kwaliteit van de leerresultaten en de leeractiviteiten,
	conclusies trekken).
  \end{itemize}
\end{itemize}

Docenten beperken zich vaak tot de keuze van leerinhoud/leerstof en tot de
keuze van studentactiviteiten en didactische werkvormen (vooral dus de
verwervingsactiviteiten). Men gaat er vaak van uit dat leer- en cognitieve
strategie\"en vanzelf aangeleerd worden. In onderwijs dat gericht is op het
leerproces met het doel het vermogen tot zelfstandig leren te verbeteren, is
het ook van belang dat docenten zich bezinnen op de doelstellingen met
betrekking tot de regulatieactiviteiten: naast kennisdoelen worden immers ook
doelen nagestreefd op het gebied van zelfstandig afstemmen, bewaken en
evalueren van het leerproces. Verder is het in het kader van het vergroten van
de zelfstandigheid belangrijk dat de verantwoordelijkheid voor het leerproces
geleidelijk wordt overgedragen aan de student. Docenten moeten overwegen welke
taken en rollen ze in de beginfase van het leerproces op zich nemen, en wanneer
en hoe ze deze taken en rollen aan de studenten zullen overdragen.

Het is wederom belangrijk om hierbij te onderkennen dat door de inherente
dynamiek van het Informatiekunde vakgebied als gevolg van ontwikkelingen
in de ICT en het maatschappelijke omveld, Informatiekundigen 
`nooit uitgeleerd raken'.

Vertaald naar het onderwijs levert dit de behoefte op aan handvatten waarmee 
we dit proces van zelfstandigheid en verantwoordelijkheid nemen door de 
student kunnen initi\"eren en geleidelijk aan laten groeien (groeiende 
zelfsturing). In het bijzonder:
\begin{itemize}
   \item handvatten om de student vanaf het begin (mede) verantwoordelijkheid 
         te geven voor, en te betrekken bij het eigen leerproces; 
   \item handvatten voor onderwijs met aandacht voor de regulatie-activiteiten,
         waarbij de rol van de docent verandert van sturend naar begeleidend 
         naar overdragend (met andere woorden, een afnemende sturing);
   \item handvatten om de student te leren reflecteren op het eigen 
         leervermogen en -proces (hierbij is \'o\'ok van belang het aankweken 
         van een andere houding -- zowel bij docent als student -- ten aanzien 
         van fouten!);\footnote{Zie ook paragraaf~\ref{p:goedfout}.} 
   \item handvatten om op flexibele wijze om te kunnen gaan met inhoudelijke 
         leerdoelen die in relatie kunnen worden gebracht met een leerproces 
         dat ruimte krijgt. 
\end{itemize}

Niet zozeer `leren leren' is onze taak, maar veeleer het leren
\emph{faciliteren} en, daarnaast, het leerproces helpen bewust te maken.  Het
initiatief en de verantwoordelijkheid voor het eigen leerproces wordt hiermee
teruggebracht bij de leerder zelf (waar het uiteindelijk ook ligt). 

Bovenstaande uiteenzetting leidt tot het volgende principe:
\begin{description}
   \item[Aandacht voor leerprocessen --] In de opleiding dient aandacht besteed
   te worden aan leerprocessen. Hierbij is het belangrijk om studenten te
   \emph{helpen ontdekken} hoe zij \emph{zichzelf}  kunnen ontplooien, met
   andere woorden, hoe zij hun eigen (en eventueel andermans) capaciteiten optimaal
   kunnen benutten en onder welke omstandigheden.    
\end{description}

\section{Onderwijsstrategie\"en}
\label{s:Strategie}
Gegeven de bovenstaande achtergronden, zijn er een aantal relevante
onderwijsstrategie\"en die we als inspiratie kunnen gebruiken
bij de inrichting van de Informatiekunde opleiding. In de komende
jaren zal er binnen het informatiekunde curriculum gericht worden
ge\"expirimenteerd met de diverse onderwijsvormen. Hiermee willen we binnen
de informatiekunde opleiding ervaring opdoen met nieuwe onderwijsvormen,
en daarmee vooruitdenken.

\subsection{Probleem-gestuurd onderwijs}
De hierboven besproken visie op onderwijs en leren sluit sterk aan bij een
klassieke methode die enige jaren geleden is herontdekt aan de universiteit van
Maastricht, en die momenteel weer sterk in opkomst is: het probleemgestuurd
onderwijs (PGO).\footnote{Deze paragraaf is grotendeels gebaseerd op een
presentatie van Arjan de Jager van het \IICD\ in het kader van het Capacity
Development Programme. Een betere verwijzing naar de achterliggende
onderwijskundige theorie volgt nog.} 

PGO heeft onder andere als doelstellingen: 
\begin{itemize}
   \item het ontwikkelen van wetenschappelijk inzicht aan de hand van
         `real-world' cases;
   \item het ontwikkelen van leer- en redeneerstrategie\"en;
   \item het ontwikkelen van zelfsturend leren;
   \item het stimuleren van onafhankelijk en innovatief denken
\end{itemize}
Een belangrijk uitgangspunt van PGO is dat leren begint met een 
\emph{probleem} dat moet worden opgelost, in plaats van met een
hoeveelheid (op zichzelf staande) kennis 
die moet worden geleerd. Dit houdt in dat het leren altijd in een \emph{context} 
plaatsvindt. Dit uitgangspunt komt in belangrijke mate tegemoet 
aan een probleem dat in het onderwijs welbekend is: het transfer-probleem. 
Wanneer leren plaatsvindt in de context van een probleem dat moet worden 
opgelost, 
is er een grotere kans dat het geleerde later ook in vergelijkbare
contexten wordt toegepast, alsdus de uitgangspunten van het PGO. 

Tevens is PGO student (leerder)-gericht. Dat wil zeggen dat de controle en 
verantwoordelijkheid voor het leerproces ligt bij de leerder, niet bij de 
docent. 
Studenten fomuleren hun eigen leerdoelen en zoeken naar de meest geschikte 
middelen 
en methodes om deze doelen te bereiken. De docent heeft een niet-sturende, 
aanvullende 

PGO komt tegemoet aan een aantal bezwaren die het traditionele onderwijs kent.
Zo biedt het een natuurlijke context (namelijk die van het desbetreffende
probleem) waarin de verbanden tussen verschillende vakken of disciplines
zichtbaar kunnen worden. Ook worden studenten al vroeg geconfronteerd met de
praktijk van hun latere vakgebied en komen daarin alle aspecten (cognitieve en
andere) tegen die ook in hun latere carri\`ere van belang zijn. De context van
het probleem zorgt er voor dat de feiten die studenten leren, geen op zichzelf
staande, betekenisloze details zijn maar relevantie hebben, waardoor ook het
begrip van studenten wordt vergroot. En tot slot zorgt de bewuste aandacht voor
het leerproces ervoor dat de opgedane kennis en vaardigheden ook in andere
situaties beter beschikbaar zullen zijn. 

PGO is vooral toepasbaar in die situaties waarin leren het karakter heeft van 
ontdekken en herstructuren: het gebruiken van eerder opgedane kennis om 
nieuwe situaties en nieuwe informatie te begrijpen. We moeten echter niet de 
fout maken te denken dat er \'e\'en vorm van onderwijs is die voor \`alle 
situaties de meest geschikte is. Er zijn wel degelijk omstandigheden waarin 
het `klassieke hoorcollege', waarin de docent zijn 
studenten stap voor stap door een stuk nieuwe materie heenloodst, van grote
waarde is. En het is helemaal niet altijd mogelijk het initiatief en de 
verantwoordelijkheid volledig bij de student te leggen. Het is daarom van 
belang te zoeken naar de juiste balans tussen de rollen van student en docent 
in het scala aan leeromstandigheden dat in een leertraject naar voren
komt. Soms zal een probleem-gebaseerde benadering hiervoor ideaal zijn.
Op andere momenten moet een docent meer sturing kunnen bieden. Bij het
ontwerpen van een opleiding gaat het erom hiermee flexibel om te kunnen gaan,
en hierin een weloverwogen, beargumenteerde keuze te maken. Idealiter
zouden we dan kunnen toewerken naar een situatie waarin `organisch
leren' mogelijk wordt.

\subsection{Organisch leermodel}
\label{p:organisch}
Ons onderwijs is van oudsher vaak lineair georganiseerd. Dat wil zeggen: 
de student doet eerst stapje $1$, dan stapje $2$, dan stapje $3$, etc. 
Dit betekent dat studenten/leerlingen afhankelijk worden; immers, iedere 
vervolgstap bestaat alleen maar uit nieuwe dingen en de student kan daar 
nog niet zelf een weg in vinden. De student/leerling kan dus niet zelf 
initiatief ontplooien, en dit is vooral voor de slimme leerling 
(geest-)dodend.
Bovendien maakt deze manier van aanbieden het integreren van stof moeizaam: 
er is geen echte tijd voor ingeruimd, en geen expliciete aandacht voor. 
Integreren vindt plaats door iets nieuws te relateren aan iets ouds; 
niet door iets nieuws te doen na (bovenop) iets ouds.

In plaats daarvan kan men zich een `organisch leermodel' voorstellen.
Dit is voor te stellen als de beweging die ontstaat wanneer men een
steen in het water gooit. 
Vanuit het midden ontstaat een cirkel. Als men nu een grotere 
steen op dezelfde plek gooit, ontstaat eerst weer dezelfde circulaire 
golfbeweging vanuit het midden, maar deze reikt nu verder. Hij begaat echter wel 
gedeeltelijk een route die al eerder is afgelegd. `Organisch' is dan: nieuw 
waarin oud inherent verweven zit (niet `bovenop' oud, als een gescheiden iets). 

In een ontwikkelingsproces moet kan het nieuwe pas aangrijpen als het oude 
een zekere vastigheid heeft bereikt, en de leerder zich vertrouwd voelt 
met dat oude en van daaruit het nieuwe durft of kan aangaan. Dit houdt echter
in dat het nieuwe niet 100\% nieuw moet zijn, maar in zichzelf het oude 
moet vervatten. Vgl.: leren kruipen bouwt voort op het hebben van spiercontrole 
en draagt die noodzaak van spiercontrole in zich. (Het is niet iets dat los 
staat van die spiercontrole).

Een dergelijke benadering lost een belangrijk probleem op dat zich in 
het onderwijs in de twintigste eeuw is gaan 
wortelen.\footnote{Zie 
  bijvoorbeeld: \MaslowURL.
  Interessant hierbij is overigens dat de behoeftenpyramide van 
  Maslow sterk op de Westerse culturen is gericht. 
  In andere culturen bestaat een andere behoeftenpyramide; zie 
  o.a. Pinto ~\cite{Pinto}, die hieraan ook consequenties verbindt in relatie tot
  de inrichting van ons onderwijs (\PintoURL).
}
de mogelijkheid tot zelf-manifestatie en zelf-ontplooiing stoelt op 
het gevoel van veiligheid, waardering, belonging en self-esteem, maar 
in ons onderwijs is het bereiken van die laatste gevoelens juist 
gekoppeld geraakt aan het goed presteren. 
Door nu het onderwijs zo in te richten dat een nieuw te onderzoeken terrein 
steeds 
het oude in zich integreert, kan de student daar een veiligheid aan ontlenen die 
hem in staat stelt nieuwe initiatieven te ontplooien. (Zonder veiligheid
kan dat niet.)

\subsection{De handelingsgerichte leercyclus}

De zusteropleiding Informatica verenigt een groot aantal van de 
hierboven genoemde uitgangspunten in de zgn.\ `handelingsgerichte 
leercyclus' \cite{ZelfstudieInformatica}. Studenten 
verwerven effectief inzicht, kennis en vaardigheden door het uitvoeren 
van studie- en leeractiviteiten met de volgende structuur.
Deze aanpak staat is ge\"ilustreerd in figuur~\ref{LeerCyclus}.
\Figure{width=12cm}{LeerCyclus}{LeerCyclus}{De handelingsgerichte leercyclus.}

De structuur van studietaken komt zoveel mogelijk overeen met de 
structuur van deze handelingsgerichte leercyclus. Het traject:
  `\emph{probleem $\rightarrow$ ontwikkelen $\rightarrow$ oplossing}' 
staat hierbij centraal. 
Door probleemoplossingsgericht handelen op basis van reeds aanwezig 
inzicht (kennis, vaardigheden) worden nieuwe kennis en vaardigheden verworven 
die met de reeds aanwezige kennis en vaardigheden worden ge{\itr}ntegreerd. 
Ook het bestuderen van theorie kan gezien worden als probleemoplossingsgericht 
handelen. 
Verwerven van inzicht gebeurt vooral tijdens de verantwoording van dit 
handelen.

Deze zienswijze kan in het onderwijs bijvoorbeeld tot uiting komen
door al vanaf het begin van de opleiding uitdagende opdrachten op te 
nemen in de vakken en projecten. 
Daarbij zou dan steeds het hele oplossingstraject aan de orde moeten
komen (en niet slechts een klein aspect) en dient bijzondere 
aandacht te worden geschonken aan de verantwoording van zowel de 
\emph{oplossing} als het oplossings\emph{proces}.  
In het oplossingstraject zou hierbij de nadruk op ontwerpen en op het 
inzetten van geavanceerde technieken (`engineering') moeten liggen. 
Naast fundamentele aspecten moeten ook het herkennen van 
toepassingssituaties en het concreet toepassen van wetenschappelijke 
technieken aan bod komen. 
Zo wordt kennis direct gekoppeld aan vaardigheden.  

\section{Samenvatting inrichtingsprincipes \& vaardigheden}
\label{s:Principes}
Tot slot van dit hoofdstuk noemen we ook hier weer even de kort de
inrichtingsprincipes voor de inrichting van de opleiding informatiekunde 
en de gewenste vaardigheden voor informatiekundigen, zoals
die volgen uit de bovenstaande visie op onderwijs en leren.

Een informatiekundige moet in staat zijn om:
\begin{description}
   \item[Kennis ontsluiten --] ... kennis- en ervaringsbronnen te ontsluiten,
   voorzover deze aansluiten bij hun reeds bestaande kennis;
   \item[Kennis inzetten --] ... voor informatiekundige problemen, relevante
   kennisgebieden aan te geven en hun mogelijke bijdrage aan de oplossing van
   het probleem te identificeren, en waar relevant deze kennis inzetten bij het
   oplossen van het probleem.
   \item[Reflectie op leren --] ... te reflecteren op het eigen leerproces (of
   dat van een college) en de daarin gebruikte leerstrategie\"en/-stijlen, en
   indien nodig, in staat op deze leerprocessen bij te sturen;
   \item[Reflectie op handelen --] ... te reflecteren op hun (of die van een
   collega) potenti\"ele rol en kunnen participeren in een maatschappelijk
   debat over kwesties die samenhangen met het eigen vakgebied.
\end{description}

Daarnaast zijn de volgende inrichtingsprincipes ge\"{\i}dentificeerd:
\begin{description}
   \item[Ondervinden van ICT --] De snelle opmars van ICT moet ook doorklinken
   in de opleiding. Studenten dienen daarom dan ook ICT aan den lijve te
   ondervinden, bijvoorbeeld door het inzetten van moderne ICT in het onderwijs
   zelf.
   \item[Aandacht voor creativiteit en verandering --] Studenten dienen
   gestimuleerd te worden om zich te ontwikkelen tot creatieve mensen die
   openstaan voor, en kunnen omgaan met veranderingen.
   \item[Rechter hersenhelft --] Naast de typische linkerhersenhelft
   benaderingswijze dient het onderwijs ook de rechter\-hersenhelft manier van
   denken, informatie verwerken en leren te stimuleren, echter zonder dat de
   wetenschappelijke werkwijze die iedere (toekomstige) academicus zich eigen
   moet maken, verloren gaat. 
   \item[Leren onder eigen verantwoordelijkheid --] Studenten dienen zelf
   verantwoordelijk gesteld te worden voor het leerproces. Een zelfstandigheid
   welke in de loop van de opleiding geleidelijk zal moeten groeien (groeiende
   zelfsturing).
   \item[Aandacht voor leerprocessen --] In de opleiding dient aandacht besteed
   te worden aan leerprocessen. Hierbij is het belangrijk om studenten te
   \emph{helpen ontdekken} hoe zij \emph{zichzelf}  kunnen ontplooien, met
   andere woorden, hoe zij hun eigen (en eventueel andermans) capaciteiten
   optimaal kunnen benutten en onder welke omstandigheden.
\end{description}

Als basis-strategie\"en voor de concrete inrichting van het onderwijs
zal een significante rol weggelegd zijn voor:
\begin{itemize}
   \item Probleemgestuurd onderwijs; projectmatige inrichting.
   \item Organisch leren.
   \item De handelingsgerichte leercyclus.
\end{itemize}

%% file: waarmee-beoordeling.tex
\chapter{Visie op beoordeling}
\label{h:waarmee-beoordeling}

Een belangrijk probleem dat bij het invoeren van vernieuwingen vaak speelt, is
dat deze vaak moeilijk of niet realiseerbaar zijn als niet ook in
andere, gerelateerde aspecten veranderingen worden doorgevoerd. Meestal is het
niet mogelijk een bepaald aspect van een geheel te veranderen zonder dat dat
ook in het grotere kader bepaalde veranderingen vereist. Datzelfde probleem
komen we tegen wanneer we ons onderwijs willen inrichten op een manier die aan
de student meer zelfstandige rol toewijst. Het is moeilijk om aan de ene kan de
student meer verantwoordelijkheid voor het eigen leerproces te geven, en aan de
andere kant bepaalde gewoontes in stand te houden die aan deze
verantwoordelijkheid voorbij gaan, of deze bemoeilijken. E\'en van de gebieden
waar dit aspect relevant is, is toetsing. 

In dit hoofdstuk staan we eerst stil bij een aantal problemen die de huidige
manier van toetsen met zich meebrengt (paragraaf~\ref{s:problemen}). Dit doen
we vanuit het perspectief dat studenten zelf ook een duidelijke eigen
verantwoordelijkheid dienen te dragen. Vervolgens bespreken we onze visie om de
ge\"{\i}dentificeerde problemen te verminderen, wat leidt tot een aantal
inrichtingsprincipes voor toetsing en beoordeling.

Merk op dat het in de bedoeling ligt dat dit hoofdstuk op termijn wordt
ge\"{\i}ntegreerd tot een \NIII\ brede visie op beoordeling. Dit hoofdstuk
gaat met name in op aspecten die voor informatiekunde relevant zijn.

\section{Problemen met de bestaande manier van toetsen}
\label{s:problemen}

\subsection{Consequenties van het niet-halen van een tentamen}
\label{p:nietgehaald}
Ons toetsingssysteem kent een aantal gewoontes die het mogelijk maken dat bij
de student een houding ontstaat die niet bevorderlijk is voor het nemen van de
eigen verantwoordelijkheid. E\'en van deze gewoontes betreft de consequenties
die verbonden zijn aan het niet behalen of niet afleggen van een tentamen. Over
het algemeen geldt dat er met betrekking tot de rest van de studie geen
consequenties zijn, anders dan dat het betreffende tentamen opnieuw afgelegd
moet worden.  Op papier bestaan bepaalde reglementen die aangeven dat een
student, bij het niet behalen van een tentamen of practicum, binnen het jaar
waarin het betreffende vak is gevolgd het vak opnieuw moet volgen\footnote{Zie
bijvoorbeeld de facultaire richtlijnen voor de OER.}. In de praktijk wordt hier
echter heel soepel mee omgegaan. Andere consequenties, waarbij een student
bijvoorbeeld pas toegang krijgt tot een vak $Y$ wanneer hij tentamen $X$ heeft
behaald, zijn er nauwelijks. 

In eerste instantie lijkt het erop dat hierdoor een grotere flexibiliteit wordt
gecre\"eerd ten aanzien van de student. Niettemin werkt dit systeem op een
aantal manieren negatief door. Zo kan het gebeuren dat studenten vakken volgen
terwijl zij niet voldoen aan de ingangsvoorwaarden die het mogelijk maken om deze
vakken met enig nut te volgen (eenvoudigweg omdat zij de voorkennis, die in
andere vakken wordt overgebracht, onvoldoende beheersen). Ter illustratie,
wanneer studenten een basisvak programmeervaardigheden uit de propedeuse nog in
hun derde jaar moeten afronden, is de kans groot dat zij in de tussentijd bij
verschillende andere vakken situaties zijn tegengekomen waarin deze
programmeervaardigheden van belang zijn geweest. Het doen en volgen van
die vakken k\`an dan een houding in het leven roepen waarin de
student min of meer lukraak probeert de tentamens voor die vakken te halen. 

Het feit dat het in ons huidige systeem (meestal) mogelijk is dat studenten
gewoon doorgaan in hun studietraject zonder dat aan het niet behalen van
bepaalde tentamens consequenties worden verbonden, heeft dan ook een aantal
ongewenste bijwerkingen die van belang zijn in relatie tot het cre\"eren van
een systeem waarin de student meer verantwoordelijkheid voor het eigen
leerproces draagt: 

\bi
\item het vergroot de indruk dat vakken min of meer op zichzelf staan, en
leidt ertoe dat studenten ze ook als zodanig beschouwen; 
\item het stimuleert een houding waarin de student `op de gok' mee kan doen aan
tentamens van vervolgvakken wanneer hij de daaraan voorafgaande stof nog
onvoldoende beheerst; 
\item het cre\"eert de indruk dat voorkennis niet of nauwelijks serieus wordt genomen;
\item het houdt de situatie in stand waarin de student de verantwoordelijkheid
voor zijn eigen leerproces kan blijven ontwijken: hij ziet immers niet dat aan 
    bepaalde gedragingen of situaties consequenties verbonden zijn.  
\ei

Met andere woorden, terwijl wij misschien denken de student met deze soepelheid 
een dienst te bewijzen, blijkt uit het bovenstaande dat dat juist niet
het geval is wanneer wij als doel hebben de student meer verantwoordelijkheid 
voor het eigen leerproces te laten nemen. Het in stand houden van de
huidige gang van zaken lijkt dan ook te conflicteren met dit streven naar
een grotere zelfstandigheid. (N.B.: Dit betekent echter niet dat we hier pleiten 
voor een heel rigide systeem; zie paragraaf~\ref{p:flexibel}.)

\subsection{Goed en fout}
\label{p:goedfout}

Een andere, nog belangrijkere factor in ons toetsingssysteem die een
belemmering vormt voor het vergroten van de eigen verantwoordelijkheid van de
student, is het enorme gewicht dat -- helaas -- in ons onderwijs (en misschien
wel in ons hele leven) is komen te hangen aan de labeltjes `goed' en `fout'.
Het gewicht van deze labeltjes, en met name het feit dat het iemand \emph{van
buitenaf} is die ze aan onze prestaties toekent, maakt dat het natuurlijk
leermechanisme volledig doorbroken wordt. Ter illustratie: in een natuurlijk
leermechanisme (bijvoorbeeld een baby die leert grijpen, een kind dat leert
lopen of praten) is het de leerder \emph{zelf} die, op basis van observaties of
ervaringen, tot de conclusie komt dat een bepaald gedrag wel of niet
functioneel is, wel of niet `werkt'. Er is een van binnenuit gesteld doel en
een (intrinsieke) motivatie om dat doel te bereiken. Op basis van
`trial-and-error' past de leerder zijn gedrag aan, en komt er uiteindelijk achter
hoe het betreffende gegeven `werkt' (dit is te vergelijken met een
`servo-mechanisme'). Nu kunnen we weliswaar zeggen dat ons toetsingssysteem ook
bedoeld is om de feedback te geven die de leerder in staat stelt te beoordelen
of hij de betreffende materie beheerst; feit is, helaas, dat aan de labeltjes
`goed' of  `fout' over het algemeen een andere lading hangt (die er natuurlijk
al heel vroeg in ons onderwijs en daarbuiten aan is komen te hangen). Van
'trial-and-error' kan geen sprake zijn, omdat er veel te veel afhangt van een
goed resultaat: gevoel van eigenwaarde bijvoorbeeld (zie ook
paragraaf~\ref{p:organisch}), of een studiebeurs. Het systeem werkt aldus in de
hand dat de leerder buiten zichzelf gaat zoeken voor de juiste antwoorden, en
niet meer bij zichzelf te rade gaat. Daarnaast wordt door het systeem van
`externe beloning' de intrinsieke motivatie juist
ondermijnd!~\cite{Kohn}.\footnote{Kohn beargumenteert dat belonen en straffen
twee zijden van dezelfde medaille zijn, die beide tot gevolg hebben dat
intrinsieke belangstelling in extrinsieke belangstelling wordt omgezet, en dat
de belangstelling voor de taak zelf vermindert. Om de negatieve consequenties
van `oordelend' evalueren te voorkomen is het niet voldoende meer of minder te
belonen, dan wel te straffen, maar is het nodig een geheel \emph{andere}
dimensie aan te spreken, en niet langer in deze dimensie van goed en fout te
denken. Zie ook punt 1 en punt 3 hierna.}
    
Deze verandering in attitude levert een aantal hardnekkige problemen op wanneer
we de leerder meer verantwoordelijkheid willen laten nemen voor het eigen
leerproces. Het heeft, ten eerste, tot gevolg dat hij denkt dat de juiste
antwoorden altijd bestaan (zie ook het probleem dat in het hiervoor genoemde
citaat van Carl Rogers wordt gesignaleerd). Ten tweede leidt het ertoe dat de
leerder niet beseft hoe hij die juiste antwoorden -- als ze bestaan -- ook zelf
(of in zichzelf) zou kunnen vinden.  Ten derde gaat een gerichtheid op de
juiste \emph{antwoorden} volkomen voorbij aan het reeds eerder genoemde belang
van het leren stellen van goede \emph{vragen}. 

Een en ander leidt tot de conclusie dat het huidige toetsingssysteem het nemen
van meer eigen verantwoordelijkheid door de leerder niet stimuleert. Zowel door
de praktische procedures als door de attitude die de gangbare wijze van
toetsing in ons onderwijs heeft bewerkstelligd, wordt een houding bij de
leerder bevorderd die in sommige opzichten moeilijk te combineren is met een
meer verantwoordelijke rol. Wanneer we derhalve willen streven naar een
grotere verantwoordelijkheid bij de leerder, dan zullen we daaraan ook
conseqenties moeten verbinden ten aanzien van de toetsing. De vraag is echter:
hoe?

\section{Flexibel, maar dan anders}
\label{p:flexibel}

Het is belangrijk om te zien dat het geven van meer verantwoordelijkheid aan de
student voor het eigen leerproces ook een fundamentele verandering vraagt in de
manier waarop getoetst wordt. Wanneer niet beide aspecten worden veranderd,
ontstaat voor de student een situatie waarin het niet mogelijk is de stap naar
een grotere eigen verantwoordelijkheid te nemen, omdat er tegelijkertijd teveel
afhangt van de wijze van toetsing die nog volgens de oude manier werkt. Uit de
hiervoor genoemde bezwaren tegen de gangbare wijze van toetsing kunnen we
mogelijk een aantal richtlijnen destilleren voor een opzet die meer compatibel
is met een grotere eigen verantwoordelijkheid voor de student. Idealiter zou
deze wijze van toetsing en beoordeling: 
\be
\item als \emph{middel} moeten worden gezien dat men de student in zijn leertraject
van feedback voorziet over zijn leerproces; niet als \emph{doel} op zich;
\item de leerder in staat moeten stellen zelf conclusies te trekken ten aanzien
van het eigen leertraject, en aan deze conclusies stappen te verbinden voor het
vervolg van het leerproces;
\item moeten laten zien dat aan het nog niet afgerond hebben van een vak zekere
consequenties zijn verbonden.  Deze consequenties dienen zodanig te worden
vormgegeven dat zij een positieve bijdrage leveren aan de motivatie van de
student;
\item de verbondenheid van vakken moeten reflecteren.
\ee

Met betrekking tot deze punten kunnen de volgende toevoegingen worden gemaakt.

{\bf ad 1.} Dit punt impliceert dat er een onderscheid gemaakt moet worden tussen
toetsen als diagnostisch instrument, en toetsen als eind-evaluatie. In ons
onderwijssysteem worden dit onderscheid vaak niet gemaakt. Tussentijdse
deeltoetsen gelden weliswaar als diagnostisch instrument, maar worden vaak
tevens meegenomen in het eindcijfer van het totaal. Sterker nog, in sommige
richtlijnen wordt gesproken over het gebruik van tussentijdse toetsen om
regelmatig studeren te bevorderen (een extern middel om de student tot studeren
aan te zetten)! Deze benadering staat min of meer haaks op het geven van meer
verantwoordelijkheid voor het eigen leerproces aan de student. Merk op, dat dit
niet betekent dat tussentijdse toetsing achterwege zou moeten worden gelaten.
Het gaat er echter om, dat die toetsing niet als doel, maar als
(feedback-)middel zou moeten fungeren aan de hand waarvan de student zelf
aanpassingen kan doen in het eigen leertraject. Het kiezen van feedback in een
andere vorm dan cijfers zou hierin een belangrijke rol kunnen spelen. 


{\bf ad 2.} Dit punt sluit nauw aan op het voorgaande punt, maar heeft tevens te
maken met de manier waarop een leertraject wordt ingericht. Op dit moment is
het meestal de docent die bepaalt wat de volgorde is waarin bepaalde materie
wordt doorgewerkt. Het is echter goed mogelijk dat studenten (die immers ieder
een eigen uitgangssituatie en een eigen leerstijl hebben) een verschillende
route naar het einddoel bewandelen. Voor elke student geldt dat deze zelf moet
leren te beoordelen hoe bepaalde feedback zijn weerslag heeft op dit
leertraject.  Wij moeten als onderwijsgevers een manier vinden om deze
verschillende benaderingen te faciliteren, en tevens de feedback te leveren die
de leerder in staat stelt het eigen leerproces aan te passen. Hierbij is het
onderscheid tussen de \emph{inhoud} die door de leerder moet worden verworven,
en de \emph{wijze en volgorde} waarop hij deze verwerft, van belang:
\bi
\item Het verwerven van de inhoud geldt voor alle leerders; hiervoor kan dan
ook voor alle leerders feedback worden ingebouwd. Dat kan bijvoorbeeld door de
lesstof zodanig in te richten dat de opbouw van de inhoud zelf als
feedbackmechanisme fungeert. Met andere woorden: door de stof zodanig in te
richten dat in de inhoud zelf steeds wordt voortgebouwd op andere aspecten, zal
een student die bepaalde aspecten nog niet beheerst, hier vanzelf mee worden
geconfronteerd, en hierin zijn verantwoordelijkheid moeten nemen;
\item De wijze en volgorde van verwerving is meer specifiek, en hierin dient te
leerder meer ruimte te krijgen voor het eigen proces. Feedback op dit vlak zal
dan ook meer liggen op metaniveau, in die zin dat het tot taak heeft de leerder
bewust te maken van de eigen toegepaste strategie\"en en leermethodes, en te
laten onderzoeken of deze methodes hebben gewerkt of niet. 
\ei

{\bf ad 3.} Wanneer we consequenties verbinden aan het afronden van een vak,
moeten we dit zodanig inrichten, dat het alternatief dat hierdoor ontstaat de
student stimuleert in het nemen van de eigen verantwoordelijkheid. Dit kunnen
we bijvoorbeeld doen door een opzet waarin het niet behalen van een vak niet
leidt tot een fase waarin alleen maar sprake is van \emph{minder} kunnen doen
(in die zin dat de student geen toegang krijgt tot bepaalde colleges en daarmee
vertraging oploopt), maar tot een route die \emph{anders} verloopt. Het minder
kunnen doen wordt in zekere zin afgedwongen door een indeling van het
curriculum in vaste jaarprogramma's. Wanneer de student in een dergelijke opzet
een college niet gehaald heeft, en daar blokkerende consequenties aan worden
verbonden, ontstaat daardoor vertraging. Om juist een positieve insteek
mogelijk te maken moet een alternatief worden gecre\"eerd waarin de student
z\`elf aan het niet behalen van het vak op verschillende manieren consequenties
kan verbinden. Bijvoorbeeld, door zelf te bepalen welke activiteiten nodig zijn
om het geconstateerde hiaat in kennis te kunnen bijspijkeren. Of door een
andere route door te stof te volgen, die beter bij zijn eigen werkwijze past.
De sleutel bij dit alles is dat het bestaan van verschillende mogelijkheden de
student eigen \emph{keuze} geeft, en m\`et die eigen keuze ontstaat ook eigen
\emph{verantwoordelijkheid}. 

Merk overigens op dat het `niet behaald hebben' of `niet afgerond hebben' van
een vak slechts een andere manier is om te zeggen dat iemand aan het leren is.
Wanneer we het op die wijze formuleren, ontstaat ook een positiever perspectief
op dit leerproces: het is eerder een \emph{beschrijving} van wat gaande is
(namelijk een \emph{leerproces}) dan een \emph{kwalificatie} omtrent een (al
dan niet afgerond) product. Dit leidt bijna als vanzelf ook tot het gebruik van
andere taal om toetsingsresultaten weer te geven (zie ook punt 1 hierboven).

{\bf ad 4.} Wanneer in het omgaan met toetsingsresultaten de notie `voorkennis'
serieus wordt genomen, wordt beter duidelijk welke vakken met elkaar
samenhangen, en hoe sterk deze samenhang is. Dit houdt niet in dat voor elk
vak voorkennis gedefinieerd moet worden, of dat elk vak per s\`e als voorkennis
voor andere vakken moet gelden. Juist door hier gedifferentieerd mee om te
gaan, ontstaat een meer flexibel beeld. Bij sommige vakken is misschien sprake
van hele harde ingangseisen; deze eisen dienen dan ook serieus te worden
genomen bij de toelating van studenten. Bij andere vakken is er misschien meer
ruimte. De samenhang van vakken kan bij de inrichting van het curriculum worden
gereflecteerd in het formuleren van deze ingangseisen en de kwalificatie die
daarbij hoort.

Het verwerken van bovenstaande punten vereist een andere manier van inrichten
van het onderwijs. In plaats van een jaarprogramma dat voor alle studenten in
vaste volgorde en vaste opzet wordt gegeven, moet er ruimte zijn voor
verschillende trajecten en verschillende werkwijzen (aansluitend op de
verschillen in leerstijlen) binnen \een\ jaar. Een opzet waarin veel van de in
dit hoofdstuk genoemde punten zouden kunnen worden verwerkt, zou bijvoorbeeld
tot de volgende inrichtingsprincipes kunnen leiden: 


\begin{description}
\item [Grotere projecten --] Elk studiejaar kent een of meerdere grotere
projecten waarin een aantal samenhangende thema's aan bod komt.\footnote{Te
overwegen valt bijvoorbeeld om in de propedeuse met \een\ of twee van zulke
projecten te defini\"eren, en het aantal te laten toenemen in latere jaren.} De
inhoud van verschillende samenhangende vakken wordt aan deze projecten
opgehangen. De projecten hebben een concrete vraag of probleemstelling als
insteek, waarin een specifiek stuk kennis en inzicht centraal staat. Aan het
einde van het project geeft de student blijk van het feit dat hij deze kennis
en inzichten verworven heeft (bijvoorbeeld door middel van verslaglegging). 

\item [Initiatief bij de student --] Studenten worden aangemoedigd aan het
begin van een project vanuit hun eigen optiek na te denken over de manier
waarop zij de beoogde inhoudelijke leerdoelen zouden willen bereiken, en
daarnaast voor zichzelf een aantal procesgerichte leerdoelen te formuleren.
Hierbij kunnen studenten (al naar gelang hun achtergrond) verschillende
trajecten volgen.

\item [Samenwerkend leren --] Studenten worden gestimuleerd om in het kader van
de projecten samen te werken. Bij aanvang van een project kunnen zij zelf
onderling afstemmen welke taken zij zouden willen vervullen. Hierdoor kan de
ene student in een bepaald project andere vaardigheden verwerven als de andere
student. Door ervoor te zorgen dat rollen rouleren, wordt bewerkstelligd dat
studenten ook steeds verschillende vaardigheden leren, en niet alleen de dingen
doen die hen makkelijk afgaan.

\item [Ondersteunende vaklijnen--] Ter ondersteuning van de grotere, probleem-
en kennisgedreven projecten zijn er een aantal vaklijnen waarin typisch
vaardigheden centraal staan. Te denken valt hierbij bijvoorbeeld aan:
programmeervaardigheden, communicatieve vaardigheden, specificeervaardigheden
(Beweren \& Bewijzen; Formeel Denken). Door de ondersteunende lijnen te
verbinden aan de context van de lopende projecten, wordt de rol van deze
vaardigheden duidelijk.
\end{description}

Het is duidelijk dat een dergelijke insteek in een aantal (maar zeker niet in
alle) opzichten nogal afwijkt van de huidige inrichting van het onderwijs; het
is dan ook vooral bedoeld als schets van een ideaalbeeld.  Voor het toewerken
naar dit ideaalbeeld zal echter -- mede met het oog op de afstemming met de
opleiding informatica -- tevens rekening gehouden moeten worden met de huidige
inrichting van het onderwijs. Het stroomlijnen van de overgang van het huidige
onderwijs naar een beeld dat meer komt in de richting van het hierboven
genoemde, zal in hoofdstuk~\ref{h:waarbinnen-randvoorwaarden} meer expliciet
aan de orde worden gesteld.

%% file: waarmee-kwaliteit.tex
\chapter{Visie op kwaliteit}
\label{h:waarmee-kwaliteit}

De informatiekunde opleiding is op natuurlijke wijze ingebed binnen de algehele
kwaliteitszorg van het \NIII.  De visie op kwaliteit met betrekking tot de
informatiekunde opleiding is daarom volledig in samenspraak met het reeds door het
\NIII\ uitgezette beleid zoals dit is verwoord in een beleidsnota
kwaliteitszorg~\cite{KwaliteitNota}.  Om het onderhavige document een complete
weergave te laten zijn van de achtergronden en motivaties van het
informatiekunde curriculum, nemen we voor de informatiekunde belangrijkste
punten onverkort over uit deze beleidsnota. In paragraaf~\ref{p:speerpunten}
staan we stil bij de speerpunten zoals deze reeds zijn benoemd in het
\NIII\ kwaliteitsbeleid. 

Onderwijsactiviteiten op een universiteit bevinden zich in een spanningsveld
tussen onderzoek en praktijk.  De meeste docenten zullen naast het geven van
onderwijs actief zijn in onderzoek.  Daar ook voor docenten tijd kostbaar is,
zal dit tot een zeker spanningsveld leiden.  In paragraaf~\ref{p:onderwijszoek}
bespreken we kort de lijnen zoals die in de beleidsnota zijn uitgezet met
betrekking tot deze relatie.  Daarnaast wordt de inhoud van de informatiekunde
opleiding, en het bijbehorende onderzoek, sterk gedreven vanuit de behoeften
zoals deze uit de praktijk van het gebruik en de inzet van ICT naar voren
komen. Dit is wederom een spanningsveld, ditmaal tussen theorie en praktijk.
Het betekent dat het voor een informatiekunde opleiding, meer nog dan voor een
informatica opleiding, belangrijk is om zich te spiegelen aan de (langere
termijn!) behoefte vanuit die praktijk.  In paragraaf~\ref{p:praktijk}
bespreken we hoe we hier binnen de informatiekunde opleiding op hoofdlijnen mee
willen omgaan.

Aan het einde van dit hoofdstuk (paragraaf~\ref{p:kwaliteit-uit}) zijn de 
belangrijkste inrichtingsprincipes met betrekking tot de kwaliteit van de 
informatiekunde opleiding samengevat. Merk op dat we hierbij niet ingaan op
algemene \NIII\ brede aspecten van kwaliteitsbeheersing en ons dus puur
richten op de implicaties van de informatiekunde opleiding zelf.

\section{Speerpunten}
\label{p:speerpunten}
In de visie van het \NIII zijn dit essenti\"ele aspecten van een goede universitaire
opleiding:
\begin{enumerate}
  \item elke docent geeft goed onderwijs, 
  \item alle docenten geven samen goed onderwijs,
  \item de studenten en dit onderwijs passen goed bij elkaar en 
  \item docenten en studenten ervaren voor hun inspanningen ook de nodige waardering,
  zowel in het heden als in de toekomst.
\end{enumerate}
In lijn hiermee heeft het \NIII\ de volgende speerpunten benoemd ten aanzien van
kwaliteitsborging:
\begin{enumerate}
   \item Docenten geven goed onderwijs als ze er hart voor hebben en over de
   nodige vakinhoudelijke en didactische capaciteiten beschikken. Daarom is het
   wenselijk dat hun onderwijs in synergie geschiedt met hun onderzoek, dat ze
   daarbij kunnen profiteren van de ervaring van hun collega's, dat ze kunnen
   leren indien nodig en dat ze adequaat beloond worden. Daarbij moeten ze
   gestimuleerd worden een optimale combinatie van inhoud en onderwijsvorm te
   bereiken.
   \item Het onderwijs als geheel wordt goed als het zich ori\"enteert op een
   gefundeerde gemeenschappelijke visie op het wetenschapsgebied, het onderwijs
   en de beroepspraktijk, en de onderdelen goed op elkaar afgestemd zijn qua
   inhoud en studeerbaarheid. De uitvoering moet adequaat worden ondersteund
   door de onderwijsorganisatie.
   \item Het binnen halen van de juiste studenten vereist gedegen, eerlijke en
   uitdagende voorlichting. Enerzijds om zoveel mogelijk potenti\"ele
   ge\"{\i}nteresseerden te bereiken, anderzijds om uitval door verkeerd
   gewekte verwachtingen tegen te gaan. Bovendien moet een opleiding actief
   inspelen op ontwikkelingen in de instromende groep VWO-ers en HBO-studenten.
   \item Beslissers, leraren, ouders, aankomende studenten en de `man in the
   street' moeten een realistisch beeld krijgen van de informatica en de
   informatiekunde als wetenschap, en moeten informatici en informatiekundigen aan
   de universiteit en in het bedrijfsleven waarderen voor wat ze zijn en kunnen. 
   \item `Regelkringen' om kwaliteit van onderwijs en onderzoeken te borgen
   dienen te worden gestimuleerd. Dergelijke regelkringen bestaan uit
   professionals en hun primaire activiteiten, niet uit bureaucratie en
   meetinstrumenten. Dit systeem kan op zich al functioneren zonder dat veel
   tijd besteed moet worden aan definitiekwesties over het wezen van kwaliteit.
\end{enumerate}

\section{Onderzoek en onderwijs}
\label{p:onderwijszoek}
Universitair onderwijs en onderzoek zijn onafscheidelijk.  Bij de verdeling van
onderwijs- en onderzoekstaken moet ervoor worden gezorgd dat beide taken elkaar
zoveel mogelijk synergetisch kunnen versterken en elkaar niet in de weg zitten.
Idealiter inspireert het lopende onderzoek van een docent de eigen colleges,
terwijl de ervaringen uit practica wederom van invloed zijn op het onderzoek.
Daarom moet zo veel mogelijk worden vermeden dat een medewerker een college
moet geven dat vakinhoudelijk weinig te maken heeft met het lopende onderzoek
van de medewerker.

Een blijvend probleem is natuurlijk wel dat ook bij een goede synergie tussen
onderwijs en onderzoek, deze toch vaak blijven concurreren om de beperkte tijd
van de medewerker. Practica moeten worden voorbereid, werk van studenten moet
worden nagekeken, en dit laat zich niet uitstellen. Anderzijds moeten artikelen
worden geschreven of beoordeeld. Ook dit laat zich vaak niet uitstellen.
Medewerkers van een Universiteit worden traditioneel bijna uitsluitend
beoordeeld op hun onderzoeksprestaties, met name op het aantal publicaties,
terwijl er onderwijsprestaties nauwlijks of niet meetellen in de beoordeling.
Het resultaat is dan vaak dat het onderwijs het kind van de rekening is. Het
streven binnen het \NIII\ is daarom ook om onderwijskundige inspanningen, en de
resultaten hiervan, beter zichtbaar te maken en medewerkers \emph{ook} op deze
resultaten te beoordelen en te belonen.

\section{In dialoog met de praktijk}
\label{p:praktijk}
Zoals reeds eerder aangegeven wordt de inhoud van de informatiekunde opleiding,
en het bijbehorende onderzoek, sterk gedreven vanuit de behoeften zoals deze
uit de praktijk van het gebruik en de inzet van ICT naar voren
komen.  Dit betekent dat het voor een informatiekunde opleiding, meer nog dan
voor een informatica opleiding, belangrijk is om zich te spiegelen aan de
(langere termijn!) behoefte vanuit die praktijk. Binnen de informatiekunde
opleiding en het achterliggende onderzoek voorzien we vijf strategie\"en om
dit te bewerkstelligen:
\begin{enumerate}
   \item Waar relevant zal er gewerkt worden met gastdocenten die werkzaam zijn 
         in de praktijk om de theorie van de colleges te voorzien van accenten 
         uit de praktijk.

   \item Er zal een actief beleid worden gevoerd met betrekking tot het 
         aantrekken van bijzondere hoogleraren op gebieden die relevant zijn 
         voor de informatiekunde. Meer concreet wordt voorzien in bijzondere 
         hoogleraren op het gebied van:
         \begin{enumerate}
            \item Architectuurgedreven systeemontwikkeling.
            \item Aanbesteding systeemontwikkelingsprojecten.
            \item Management van systeemontwikkelingsprojecten.
         \end{enumerate}

   \item Waar mogelijk worden onderzoeksprojecten gemeenschappelijk met 
         vertegenwoordigers uit de praktijk uitgevoerd.

   \item Waar mogelijk zal bij opgaven/werkstukken met cases uit de praktijk 
         gewerkt worden. Denk hierbij zoals aan het evalueren van 
         praktijksituaties aan de hand van theorie\"en zoals deze in de 
         colleges zijn aangedragen.

   \item De kwaliteit van de opleiding en het onderzoek vanuit een academisch 
         perspectief wordt onder andere gewaarborgd door regelmatige visitaties 
         vanuit de (internationale) academische context.

         Hier willen we voor de informatiekunde opleiding en het informatiekunde 
         onderzoek een praktijkgericht perspectief naast plaatsen door een ``raad 
         van advies'' samen te stellen met vertegenwoordigers uit het werkveld.  
         Met andere woorden, vertegenwoordigers uit de vier beroepsrichtingen: 
         onderzoek, onderwijs, vakman en management. Hierbij wordt bijvoorbeeld 
         gedacht aan vertegenwoordigers van:
         \begin{enumerate}
            \item Zuster faculteiten van Universitaire instellingen.
            \item Zuster faculteiten van HBO instellingen.
            \item ICT dienstverlenende organisaties.
            \item Gebruikers van ICT.
         \end{enumerate}
	 Het ligt hierbij voor de hand dat een dergelijke verbinding wordt
	 opgezet binnen het kader van het Nederlands Architectuurforum (NAF).
	 Dit is een gremium waar het \NIII\ reeds in deelneemt, en dat een
	 overkoepelende organisatie vormt van leveranciers, gebruikers en
	 kennisinstellingen op het gebied van architectuur van
	 informatiesystemen.
\end{enumerate}

\section{Inrichtingsprincipes}
\label{p:kwaliteit-uit}
Als afronding van de bovenstaande bespreking van de kwaliteitsborging van de
informatiekunde opleiding, geven we hier de inrichtingsprincipes weer zoals die
uit deze bespreking voortvloeien. Merk wederom op dat we hierbij niet ingaan op
algemene \NIII\ brede aspecten van kwaliteitsbeheersing en ons dus puur richten
op de implicaties van de informatiekunde opleiding zelf.
\begin{description}
   \item[Synergie onderwijs \& onderzoek --] Er dient synergie nagestreefd te
   worden tussen onderwijs en onderzoek.  Concreet, dient het onderwijs moet
   gefundeerd zijn op een gemeenschappelijke visie op het wetenschapsgebied,
   het onderwijs en de beroepspraktijk, en de onderdelen goed op elkaar
   afgestemd zijn qua inhoud en studeerbaarheid.
   \item[Voorlichting --] 
   Potenti\"ele studenten, collega wetenschappers en beroepsbeoefenaars dienen over
   de opleiding te worden voorgelicht middels gedegen, eerlijke en uitdagende 
   voorlichting. Dit begint met het hebben van een duidelijke visie op het 
   vakgebied, en de rol die afgestudeerde informatiekundigen hierin kunnen
   vervullen.
   \item[Dialoog met de praktijk --] Er dient een dialoog te worden aangegaan
   met de praktijk, middels: gastdocenten, bijzondere hoogleraren,
   gemeenschappelijke onderzoeksprojecten, gebruik van praktijkcasussen in het
   onderwijs, en een raad van advies van vertegenwoordigers van de drie
   beroepsvelden.
\end{description}

%% file: brein.tex
\chapter{Informatieverwerking door linker- en rechterhersenhelft}
\label{h:brein}

(Onderstaande classificatie is ontleend aan 
\cite{Book:89:Edwards:DrawingontheRightSideoftheBrain}, p.\ 40.

\subsection*{\bf{L}-Mode}
\begin{description}
\item[Verbal:] Using words to name, describe, define. 
\item[Analytic:] Figuring things out step-by-step and part-by-part. 
\item[Symbolic:] Using a symbol to \emph{stand for} something. For example,
the drawn form {\%plaatje van oog} stands for \emph{eye}, the sign $+$
stands for the process of addition.
\item[Abstract:] Taking out a small bit of information and using it to
represent the whole thing.
\item[Temporal:] Keeping track of time, sequencing one thing after
another; doing first things first, second things second,  etc.
\item[Rational:] Drawing conclusions based on \emph{reason} and
\emph{facts}.
\item[Digital:] Using numbers as in counting.
\item[Logical:] Drawing conclusions based on logic; one thing following
another in logical order. For example: a mathematical theorem or a well-
stated argument.
\item[Linear:] Thinking in terms of linked ideas, one thought directly
following another, often leading to a convergent solution.
\end{description}

\subsection*{\emph{R}-Mode}
\begin{description}
\item[Nonverbal:] Awareness of things, but minimal connection with
words.
\item[Synthetic:] Putting things together to form wholes.
\item[Concrete:] Relating to things as they are, at the present moment.
\item[Analogic:] Seeing likenesses between things; understanding
metaphoric relationships.
\item[Nontemporal:] Without a sense of time.
\item[Nonrational:] Not requiring a basis of reason or facts;
willingness to suspend judgement.
\item[Spatial:] Seeing where things are in relation to other things, and
how parts go together to form a whole.
\item[Intuitive:] Making leaps of insight, often based on incomplete
patterns, hunches, feelings or visual images.
\item[Holistic:] Seeing whole things all at once; perceiving the overall
patterns and structures, often leasing to divergent conclusions.
\end{description}

%% file: main.bbl
\begin{thebibliography}{10}

\bibitem{Book:56:Ashby:Cybernetics}
W.R. Ashby.
\newblock {\em An Introduction to Cybernetics}.
\newblock Chapman \& Hall, London, England, 1956.

\bibitem{Blase}
K.~Blase, H.~Altijng, J.~Stenvers, I.~van Moorsel, en M.~Sluijter.
\newblock In {\em Leren met hoofd, hart en handen.} APS/Educare, Utrecht, 2002.

\bibitem{BS93}
M.~Boekaerts en P.~Simons.
\newblock {\em Leren en instructie: psychologie van de leerling en het
  leerproces}.
\newblock Dekker \& Van de Vegt, Assen, 1993.

\bibitem{Book:96:Buzan:MindMapBook}
T.~Buzan en B.~Buzan.
\newblock {\em The Mind-Map Book}.
\newblock Plume, paperpack reprint edition edition, 1996.

\bibitem{Book:98:Carter:HetBreininKaart}
R.~Carter.
\newblock {\em Het Brein in Kaart}.
\newblock Uniepers i.s.m. Segment B.V./Natuur en Techniek, Abcoude, 1998.

\bibitem{Boer93}
B.~de~Boer, F.~Reubel, R.~Reinards, en J.~van~der Sanden.
\newblock {\em Zelfstandig leren in beroepsopleidingen}.
\newblock Wolters Noordhoff, Groningen, 1993.

\bibitem{Book:90:deBono:LateralThinking}
E.~de~Bono.
\newblock {\em Lateral Thinking: Creativity Step-by-Step}.
\newblock Harper Collins, 1990.

\bibitem{Book:97:deBono:LeerUzelfDenken}
E.~de~Bono.
\newblock {\em Leer Uzelf Denken}.
\newblock Element Uitgevers, Naarden, tweede edition, 1997.

\bibitem{Book:89:Edwards:DrawingontheRightSideoftheBrain}
B.~Edwards.
\newblock {\em Drawing on the Right Side of the Brain}.
\newblock J.P. Tarcher, revised edition, 1989.

\bibitem{Book:99:Edwards:TheNewDrawingontheRightSideoftheBrain}
B.~Edwards.
\newblock {\em The New Drawing on the Right Side of the Brain}.
\newblock J.P. Tarcher, revised en expanded edition, 1999.

\bibitem{Elkind}
D.~Elkind.
\newblock {\em Kinderen en Adolescenten}.
\newblock Ambo, Baarn, tweede druk edition, 1970, 1974.

\bibitem{Book:93:Gardner:MultipleIntelligences}
H.E. Gardner.
\newblock {\em Multiple Intelligences: The Theory in Practice}.
\newblock Basic Books, New York, 1993.

\bibitem{Report:99:IEEE:Architecture}
IEEE Standards Department, The Architecture Working Group of the Software
  Engineering Committee.
\newblock {\em {Recommended Practice for Architectural Description of Software
  Intensive Systems}}, September 2000.
\newblock \url{http://www.ieee.org}.

\bibitem{InformaticaVerkenning}
Verkenningscommissie Informatica.
\newblock {\em Geen toekomst zonder Informatica -- Toekomstverkenning
  Informatica}, Juni 1996.

\bibitem{Curriculum2003}
Curriculumcommissie Informatiekunde.
\newblock {\em Informatiekunde -- Curriculum 2003}, 2003.

\bibitem{Kohn}
A.~Kohn.
\newblock {\em Punished by Rewards: The Trouble with Gold Stars, Incentive
  Plans, A's, Praise, en Other Bribes}.
\newblock Houghton Mifflin Co, paperback edition, 1973, 1999.

\bibitem{Book:02:Maier:ArtOfSystemArchitecture}
M.W. Maier en R.~Rechtin.
\newblock {\em The Art of System Architecting}.
\newblock CRC Press, Boca Raton, Florida, 2nd edition, 2002.

\bibitem{BeleidsbriefInternationaal}
Cultuur en~Wetenschappen Ministerie~van Onderwijs.
\newblock {\em Beleidsbrief Internationalisering van het Onderwijs}.
\newblock Zoetermeer, 1999.

\bibitem{KwaliteitNota}
NIII Onderwijsdirectie.
\newblock {\em Beleidsnota -- Kwaliteitszorg}, Januari 2002.

\bibitem{Pinto}
D.~Pinto.
\newblock {\em Interculturele Communicatie}.
\newblock Bohn Stafleu Van Loghum, Houten, tweede herziene druk edition, 1994.

\bibitem{Book:91:Rechtin:SystemArchitecture}
E.~Rechtin.
\newblock {\em Systems architecting: creating en building complex systems}.
\newblock Prentice-Hall PTR, Upper Saddle River, New Jersey, 1991.

\bibitem{rogers}
C.R. Rogers.
\newblock {\em Leren in Vrijheid}.
\newblock De Toorts, Haarlem, tiende druk, 1993 edition, 1973/1993.

\bibitem{Sperry}
R.W. Sperry.
\newblock Lateral specialization of cerebral function in the surgically
  separated hemispheres.
\newblock In {\em The Psychophysiology of Thinking}, pages 209--229. Academic
  Press, New York, 1973.

\bibitem{kindzijn}
J.B. Thomson, T.~Kahn, M.~Masheder, L.~Oldfield, M.~Gl{\"o}ckler, and
  R.~Meighan.
\newblock {\em Gewoon kind zijn}.
\newblock Christofoor, Zeist, 1998.

\bibitem{Book:98:Brandhof:GebruikjeHersens}
J.-W. van~den Brandhof.
\newblock {\em Gebruik je hersens}.
\newblock Uitgeverij Verba, Hoevelaken, 1998.

\bibitem{Rijswijk92}
F.~van Rijswijk en J.~van~der Sanden.
\newblock {\em Leren kun je leren: didactische uitgangspunten voor de
  verbetering van het zelfstandig leervermogen}.
\newblock SVE, Amersfoort, 1992.

\bibitem{Vermunt92}
J.D.H.M. Vermunt.
\newblock {\em Leerstijlen en sturen van leerprocessen in het hoger onderwijs:
  naar procesgerichte instructie in zelfstandig denken}.
\newblock Swets en Zeitlinger, Amsterdam, 1992.

\bibitem{ZelfstudieInformatica}
Nijmeegs~Instituut voor Informatica \&~Informatiekunde.
\newblock {\em Zelfstudie Informatica}.
\newblock Nijmegen, 2000.

\end{thebibliography}
